\documentclass {book}
\usepackage {latexsym, amsmath}
\usepackage{amssymb}
\usepackage{makeidx}
\makeindex

\begin{document}

\title {Neutrality and Many-Valued Logics}
\author {Schumann A., Smarandache F.}
\pagestyle {myheadings}

\begin {titlepage}

\vspace {30 pt} \begin {center} \LARGE Andrew Schumann \&
Florentin Smarandache
 \end {center}

\begin {center}
\vspace {24 pt} \LARGE{\textbf{Neutrality and Many-Valued Logics}}
\end {center}
\begin {center}
\vspace {470 pt} July, 2007
\end {center}

\end {titlepage}

\chapter*{Preamble}

This book written by A. \textsc{Schumann} $\&$ F.
\textsc{Smarandache} is devoted to advances of non-Archimedean
multiple-validity idea and its applications to logical reasoning.
\textsc{Leibnitz} was the first who proposed \textsc{Archimedes}'
axiom to be rejected. He postulated infinitesimals (infinitely
small numbers) of the unit interval $[0,1]$ which are larger than
zero, but smaller than each positive real number.
\textsc{Robinson} applied this idea into modern mathematics in
\cite{Robin} and developed so-called non-standard analysis. In the
framework of non-standard analysis there were obtained many
interesting results examined in \cite{Cutland}, \cite{Davis},
\cite{Hurd}, \cite{Robin}.\\

There exists also a different version of mathematical analysis in
that \textsc{Archi\-medes}' axiom is rejected, namely, $p$-adic
analysis (e.g., see:  \cite{Bach}, \cite{Kobl}, \cite{Mahler},
\cite{Rober}). In this analysis, one investigates the properties
of the completion of the field $\mathbf{Q}$ of rational numbers
with respect to the metric $\rho_p(x, y) = |x - y|_p$, where the
norm $|\cdot|_p$ called $p$-adic is defined as follows:

\begin{itemize}
    \item $|y|_p=0\leftrightarrow y =0$,
    \item $| x\cdot y|_p=|x|_p \cdot |y|_p$,
    \item $|x+y|_p\leqslant\max(|x|_p, |y|_p)$ (non-Archimedean triangular inequality).
\end{itemize}

That metric over the field $\mathbf{Q}$ is non-Archimedean,
because $|n\cdot 1|_p\leqslant 1$ for all $n\in \mathbf{Z}$. This
completion of the field $\mathbf{Q}$ is called the field
$\mathbf{Q}_p$ of $p$-adic numbers. In
$\mathbf{Q}_p$ there are infinitely large integers.\\

Nowadays there exist various many-valued logical systems (e.g.,
see \textsc{Malinowski}'s book \cite{mal}). However,
non-Archimedean and $p$-adic logical multiple-validities were not
yet systematically regarded. In this book, \textsc{Schumann} $\&$
\textsc{Smarandache} define such multiple-validities and describe
the basic properties of non-Archimedean and $p$-adic valued
logical systems proposed by them in \cite{schu1}, \cite{schu2},
\cite{schu3}, \cite{schu4}, \cite{schu7}, \cite{Smaran2},
\cite{Smaran3}. At the same time, non-Archimedean valued logics
are constructed on the base of t-norm approach as fuzzy ones and
$p$-adic
valued logics as discrete multi-valued systems.\\

Let us remember that the first logical multiple-valued system is
proposed by the Polish logician Jan \textsc{{\L}ukasiewicz} in
\cite{Lukas2}. For the first time he spoke about the idea of
logical many-validity at Warsaw University on 7 March 1918
(\textit{Wyk{\l}ad po\.{z}egnalny wyg{\l}oszony w auli
Uniwersytetu Warszawskiego w dniu 7 marca 1918 r}., page 2).
However \textsc{{\L}ukasiewicz} thought already about such a logic
and rejection of the Aristotelian principle of contradiction in
1910 (\textit{O zasadzie sprzeczno\`{s}ci u Arystotelesa},
Krak\'{o}w 1910). Creating many-valued logic,
\textsc{{\L}ukasiewicz} was inspired philosophically. In the
meantime, \textsc{Post} designed his many-valued logic in 1921 in
\cite{Post} independently and for
combinatorial reasons as a generalization of Boolean algebra.\\

The logical multi-validity that runs the unit interval $[0,1]$
with infinitely small numbers for the first time was proposed  by
\textsc{Smarandache} in \cite{Smaran2}, \cite{Smaran3},
\cite{Smaran4}, \cite{Smar1}, \cite{Smar1a}. The neutrosophic
logic, as he named it, is conceived for a philosophical
explication of the neutrality concept. In this book, it is shown
that neutrosophic logic is a generalization of non-Archimedean and
$p$-adic valued
logical systems.\\

In this book non-Archimedean and $p$-adic multiple-validities idea
is regarded as one of possible approaches to explicate the
neutrality concept.\\

\hfill K. Trz\c{e}sicki

\hfill Bia{\l}ystok, Poland

\chapter*{Preface}

In this book, we consider various many-valued logics: standard,
linear, hyperbolic, parabo\-lic, non-Archimedean, $p$-adic,
interval, neutrosophic, etc. We survey also results which show the
tree different proof-theoretic frameworks for many-valued logics,
e.g.\ frameworks of the following deductive calculi:
\textsc{Hilbert}'s style, sequent, and hypersequent. Recall that
hypersequents are a natural generalization of \textsc{Gent\-zen}'s
style sequents that was introduced independently by \textsc{Avron}
and \textsc{Pottinger}. In particular, we examine
\textsc{Hilbert}'s style, sequent, and hypersequent calculi for
infinite-valued logics based on the three fundamental continuous
t-norms: \textsc{{\L}ukasiewicz}'s, \textsc{G\"{o}del}'s, and
Product
logics.\\

We present a general way that allows to construct systematically
analytic calculi for a large family of non-Archimedean many-valued
logics: hyperrational-valued, hyperreal-valued, and $p$-adic
valued logics characterized by a special format of semantics with
an appropriate rejection of \textsc{Archimedes}' axiom. These
logics are built as different extensions of standard many-valued
logics (namely, \textsc{{\L}ukasiewicz}'s, \textsc{G\"{o}del}'s,
Product, and
\textsc{Post}'s logics).\\

The informal sense of \textsc{Archimedes}' axiom is that anything
can be measured by a ruler. Also logical multiple-validity without
\textsc{Archimedes}' axiom consists in that the set of truth
values is infinite and it
is not well-founded and well-ordered.\\

We consider two cases of non-Archimedean multi-valued logics: the
first with many-validity in the interval $[0,1]$ of hypernumbers
and the second with many-validity in the ring $\mathbf{Z}_p$ of
$p$-adic integers. Notice that in the second case we set discrete
infinite-valued logics. The following logics are investigated:

\begin{itemize}
    \item hyperrational valued \textsc{{\L}u\-kasiewicz}'s,
\textsc{G\"{o}\-del}'s, and Product logics,
    \item hyperreal valued
\textsc{{\L}u\-kasiewicz}'s, \textsc{G\"{o}\-del}'s, and Product
logics,
    \item $p$-adic valued \textsc{{\L}u\-kasie\-wicz}'s,
\textsc{G\"{o}\-del}'s, and \textsc{Post}'s logics.
\end{itemize}

In \cite{Haj} \textsc{H\'{a}jek} classifies truth-functional fuzzy
logics as logics whose conjunction and implication are interpreted
via continuous t-norms and their residua. Fundamental logics for
this classification are \textsc{{\L}ukasiewicz}'s,
\textsc{G\"{o}del}'s and Product ones. Further, \textsc{H\'{a}jek}
proposes basic fuzzy logic $BL$ which has validity in all logics
based on continuous t-norms. In this book, for the first time we
consider hypervalued and $p$-adic valued extensions of basic
fuzzy logic $BL$.\\

On the base of non-Archimedean valued logics, we construct
non-Archimede\-an valued interval neutrosophic logic
$\mathbf{INL}$ by which we can describe neutrality phenomena. This
logic is obtained by adding to the truth valuation a truth triple
$t$, $i$, $f$ instead of one truth value $t$, where $t$ is a
truth-degree, $i$ is an indeterminacy-degree, and $f$ is a
falsity-degree. Each parameter of this triple runs either the unit
interval $[0,1]$ of hypernumbers or the ring $\mathbf{Z}_p$ of
$p$-adic integers.\\

\hfill A. Schumann $\&$ F. Smarandache

\tableofcontents

\chapter{Introduction}

\section{Neutrality concept in logic}

Every point of view $A$ tends to be neutralized, diminished,
balanced by \textsl{Non}-$A$. At the same time, in between $A$ and
\textsl{Non}-$A$ there are infinitely many points of view
\textsl{Neut}-$A$. Let's note by $A$ an idea, or proposition,
theory, event, concept, entity, by \textsl{Non}-$A$ what is not
$A$, and by \textsl{Anti}-$A$ the opposite of $A$.
\textsl{Neut}-$A$ means what is neither $A$ nor \textsl{Anti}-$A$,
i.e. neutrality\index{neutrality} is also in between the two
extremes.\\

The classical logic\index{classical logic}, also called Boolean
logic by the name of British mathematician G. \textsc{Boole}, is
two-valued. Thus, neutralities are ignored in this logic.
\textsc{Peirce}, before 1910, developed a semantics for
three-valued logic in an unpublished note, but \textsc{Post}'s
dissertation \cite{Post} and \textsc{{\L}ukasiewicz}'s work
\cite{Lukas2} are cited for originating the three-valued logic.
Here 1 is used for truth, $\frac{1}{2}$ for indeterminacy, and 0
for falsehood. These truth values can be understood as $A$,
\textsl{Neut}-$A$, \textsl{Non}-$A$ respectively. For example, in
three valued \textsc{{\L}ukasiewicz}'s logic the negation of
$\frac{1}{2}$ gives $\frac{1}{2}$ again, i.e. the neutrality
negation is the neutrality again.\\

However, we can consider neutralities as degrees between truth and
falsehood. In this case we must set multiple-valued logics.\\

The $n$-valued logics were developed by \textsc{{\L}ukasiewicz} in
\cite{Lukas1}, \cite{Lukas2}. The practical applications of
infinite valued logic, where the truth-value may be any number in
the closed unit interval $[0, 1]$, are regarded by \textsc{Zadeh}
in \cite{Zadeh1}. This logic is called fuzzy one.\\

In the meantime, the ancient Indian logic (ny\={a}ya) considered
four possible values of a statement: `true (only)', `false
(only)', `both true and false', and `neither true nor false'. The
Buddhist
logic added a fifth value to the previous ones, `none of these'.\\

As we see, we can get the neutralities\index{neutrality} in the
framework of many-valued logics. There is also an other way, when
we set neutralities as the main property of two-valued logical
calculi. Namely, it is possible to develop systems where the
principle of classical logic, which entails that from
contradictory premises any formula can be derived, in symbols:
$\alpha\wedge\neg\alpha\vdash\beta$, is violated. It is called the
\textsc{Duns Scotus} law\index{\textsc{Duns Scotus} law}, which is
valid not only in classical logic, but in almost all the known
logical systems, like intuicionistic logic.\\

For the first time the Russian logician N. \textsc{Vasil'\`{e}v}
proposed in \cite{Vasil'ev1} and \cite{Vasil'ev2} to violate the
\textsc{Duns Scotus} law, who perceived that the rejection of the
law of non-contradiction could lead to a non-Aristotelian logic in
the same way as the violation of the parallel postulate of
Euclidean geometry had conduced to non-Euclidean geometry.\\

The other logician who discussed the possibility of violating the
ancient Aristotelian principle of contradiction was the Polish
logician J. \textsc{{\L}ukasiewicz}, but he did not elaborate any
logical system to cope with his intuitions. His idea was developed
by S. \textsc{Ja\`{s}kowski} in \cite{Jask}, who constructed a
system of propositional paraconsistent logic, where he
distinguished between contradictory (inconsistent) systems and
trivial ones.\\

\textsc{Vasil'\`{e}v} and \textsc{{\L}ukasiewicz} were the
forerunners of paraconsistent logic in which it is devoted to the
study of logical systems which can base on inconsistent theories
(i.e., theories which have contradictory theses, like $\alpha$ and
$\neg\alpha$) but which are not trivial (in the sense that not
every well formed formula of their languages are also axioms).\\

In this book we will investigate so-called non-Archimedean
multiple-valued logics, especially non-Archimedean valued fuzzy
logics, in which we have the violation the \textsc{Duns Scotus}
law too.

\section{Neutrality and non-Archimedean logical mul\-tiple-validity}

The development of fuzzy logic and fuzziness was motivated in
large measure by the need for a conceptual framework which can
address the issue of uncertainty\index{uncertainty} and lexical
imprecision. Recall that fuzzy logic was introduced by Lotfi
\textsc{Zadeh} in 1965 (see \cite{Zadeh1}) to represent data and
information possessing nonstatistical uncertainties. Florentin
\textsc{Smarandache} had generalized fuzzy logic\index{generalized
fuzzy logic} and introduced two new concepts (see \cite{Smaran2},
\cite{Smaran3}, \cite{Smaran4}):

\begin{enumerate}
\item  neutrosophy\index{neutrosophy}  as study of neutralities;
\item  neutrosophic logic  and  neutrosophic
probability\index{neutrosophic probability}  as a mathematical
model of uncertainty\index{uncertainty}, vagueness, ambiguity,
imprecision, undefined, unknown,
incompleteness\index{incompleteness},
inconsistency\index{inconsistency}, redundancy,
contradiction\index{contradiction}, etc.
\end{enumerate}

Neutrosophy\index{neutrosophy} proposed by \textsc{Smarandache} in
\cite{Smaran4} is a new branch of philosophy, which studies the
nature of neutralities, as well as their logical applications.
This branch represents a version of paradoxism studies. The
essence of paradoxism studies is that there is a neutrality for
any two extremes. For example, denote by $A$ an idea (or
proposition, event, concept), by \textsl{Anti}-$A$ the opposite to
$A$. Then there exists a neutrality \textsl{Neut}-$ A$ and this
means that something is neither $A$ nor \textsl{Anti}-$ A$. It is
readily seen that the paradoxical reasoning can be modelled if
some elements $\theta_i$ of a frame $\Theta$ are not exclusive,
but non-exclusive, i.e., here $\theta_i$ have a non-empty
intersection. A mathematical model that has such a property is
called the \textsc{Dezert-Smarandache}
model\index{\textsc{Dezert-Smarandache} model} (DSm model). A
theory of plausible and paradoxical reasoning that studies DSm
models\index{DSm models} is called the \textsc{Dezert-Smarandache}
theory (DSmT)\index{\textsc{Dezert-Smarandache} theory (DSmT)}. It
is totally different from those of all existing approaches
managing uncertainties and fuzziness. In this book, we consider
plausible reasoning on the base of particular case of infinite DSm
models\index{DSm models}\index{\textsc{Dezert-Smarandache} model},
namely, on the base of non-Archimedean structures\index{non-Archimedean structure}.\\

Let us remember that \textsc{Archimedes}'
axiom\index{\textsc{Archimedes}' axiom} is the formula of infinite
length that has one of two following notations:
\begin{itemize}
\item for any $\varepsilon$ that belongs to the interval $[0,1]$,
we have
\begin{equation}
(\varepsilon > 0) \rightarrow [(\varepsilon \geq 1) \vee
(\varepsilon + \varepsilon \geq 1) \vee (\varepsilon + \varepsilon
+ \varepsilon \geq1) \vee \ldots ], \label{SchumannEq1}
\end{equation}
\item  for any positive integer $\varepsilon$, we have
\begin{equation}
[(1 \geq \varepsilon) \vee (1 + 1 \geq \varepsilon) \vee (1 + 1 +
1\geq \varepsilon) \vee \ldots ]. \label{SchumannEq2}
\end{equation}
\end{itemize}

Formulas \eqref{SchumannEq1} and \eqref{SchumannEq2} are valid in
the field $\mathbf{Q}$ of rational numbers and as well as in the
field $\mathbf{R}$ of real numbers. In the ring $\mathbf{Z}$ of
integers, only formula \eqref{SchumannEq2} has a nontrivial sense,
because $\mathbf{Z}$ doesn't contain numbers of the open
interval $(0,1)$.\\

Also, \textsc{Archimedes}' axiom\index{\textsc{Archimedes}' axiom}
affirms the existence of an integer multiple of the smaller of two
numbers which exceeds the greater: for any positive real or
rational number $\varepsilon$, there exists a positive integer $n$
such that $\varepsilon \geq \frac{1}{n}$ or
$n\cdot \varepsilon\geq 1$.\\

The negation of \textsc{Archimedes}' axiom has one of two
following forms:

\begin{itemize}
\item there exists $\varepsilon$ that belongs to the interval
$[0,1]$ such that
\begin{equation}
(\varepsilon > 0) \wedge [(\varepsilon < 1) \wedge (\varepsilon +
\varepsilon < 1) \wedge (\varepsilon + \varepsilon + \varepsilon <
1) \wedge \ldots ], \label{SchumannEq3}
\end{equation}
\item there exists a positive integer $\varepsilon$ such that
\begin{equation}
[(1 < \varepsilon) \wedge (1 + 1 < \varepsilon) \wedge (1 + 1 + 1
< \varepsilon) \wedge \ldots ]. \label{SchumannEq4}
\end{equation}
\end{itemize}
Let us show that \eqref{SchumannEq3} is the negation of
\eqref{SchumannEq1}. Indeed,

\begin{align*}
\neg\forall\varepsilon\,[(\varepsilon > 0) \rightarrow
[(\varepsilon \geq 1) \vee (\varepsilon + \varepsilon \geq 1) \vee
(\varepsilon +
\varepsilon + \varepsilon \geq1) \vee \ldots ]]& \leftrightarrow \\
\exists\varepsilon\,\neg \neg[(\varepsilon > 0) \wedge \neg
[(\varepsilon \geq 1) \vee (\varepsilon + \varepsilon \geq 1) \vee
(\varepsilon +
\varepsilon + \varepsilon \geq1) \vee \ldots ]]& \leftrightarrow \\
\exists\varepsilon\,[(\varepsilon > 0) \wedge [\neg(\varepsilon
\geq 1) \wedge \neg (\varepsilon + \varepsilon \geq 1) \wedge \neg
(\varepsilon + \varepsilon + \varepsilon \geq1) \wedge \ldots ]]&
\leftrightarrow
\\
\exists\varepsilon\,[(\varepsilon > 0) \wedge [(\varepsilon < 1)
\wedge (\varepsilon + \varepsilon < 1) \wedge (\varepsilon +
\varepsilon + \varepsilon < 1) \wedge \ldots ]]&
\end{align*}

It is obvious that formula  \eqref{SchumannEq3} says that there
exist \textit{infinitely small numbers} (or
\textit{infinitesimals}\index{infinitesimals}), i.e., numbers that
are smaller than all real or rational numbers of the open interval
$(0,1)$. In other words, $\varepsilon$ is said to be an
infinitesimal if and only if, for all positive integers $n$, we
have $|\varepsilon| < \frac{1}{n}$. Further, formula
\eqref{SchumannEq4} says that there exist \textit{infinitely large
integers}\index{infinitely large integers} that are greater than
all positive integers. Infinitesimals and infinitely large
integers are called \textit{nonstandard numbers}\index{nonstandard
numbers} or \textit{actual infinities}\index{actual
infinities}.\\

The field that satisfies all properties of $\mathbf{R}$ without
\textsc{Archimedes}' axiom is called the field of {\it{hyperreal
numbers}}\index{hyperreal numbers} and it is denoted by
${}^*\mathbf{R}$. The field that satisfies all properties of
$\mathbf{Q}$ without \textsc{Archimedes}' axiom is called the
field of hyperrational numbers\index{hyperrational numbers} and it
is denoted by ${}^*\mathbf{Q}$. By definition of field, if
$\varepsilon \in \mathbf{R}$ (respectively $\varepsilon \in
\mathbf{Q}$), then $1/\varepsilon\in \mathbf{R}$ (respectively
$1/\varepsilon \in \mathbf{Q}$). Therefore ${}^*\mathbf{R}$ and
${}^*\mathbf{Q}$ contain simultaneously infinitesimals and
infinitely large integers: for an infinitesimal $\varepsilon$, we
have $N = \frac{1}{\varepsilon}$, where $N$ is an infinitely large
integer.\\

The ring that satisfies all properties of $\mathbf{Z}$ without
\textsc{Archimedes}' axiom is called the ring of hyperintegers and
it is denoted by ${}^*\mathbf{Z}$. This ring includes infinitely
large integers. Notice that there exists a version of
${}^*\mathbf{Z}$ that is called the ring of $p$-adic
integers\index{$p$-adic integers} and is denoted by
$\mathbf{Z}_p$.\\

We will show that nonstandard numbers (actual infinities) are
non-exclusive elements\index{non-exclusive elements}. This means
that their intersection isn't empty with some other elements.
Therefore non-Archimedean structures\index{non-Archimedean
structure} of the form ${}^*\mathbf{S}$ (where we obtain
${}^*\mathbf{S}$ on the base of the set $\mathbf{S}$ of exclusive
elements) are particular case of the DSm model\index{DSm
models}\index{\textsc{Dezert-Smarandache} model}. These structures
satisfy the properties:

\begin{enumerate}
  \item all members of $\mathbf{S}$ are exclusive and $\mathbf{S}\subset
  {}^*\mathbf{S}$,
  \item all members of ${}^*\mathbf{S}\backslash\mathbf{S}$ are non-exclusive,
  \item if a member $a$ is non-exclusive, then there exists an exclusive
  member $b$ such that $a \cap b \neq \emptyset$,
  \item there exist non-exclusive members $a$, $b$ such that $a \cap b \neq \emptyset$,
  \item each positive non-exclusive member is greater (or less) than each
  positive exclusive member.
\end{enumerate}

We will consider the following principal versions of the logic on
non-Archi\-me\-dean structures\index{non-Archimedean structure}:

\begin{itemize}
    \item hyperrational valued \textsc{{\L}u\-kasiewicz}'s,
\textsc{G\"{o}\-del}'s, and Product logics\index{hyperrational
valued \textsc{{\L}u\-kasiewicz}'s logic}\index{hyperrational
valued \textsc{G\"{o}\-del}'s logic}\index{hyperrational valued
Product logic},
    \item hyperreal valued
\textsc{{\L}u\-kasiewicz}'s\index{hyperreal valued
\textsc{{\L}u\-kasiewicz}'s logic},
\textsc{G\"{o}\-del}'s\index{hyperreal valued
\textsc{G\"{o}\-del}'s logic}, and Product logics\index{hyperreal
valued Product logic},
    \item $p$-adic valued \textsc{{\L}u\-kasie\-wicz}'s\index{$p$-adic
valued  \textsc{{\L}u\-kasiewicz}'s logic},
\textsc{G\"{o}\-del}'s\index{$p$-adic valued
\textsc{G\"{o}\-del}'s logic}, and \textsc{Post}'s
logics\index{$p$-adic valued \textsc{Post}'s logic}.
\end{itemize}

For the first time non-Archimedean logical multiple-validities
were regarded by \textsc{Schumann} in \cite{schu1}, \cite{schu2},
\cite{schu3}, \cite{schu4}, \cite{schu7}, \cite{schu8}.\\

The non-Archimedean structures are not well-founded and
well-ordered. Also, \textbf{the logical neutrality may be examined
as non-well-ordered multiple validity, i.e.\ as non-Archimedean
one}. Two elements are neutral\index{neutrality} in a true sense
if both are incompatible by the ordering relation.

\section{Neutrality and neutrosophic logic}

The non-Archimedean valued logic can be generalized to a
\textit{neutrosophic logic}\index{neutrosophic logic}, where the
truth values of ${}^*[0, 1]$ are extended to truth triples of the
form $\langle t, i, f\rangle\subseteq ({}^*[0, 1])^3$, where $t$
is the truth-degree, $i$ the indeterminacy-degree, $f$ the
falsity-degree and they are approximated by non-standard subsets
of ${}^*[0, 1]$, and these subsets may overlap and exceed the unit
interval in the sense of
the non-Archimedean analysis.\\

Neutrosophic logic\index{neutrosophic logic} was introduced by
\textsc{Smarandache} in \cite{Smaran2}, \cite{Smaran3}. It is an
alternative to the existing logics, because it represents a
mathematical model of uncertainty on non-Archimedean
structures\index{non-Archimedean structure}. It is a non-classical
logic in which each proposition is estimated to have the
percentage of truth in a subset $t\subseteq{}^*[0, 1]$, the
percentage of indeterminacy in a subset $i\subseteq{}^*[0, 1]$,
and the percentage of falsity in a subset $f\subseteq{}^*[0, 1]$.
Thus, neutrosophic logic is a formal frame trying to measure the
truth, indeterminacy, and falsehood simultaneously, therefore it
generalizes:

\begin{itemize}
    \item Boolean logic ($i=\emptyset$, $t$ and $f$ consist of either 0 or
1);
    \item $n$-valued logic ($i=\emptyset$, $t$ and $f$ consist of members
    0, 1, \dots, $n-1$);
    \item fuzzy logic ($i=\emptyset$, $t$ and $f$ consist of members
     of $[0,1]$).
\end{itemize}

In simple neutrosophic logic\index{neutrosophic logic}, where $t$,
$i$, $f$ are singletons, the tautologies have the truth value
$\langle {}^*1, {}^*0, {}^*0\rangle$, the contradictions the value
$\langle {}^*0, {}^*1, {}^*1\rangle$. While for a paradox, we have
the truth value $\langle {}^*1, {}^*1, {}^*1\rangle$. Indeed, the
paradox is the only proposition true and false in the same time in
the same world, and indeterminate as well! We can assume that some
statements are indeterminate in all possible worlds, i.e. that
there exists ``absolute
indeterminacy'' $\langle {}^*1, {}^*1, {}^*1\rangle$.\\

The idea of tripartition (truth, indeterminacy, falsehood)
appeared in 1764 when J. H. \textsc{Lambert} investigated the
credibility of one witness affected by the contrary testimony of
another.  He generalized \textsc{Hooper}'s rule of combination of
evidence (1680s), which was a non-Bayesian approach to find a
probabilistic model. \textsc{Koopman} in 1940s introduced the
notions of lower and upper probability, followed by \textsc{Good},
and \textsc{Dempster} (1967) gave a rule of combining two
arguments. \textsc{Shafer} (1976) extended it to the
\textsc{Dempster-Shafer} theory of belief functions by defining
the belief and plausibility functions and using the rule of
inference of \textsc{Dempster} for combining two evidences
proceeding from two different sources. Belief function is a
connection between fuzzy reasoning and probability.\\

The \textsc{Dempster-Shafer} theory of belief functions is a
generalization of the Bayesian probability (\textsc{Bayes} 1760s,
\textsc{Laplace} 1780s); this uses the mathematical probability in
a more general way, and is based on probabilistic combination of
evidence in artificial intelligence. In \textsc{Lambert}'s
opinion, ``there is a chance $p$ that the witness will be faithful
and accurate, a chance $q$ that he will be mendacious, and a
chance $1-p-q$ that he will simply be careless''. Therefore we
have three components: accurate, mendacious, careless, which add
up to 1.\\

\textsc{Atanassov} (see \cite{Atanass1}, \cite{Atanass2}) used the
tripartition to give five generalizations of the fuzzy set,
studied their properties and applications to the neural networks
in medicine:

\begin{itemize}
    \item \textit{Intuitionistic Fuzzy Set}\index{intuitionistic fuzzy set} (IFS): given an universe $U$, an IFS
    $A$ over $U$ is a set of ordered triples $\langle$universe element, degree of
membership $M$, degree of non-membership $N\rangle$ such that
$M+N\leqslant 1$ and $M, N \in [0, 1]$.  When $M + N = 1$ one
obtains the fuzzy set, and if $M + N < 1$ there is an
indeterminacy $I = 1-M-N$.
    \item \textit{Intuitionistic $\mathfrak L$-Fuzzy Set}\index{intuitionistic $\mathfrak L$-fuzzy set} (ILFS):
Is similar to IFS, but $M$ and $N$ belong to a fixed lattice
$\mathfrak L$.
    \item \textit{Interval-valued Intuitionistic Fuzzy Set}\index{interval-valued intuitionistic fuzzy set} (IVIFS):
Is similar to IFS, but $M$ and $N$ are subsets of $[0, 1]$ and
$\max M + \max N \leqslant 1$.
    \item \textit{Intuitionistic Fuzzy Set of Second Type} (IFS2):
Is similar to IFS, but $M^2 + N^2 \leqslant 1$. $M$ and $N$ are
inside of the upper right quarter of unit circle.
    \item \textit{Temporal IFS}:
Is similar to IFS, but $M$ and $N$ are functions of the
time-moment too.
\end{itemize}

However, sometimes a too large  generalization may have no
practical impact.  Such unification theories, or attempts, have
been done in the history of sciences. \textsc{Einstein} tried in
physics to build a Unifying Field Theory that seeks to unite the
properties of gravitational, electromagnetic, weak, and strong
interactions so that a single set of equations can be used to
predict all their characteristics; whether such a theory may be
developed it is not known at the present.\\

\textsc{Dezert} suggested to develop practical applications of
neutrosophic logic (see \cite{Smar1}, \cite{Smar1a}), e.g.\ for
solving certain practical problems posed in the domain of research
in Data/Information fusion.

\chapter{First-order logical language}
\section{Preliminaries}
Let us remember some basic logical definitions.
\newtheorem{definitions} {Definition}
\begin {definitions} A first-order logical language\index{first-order logical language} $\mathcal{L}$
consists of the following symbols:
\begin{enumerate}
    \item Variables: \emph{(i)} Free variables\index{free variables}: $a_0, a_1, a_2, \ldots , a_j , \ldots$
($j \in \omega$). \emph{(ii)} Bound variables\index{bound
variables}: $x_0, x_1, x_2, \ldots , x_j , \ldots$ ($j \in
\omega$) \item Constants: \emph{(i)} Function
symbols\index{function symbols} of arity $i$ ($i \in \omega$):
$F^i_0 , F^i_1 , F^i_2 , \ldots , F^i_ j , \ldots$ ($j \in \omega
$). Nullary function symbols are called individual constants.
\emph{(ii)} Predicate symbols\index{predicate symbols} of arity
$i$ ($i \in \omega$): $P^i_0 , P^i_1 , P^i_2 , \ldots , P^i_ j ,
\ldots$ ($j \in \omega $). Nullary predicate symbols are called
truth constants.\item Logical symbols: \emph{(i)} Propositional
connectives\index{propositional connectives} of arity $n_j$ :
$\square ^{n_0}_0 $, $\square ^{n_1}_1$, \dots, $\square
^{n_r}_r$. \emph{(ii)} Quantifiers\index{quantifiers}:
$\mathrm{Q}_0, \mathrm{Q}_1, \ldots , \mathrm{Q}_q$. \item
Auxiliary symbols: $($, $)$, and , (comma).\end{enumerate}
\end {definitions}

\textit{Terms}\index{terms} are inductively defined as follows:

\begin{enumerate}
    \item Every individual constant is a
term. \item Every free variable is a term. \item If $F^n$ is a
function symbol of arity $n$, and $t_1, \ldots , t_n$ are terms,
then $F^n(t_1,\ldots, t_n)$ is a term.
\end{enumerate}

\textit{Formulas}\index{formulas} are inductively defined as
follows:

\begin{enumerate}
    \item If $P^n$ is a predicate symbol of arity $n$, and $t_1, \ldots
, t_n$ are terms, then $P^n(t_1, \ldots , t_n)$ is a formula. It
is called atomic or an atom. It has no outermost logical symbol.
    \item If $\varphi_1, \varphi_2, \ldots , \varphi_n$ are formulas  and
$\square ^n$ is a propositional connective of arity $n$, then
$\square ^n(\varphi_1,\varphi_2,\ldots,\varphi_n)$ is a formula
with outermost logical symbol $\square ^n$.
    \item If $\varphi$ is a formula not containing the bound
variable $x$, $a$ is a free variable and $\mathrm{Q}$ is a
quantifier, then $\mathrm{Q}x \varphi(x)$, where $\varphi(x)$ is
obtained from $\varphi$ by replacing $a$ by $x$ at every
occurrence of $a$ in $\varphi$, is a formula. Its outermost
logical symbol is $\mathrm{Q}$.
\end{enumerate}

A formula is called \textit{open} if it contains free variables,
and \textit{closed} otherwise. A formula without quantifiers is
called \textit{quantifier-free}. We denote the set of formulas of
a language $\mathcal{L}$ by $L$.\\

We will write $\varphi(x)$ for a formula possibly containing the
bound variable $x$, and $\varphi(a)$ (resp. $\varphi(t)$) for the
formula obtained from $\varphi$ by replacing every occurrence of
the variable $x$ by the free variable $a$ (resp. the term $t$).\\

Hence, we shall need meta-variables for the symbols of a language
$\mathcal{L}$. As a notational convention we use letters $\varphi,
\phi, \psi, \ldots$ to denote formulas, letters
$\Gamma,\Delta,\Lambda, \ldots$ for sequences and sets of
formulas.

\begin {definitions}  A matrix\index{matrix}, or matrix logic\index{matrix logic}, $\mathfrak{M}$ for a language
$\mathcal{L}$ is given by:
\begin{enumerate}
    \item a nonempty set of truth values $V$ of cardinality $|V|=m$,
    \item a subset $V_+ \subseteq V$ of designated truth values,
    \item an algebra with domain $V$ of appropriate type: for
every $n$-place connective $\square$ of $\mathcal{L}$ there is an
associated truth function $\widetilde{\square}$: $V^n \to V$ , and
    \item for every quantifier $\mathrm{Q}$, an associated truth function $\widetilde{\mathrm{Q}}$:
$\wp(V)\backslash \emptyset \to V$
\end{enumerate}
\end {definitions}

Notice that a truth function for quantifiers is a mapping from
nonempty sets of truth values to truth values: for a non-empty set
$M \subseteq V$, a quantified formula $\mathrm{Q}x \varphi(x)$
takes the truth value $\widetilde{\mathrm{Q}}(M)$ if, for every
truth value $v \in V$, it holds that $v \in M$ iff there is a
domain element $d$ such that the truth value of $\varphi$ in this
point $d$ is $v$ (all relative to some interpretation). The set
$M$ is called the distribution of $\varphi$. For example, suppose
that there are only the universal quantifier $\forall$ and the
existential quantifier $\exists$ in $\mathcal{L}$. Further, we
have the set of truth values $V = \{\top, \bot\}$, where $\bot$ is
false and $\top$ is true, i.e. the set of designated truth values
$V_+ = \{\top\}$. Then we define the truth functions for the
quantifiers $\forall$ and $\exists$ as follows:
\begin{center}$\widetilde{\forall}(\{\top\}) = \top$,

$\widetilde{\exists}(\{\bot\}) = \bot$,

$\widetilde{\forall}(\{\top, \bot\}) =
\widetilde{\forall}(\{\bot\}) = \bot$,

$\widetilde{\exists}(\{\top, \bot\}) =
\widetilde{\exists}(\{\top\}) = \top$
\end{center}
Also, a matrix logic\index{matrix logic} $\mathfrak{M}$ for a
language $\mathcal{L}$ is an algebraic system $ \mathfrak {M} =
\langle V$, $\widetilde{\square}_0$, $\widetilde{\square}_1$,
\dots, $\widetilde{\square}_r$, $\widetilde{\mathrm{Q}}_0$,
$\widetilde{\mathrm{Q}}_1$, \dots, $\widetilde{\mathrm{Q}}_q$,
$V_+ \rangle  $, where
\begin {enumerate}
    \item $V $ is a nonempty set of truth values for well-formed formulas of $\mathcal{L}$,
    \item $\widetilde{\square}_0$, $\widetilde{\square}_1$, \ldots, $\widetilde{\square}_r $ are a set of matrix operations defined on the set $V $ and assigned to corresponding
    propositional connectives $\square
^{n_0}_0 , \square ^{n_1}_1, \ldots , \square ^{n_r}_r$ of
$\mathcal{L}$,
    \item $\widetilde{\mathrm{Q}}_0, \widetilde{\mathrm{Q}}_1, \ldots,
\widetilde{\mathrm{Q}}_q$ are a set of matrix operations defined
on the set $V $ and assigned to corresponding quantifiers
$\mathrm{Q}_0, \mathrm{Q}_1, \ldots , \mathrm{Q}_q$ of
$\mathcal{L}$,
    \item $V_+$ is a set of designated truth values such that $V_+ \subseteq V $.
\end {enumerate}
A \textit{structure}\index{structure} $\mathcal{M} = \langle D,
\Phi\rangle$ for a language $\mathcal{L}$ (an
$\mathcal{L}$-structure\index{$\mathcal{L}$-structure}) consists
of the following:
\begin{enumerate}
    \item A non-empty set $D$, called the domain (elements of $D$ are
called individuals).
    \item A mapping $\Phi$ that satisfies the following:
\begin{enumerate}
    \item Each $n$-ary function symbol $F$ of $\mathcal{L}$ is mapped
to a function $\widetilde{F}$: $D^n \to D$ if $n > 0$, or to an
element of $D$ if $n = 0$.
    \item Each $n$-ary predicate symbol $P$ of $\mathcal{L}$ is
mapped to a function $\widetilde{P}$: $D^n \to V$ if $n
> 0$, or to and element of $V$ if $n = 0$.
\end{enumerate}
\end{enumerate}
Let $\mathcal{M}$ be an $\mathcal{L}$-structure. An
assignment\index{assignment} $s$ is a mapping from the free
variables of $\mathcal{L}$ to individuals.

\begin {definitions}  An $\mathcal{L}$-structure $\mathcal{M} = \langle D,
\Phi\rangle $ together with an assignment $s$ is said to be an
interpretation\index{interpretation} $I = \langle \mathcal{M},
s\rangle $.
\end {definitions}

Let $I = \langle \langle D , \Phi\rangle, s\rangle $ be an
interpretation. Then we can extend the mapping $\Phi$ to a mapping
$\Phi_I$ from terms to individuals:
\begin{itemize}
    \item If $a$ is a free variable, then $\Phi_I(a) = s(a)$.
    \item If $t$ is of the form $F(t_1,\ldots, t_n)$, where $F$ is a
function symbol of arity $n$ and $t_1$, \ldots, $t_n$ are terms,
then $\Phi_I(t) = \Phi(f)\Phi_I(t_1)\ldots\Phi_I(t_n))$.
\end{itemize}

\begin {definitions} Given an interpretation $I = \langle
\mathcal{M}, s\rangle $, we define the valuation\index{valuation}
$\mathrm{val}_I$ to be a mapping from formulas $\varphi$ of
$\mathcal{L}$ to truth values as follows:
\begin{enumerate}
    \item If $\varphi$ is atomic, i.e., of the form $P(t_1,\ldots, t_n)$,
where $P$ is a predicate symbol of arity $n$ and $t_1, \ldots ,
t_n$ are terms, then $\mathrm{val}_I(\varphi) =
\Phi(P)(\Phi_I(t_1)\ldots\Phi_I(t_n))$.
    \item If the outermost logical symbol of $\varphi$ is a propositional
connective $\square$ of arity $n$, i.e., $\varphi$ is of the form
$\square(\psi_1,\ldots, \psi_n)$, where $\psi_1, \ldots , \psi_n$
are formulas, then $\mathrm{val}_I(\varphi) = \widetilde{\square}(
\mathrm{val}_I(\psi_1),\ldots, \mathrm{val}_I(\psi_n))$.
    \item If the outermost logical symbol of $\varphi$ is a quantifier
$\mathrm{Q}$, i.e., $\varphi$ is of the form
$\mathrm{Q}x~\psi(x)$, then
$$\mathrm{val}_I(\varphi) = \widetilde{\mathrm{Q}}(\bigcup\limits_{d \in D} \mathrm{val}_I (\psi(d))).$$
\end{enumerate}\end {definitions}

Suppose $|V|\geqslant 2$. A formula $\varphi$ is said to be
\textit{logically valid}\index{logically validity} (e.g.\ it is a
\textit{many-valued tautology}\index{many-valued tautology}) iff,
for every interpretation $I$, it holds that
$\mathrm{val}_I(\varphi) \in V_+$. Sometimes, a logically valid
formula $\psi$ is denoted by $\models \psi$.\\

Suppose $\mathcal{M}$ is a structure for $\mathcal{L}$,  and
$\varphi$ a formula of $\mathcal{L}$. A formula $\varphi$ is
called \textit{satisfiable}\index{satisfiable} iff there is an
interpretation $I$ such that $\mathrm{val}_I(\varphi) \in V_+$. A
satisfiable formula is denoted by $\mathcal{M} \models \varphi$.
This means that $\varphi$ is satisfiable on $\mathcal{M}$ for
every assignment $s$. In this case $\mathcal{M}$ is called a {\em
model\/}\index{model} of
$\varphi$.\\

If $\Gamma$ is a set of formulas, we will write $\mathcal{M}
\models \Gamma$ if $\mathcal{M} \models \gamma$ for every formula
$\gamma \in \Gamma$, and say that $\mathcal{M}$ is a {\em
model\/}\index{model} of $\Gamma$ or that $\mathcal{M}$ {\em
satisfies\/}\index{satisfies} $\Gamma$.\\

Suppose $\Gamma$ is a set of formulas of $\mathcal{L}$ and $\psi$
is a formula of $\mathcal{L}$.  Then $\Gamma$ {\em
implies\/}\index{implies} $\psi$,  written as $\Gamma \models
\psi$,  if $\mathcal{M} \models \psi$ whenever $\mathcal{M}
\models \Gamma$ for every structure $\mathcal{M}$ for
$\mathcal{L}$. If $\Gamma$ and $\Delta$ are sets of formulas of
$\mathcal{L}$, then $\Gamma$ {\em implies\/} $\Delta$,  written as
$\Gamma \models \Delta$,  if $\mathcal{M} \models \Delta$ whenever
$\mathcal{M} \models \Gamma$ for every structure
$\mathcal{M}$ for $\mathcal{L}$.\\

Suppose $x$ is a variable, $t$ is a term, and $\varphi$ is a
formula.  Then the fact that {\em $t$ is
substitutable\index{substitutable} for $x$ in $\varphi$\/} is
defined as follows:

\begin{enumerate}
\item If $\varphi$ is atomic, then $t$ is substitutable for $x$ in
$\varphi$. \item If $\varphi$ is $\square(\psi_1,\ldots, \psi_n)$,
then $t$ is substitutable for $x$ in $\varphi$ if and only if $t$
is substitutable for $x$ in $\psi_1$, $t$ is substitutable for $x$
in $\psi_n$, etc. \item If $\varphi$ is $\forall y \, \psi$, then
$t$ is substitutable for $x$ in $\varphi$ if and only if either
 \begin{enumerate}
  \item $x$ does not occur free in $\varphi$,  or
  \item if $y$ does not occur in $t$ and $t$ is substitutable for $x$ in $\psi$.
  \end{enumerate}
\end{enumerate}

\section{Hilbert's type calculus for classical logic}

\begin{definitions} Classical logic\index{classical logic} is built in the framework of the language
$\mathcal{L}$ and the valuation $\mathrm{val}_I$ of its formulas
is a mapping to truth values $0$, $1$ that is defined as follows:

\begin{itemize}
    \item $\mathrm{val}_I(\neg \alpha)= 1 - \mathrm{val}_I(\alpha),$
    \item $\mathrm{val}_I(\alpha\vee\beta)= \max(\mathrm{val}_I(\alpha),\mathrm{val}_I(\beta)),$
    \item $\mathrm{val}_I(\alpha\wedge\beta)= \min(\mathrm{val}_I(\alpha),\mathrm{val}_I(\beta)),$
    \item $\mathrm{val}_I(\alpha\rightarrow\beta)=
\left\{%
\begin{array}{ll}
   0, & \hbox{if $\mathrm{val}_I(\alpha)=1$ and $\mathrm{val}_I(\beta)=0$} \\
  1, & \hbox{otherwise.} \\
\end{array}%
\right.$

\end{itemize}

\end{definitions}

Consider \textsc{Hilbert}'s calculus for classical
logic\index{\textsc{Hilbert}'s calculus for classical logic} (see
\cite{Hilbe1}, \cite{Hilbe2}). Its axiom schemata are as follows:

\begin{equation}
    \psi\rightarrow(\varphi\rightarrow\psi)\label{shumannA1},
\end{equation}
\begin{equation}
(\psi\rightarrow\varphi)\rightarrow((\psi\rightarrow(\varphi\rightarrow\chi))\rightarrow(\psi\rightarrow\chi))\label{shumannA2},
\end{equation}
\begin{equation}
(\varphi\wedge\psi)\rightarrow\varphi\label{shumannA3},
\end{equation}
\begin{equation}
(\varphi\wedge\psi)\rightarrow\psi\label{shumannA4},
\end{equation}
\begin{equation}
(\chi\rightarrow\varphi)\rightarrow((\chi\rightarrow\psi)\rightarrow(\chi\rightarrow(\varphi\wedge\psi))),\label{shumannA5}
\end{equation}
\begin{equation}
    \psi\rightarrow(\psi\vee\varphi),\label{shumannA6}
\end{equation}
\begin{equation}
    \varphi\rightarrow(\psi\vee\varphi),\label{shumannA7}
\end{equation}
\begin{equation}
(\varphi\rightarrow\chi)\rightarrow((\psi\rightarrow\chi)\rightarrow((\varphi\vee\psi)\rightarrow\chi)),\label{shumannA8}
\end{equation}
\begin{equation}
(\varphi\rightarrow\chi)\rightarrow(\varphi\rightarrow\neg\chi)\rightarrow\neg\varphi),\label{shumannA9}
\end{equation}
\begin{equation}
\neg\neg\varphi\rightarrow\varphi,\label{shumannA10}
\end{equation}
\begin {equation} \forall x \varphi (x) \rightarrow \varphi [x/t],
\end {equation}
\begin {equation} \varphi [x/t] \rightarrow \exists x \varphi(x),
\end {equation}
\\

\noindent where the formula $\varphi [x/t]$ is the result of
substituting
the term $t$ for all free occurrences of $x$ in $\varphi$.\\

In \textsc{Hilbert}'s calculus there are the following inference
rules:

\begin{enumerate}
    \item \textit {Modus ponens}\index{modus ponens}: if two formulas $
\varphi$ and $ \varphi \rightarrow \psi$ hold, then we deduce a
formula $ \psi$:

\[ \frac {\varphi,\quad \varphi \rightarrow \psi} {\psi}.
\]

    \item \textit{Universal generalization}\index{universal generalization}:

\[ \frac {\varphi \rightarrow \psi(a)} {\varphi \rightarrow\forall x
\psi(x)},
\]

where $a$ is not free in the expression $\varphi$.

    \item \textit{Existential generalization}\index{existential generalization}:

\[ \frac {\varphi(a) \rightarrow \psi} {\exists x \varphi(x) \rightarrow \psi}.
\]

where $a$ is not free in the expression $\psi$.

\end{enumerate}

\begin{definitions} \index{deduction} \index{proof}
Let $\Sigma$ be a set of formulas.  A deduction or proof from
$\Sigma$ in $\mathcal{L}$ is a finite sequence $\varphi_1
\varphi_2 \dots \varphi_n$ of formulas such that for each $k \le
n$,
\begin{enumerate}
\item $\varphi_k$ is an axiom,  or \item $\varphi_k \in \Sigma$,
or \item there are $i_1,\dots,i_n < k$ such that $\varphi_k$
follows from $\varphi_{i_1}$,\dots, $\varphi_{i_n}$ by inference
rules.
\end{enumerate}
A formula of $\Sigma$ appearing in the deduction is called a
premiss\index{premiss}. $\Sigma$ proves\index{proves} a formula
$\alpha$,  written as $\Sigma \vdash \alpha$, if $\alpha$ is the
last formula of a deduction from $\Sigma$.  We'll usually write
$\vdash \alpha$ for $\emptyset \vdash \alpha$.
\end{definitions}

A formula $\alpha$ such that $\vdash \alpha$ is called
\emph{provable}\index{provable}. It is evident that all formulas of the form \eqref{shumannA1} -- \eqref{shumannA10} are provable. \\

As an example, show that $\vdash\psi\rightarrow\psi$.\\

At the first step, take axiom schema \eqref{shumannA1}:

$$\psi\rightarrow((\varphi\rightarrow\psi)\rightarrow\psi).$$

At the second step, take axiom schema \eqref{shumannA2}:

$$(\psi\rightarrow(\varphi\rightarrow\psi))\rightarrow((\psi\rightarrow((\varphi\rightarrow\psi)\rightarrow\psi))\rightarrow(\psi\rightarrow\psi)).$$

At the third step, using axiom schema \eqref{shumannA1} and the
formula of the second step we obtain by modus ponens the following
expression:

$$(\psi\rightarrow((\varphi\rightarrow\psi)\rightarrow\psi))\rightarrow(\psi\rightarrow\psi).$$

At the last step, using the formula of the first step and the
formula of the third step we obtain by modus ponens that

$$\psi\rightarrow\psi.$$\\

Further, we will consider the proof and the provability for
various \textsc{Hilbert}'s type nonclassical calculi just in the
sense of definition 6.

\newtheorem{theorems}{Theorem}\begin{theorems}[Soundness and Completeness Theorem]
\index{completeness theorem}\index{soundness theorem} Let $\alpha$
be a formula and $\Delta$ be a set of formulas of classical logic.

$\Delta \models \alpha$ if and only if $\Delta \vdash
\alpha$.\hfill $\Box $
\end{theorems}

\section{Sequent calculus for classical logic}

Let us remember that the sequent calculus for usual two-valued
logic\index{sequent calculus for classical logic} was proposed by \textsc{Gentzen} in \cite{Gent1}.\\

By his definition, a \textit{sequent}\index{sequent} is an
expression of the form $\Gamma_1\hookrightarrow \Gamma_2$, where
$\Gamma_1=\{\varphi_1,\dots,\varphi_j\}$,
$\Gamma_2=\{\psi_1,\dots,\psi_i\}$ are finite sets of formulas of
the language $\mathcal{L}$, that has the following interpretation:
$\Gamma_1\hookrightarrow \Gamma_2$ is logically valid iff
$$\bigwedge_j\varphi_j\rightarrow\bigvee_i\psi_i$$ is logically
valid.\\

Also, we read such a sequent as: ``$\varphi_1$ and \dots and
$\varphi_j$ entails $\psi_1$ or \dots or $\psi_i$''.\\

A sequent of the form $\emptyset\hookrightarrow \Gamma$ is denoted
by $\hookrightarrow \Gamma$. A sequent of the form
$\Gamma\hookrightarrow \emptyset$ is denoted
by $\Gamma\hookrightarrow $.\\

The inference rules\index{inference rules} are expressions
containing formula variables $\varphi$, $\psi$, \dots and
multisets of formulas variables $\Gamma$, $\Delta$, \dots such
that replacing these with actual formulas and multisets of
formulas, gives ordered pairs consisting of a sequent $S$ (the
conclusion) and a finite set of sequents $S_1$, \dots, $S_n$ (the
premises, written: $\frac{S_1, \dots, S_n}{S}$). Rules where $n =
0$ are called the \textit{initial sequents}\index{initial sequent}
or
axioms.\\

The only \textit{axiom}\index{axiom} of the sequent calculus for
classical logic is as follows:
$\psi\hookrightarrow\psi$.\\

The \textit{inference rules}\index{inference rules}:

\begin{enumerate}
    \item \textbf{Structural rules}\index{structural rules}:
\begin{align*}
& \frac{\Gamma_1\hookrightarrow
\Gamma_2}{\psi,\Gamma_1\hookrightarrow \Gamma_2},& \qquad
\frac{\Gamma_1\hookrightarrow \Gamma_2}{\Gamma_1\hookrightarrow \Gamma_2,\psi} \\
\end{align*}

(the left and right weakening rules respectively),

\begin{align*}
& \frac{\psi,\psi,\Gamma_1\hookrightarrow
\Gamma_2}{\psi,\Gamma_1\hookrightarrow \Gamma_2},& \qquad
\frac{\Gamma_1\hookrightarrow \Gamma_2,\psi,\psi}{\Gamma_1\hookrightarrow \Gamma_2,\psi} \\
\end{align*}

(the left and right contraction rules respectively),

\begin{align*}
& \frac{\Gamma_1,\psi,\chi,\Delta\hookrightarrow
\Gamma_2}{\Gamma_1,\chi,\psi,\Delta\hookrightarrow \Gamma_2},&
\qquad
\frac{\Gamma_1\hookrightarrow \Gamma_2,\psi,\chi,\Delta}{\Gamma_1\hookrightarrow \Gamma_2,\chi,\psi,\Delta}. \\
\end{align*}

(the left and right exchange rules respectively).\\

    \item \textbf{Logical rules}\index{logical rules}, where $(\# \Rightarrow)$ and $(\Rightarrow \# )$ are the left and right introduction rules for a
connective $\#\in\{\neg,\rightarrow,\vee,\wedge,\forall,\exists\}$
respectively:
\begin{align*}
& \frac{\Gamma_1\hookrightarrow
\Gamma_2,\psi}{\Gamma_1,\neg\psi\hookrightarrow
\Gamma_2}~~(\neg\Rightarrow),& \qquad
\frac{\Gamma_1,\psi\hookrightarrow \Gamma_2}{\Gamma_1\hookrightarrow \Gamma_2,\neg\psi}~~(\Rightarrow\neg), \\
\end{align*}
\begin{align*}
& \frac{\psi,\chi,\Gamma_1\hookrightarrow
\Gamma_2}{\psi\wedge\chi,\Gamma_1\hookrightarrow
\Gamma_2}~~(\wedge\Rightarrow),& \qquad
\frac{\Gamma_1\hookrightarrow \Gamma_2,\psi\qquad\Gamma_1\hookrightarrow \Gamma_2,\chi}{\Gamma_1\hookrightarrow \Gamma_2,\psi\wedge\chi}~~(\Rightarrow\wedge), \\
\end{align*}
\begin{align*}
& \frac{\Gamma_1,\psi\hookrightarrow
\Gamma_2\qquad\Gamma_1,\chi\hookrightarrow
\Gamma_2}{\Gamma_1,\psi\vee\chi\hookrightarrow \Gamma_2}
~~(\vee\Rightarrow), & \qquad \frac{\Gamma_1\hookrightarrow
\Gamma_2,\psi,\chi}{\Gamma_1\hookrightarrow
\Gamma_2,\psi\vee\chi}~~(\Rightarrow\vee), \\
\end{align*}
\begin{align*} & \frac{\Gamma_1\hookrightarrow
\Gamma_2,\psi\qquad \chi,\Delta_1\hookrightarrow
\Delta_2}{\psi\rightarrow\chi,\Gamma_1,\Delta_1\hookrightarrow
\Gamma_2,\Delta_2}~~(\rightarrow\Rightarrow),& \qquad
\frac{\psi,\Gamma_1\hookrightarrow \Gamma_2,\chi}{\Gamma_1\hookrightarrow \Gamma_2,\psi\rightarrow\chi}~~(\Rightarrow\rightarrow), \\
\end{align*}
\begin{align*} & \frac{\varphi [x/t],\Gamma_1\hookrightarrow \Gamma_2}{\forall x~\varphi (x),\Gamma_1\hookrightarrow \Gamma_2}~~(\forall\Rightarrow),&
\qquad \frac{\Gamma_1\hookrightarrow
\Gamma_2,\varphi(a)}{\Gamma_1\hookrightarrow \Gamma_2,\forall x~\varphi (x)}~~(\Rightarrow\forall), \\
\end{align*}
\begin{align*} & \frac{\varphi (a),\Gamma_1\hookrightarrow \Gamma_2}{\exists x~\varphi (x),\Gamma_1\hookrightarrow \Gamma_2}~~(\exists\Rightarrow),&
\qquad \frac{\Gamma_1\hookrightarrow
\Gamma_2,\varphi [x/t]}{\Gamma_1\hookrightarrow \Gamma_2,\exists x~\varphi (x)}~~(\Rightarrow\exists), \\
\end{align*}

where the formula $\varphi [x/t]$ is the result of substituting
the term $t$ for all free occurrences of $x$ in $\varphi$, $a$ has
no occurrences in the below sequent.

\item \textbf{Cut rule}\index{cut rule}:

$$\frac{\Gamma_1\hookrightarrow \Delta_1,\psi\qquad\psi,\Gamma_2\hookrightarrow \Delta_2}{\Gamma_1,\Gamma_2\hookrightarrow \Delta_1,\Delta_2}.$$

\end{enumerate}

\begin{definitions} A proof\index{proof} (or derivation\index{derivation}) for a sequent calculus of a sequent
$S$ from a set of sequents $U$ is a finite tree such that:

\begin{itemize}
    \item $S$ is the root of the tree and is called the end-sequent.
    \item The leaves of the tree are all initial sequents
or members of $U$.
    \item Each child node of the tree is obtained from its parent
nodes by an inference rule, i.e. if $S$ is a child node of $S_1$,
\dots, $S_n$, then $\frac{S_1, \dots, S_n}{S}$ is an instance of a
rule.
\end{itemize}

If we have a proof tree with the root $S$ and $U=\emptyset$, then
$S$ is called a provable sequent. If we have a proof tree with the
root $S$ and $U\neq\emptyset$, then $S$ is called a derivable
sequent from premisses $U$.

\end{definitions}

It can be easily shown that for every formula $\psi$ of $\mathcal L$, we have that $\vdash\psi$ if and only if $\hookrightarrow\psi$ is a provable sequent.\\

Further, we will consider the proof and the provability for
various nonclassical sequent and hypersequent calculi just in the sense of definition 7.\\

As an example, using the above mentioned inference rules, show
that the following proposition $(\psi\wedge (\varphi\vee
\chi))\rightarrow ((\psi\wedge \varphi)\vee (\psi\wedge \chi))$ of
classical logic is provable:

$$1.~~~\frac{\psi\hookrightarrow \psi\qquad \varphi\hookrightarrow \varphi\qquad \chi\hookrightarrow \chi}{\psi,\varphi\hookrightarrow \psi\qquad \psi,\varphi\hookrightarrow \varphi\qquad \psi,\chi\hookrightarrow \psi\qquad \psi,\chi\hookrightarrow \chi},$$

$$2.~~~\frac{\psi,\varphi\hookrightarrow \psi\qquad \psi,\varphi\hookrightarrow \varphi\qquad \psi,\chi\hookrightarrow \psi\qquad \psi,\chi\hookrightarrow \chi}{\psi,\varphi\hookrightarrow \psi\wedge\varphi\qquad \psi,\chi\hookrightarrow \psi\wedge\chi},$$

$$3.~~~\frac{\psi,\varphi\hookrightarrow \psi\wedge\varphi\qquad \psi,\chi\hookrightarrow \psi\wedge\chi}{\psi,\varphi\hookrightarrow \psi\wedge\varphi,\psi\qquad\psi,\varphi\hookrightarrow \psi\wedge\varphi,\chi\qquad \psi,\chi\hookrightarrow \psi\wedge\chi,\psi\qquad \psi,\chi\hookrightarrow \psi\wedge\chi,\varphi},$$

$$4.~~~\frac{\psi,\varphi\hookrightarrow \psi\wedge\varphi,\psi\qquad\psi,\varphi\hookrightarrow \psi\wedge\varphi,\chi\qquad \psi,\chi\hookrightarrow \psi\wedge\chi,\psi\qquad \psi,\chi\hookrightarrow \psi\wedge\chi,\varphi}{\psi,\varphi\hookrightarrow \psi\wedge\varphi,\psi\wedge\chi\qquad \psi,\chi\hookrightarrow \psi\wedge\chi,\psi\wedge\varphi},$$

$$5.~~~\frac{\psi,\varphi\hookrightarrow \psi\wedge\varphi,\psi\wedge\chi\qquad \psi,\chi\hookrightarrow \psi\wedge\chi,\psi\wedge\varphi}{\psi,\varphi\vee\chi\hookrightarrow \psi\wedge\varphi,\psi\wedge\chi},$$

$$6.~~~\frac{\psi,\varphi\vee\chi\hookrightarrow \psi\wedge\varphi,\psi\wedge\chi}{\psi\wedge(\varphi\vee\chi)\hookrightarrow (\psi\wedge\varphi)\vee(\psi\wedge\chi)}.$$

$$7.~~~\frac{\psi\wedge(\varphi\vee\chi)\hookrightarrow (\psi\wedge\varphi)\vee(\psi\wedge\chi)}{\hookrightarrow (\psi\wedge(\varphi\vee\chi))\rightarrow ((\psi\wedge\varphi)\vee(\psi\wedge\chi))}.$$


\chapter{$n$-valued {\L}ukasiewicz's logics}
\section{Preliminaries}
For the first time the Polish logician Jan \textsc{{\L}ukasiewicz}
began to create systems of many-valued logic, using a third value
``possible'' to deal with \textsc{Aristotle}'s paradox of the sea
battle\index{\textsc{Aristotle}'s paradox of the sea battle} (see
\cite{Lukas2}, \cite{Trzesic}). Now many-valued logic has
applications in diverse fields. In the earlier years of
development of multiple-validity idea, the most promising field of
its application is artificial intelligence. This application
concerns vague notions and commonsense reasoning, e.g. in expert
systems. In this context fuzzy logic is also interesting, because
multiple-validity is modelled in artificial intelligence via fuzzy
sets and fuzzy logic.\\

Now consider $n $-valued \textsc{{\L}ukasiewicz}'s matrix
logic\index{$n$-valued \textsc{{\L}ukasiewicz}'s matrix logic} $
\mathfrak {M}_{\L_{n}}$ defined as the ordered system $\langle V _
{n}$, $\neg_L$, $\rightarrow_L$, $\vee$, $\wedge$,
$\widetilde{\exists}, \widetilde{\forall}, \{n -1\} \rangle  $ for
any $n \geqslant 2 $, $n \in \mathbf {N} $, where
\begin
{enumerate}
    \item $V_{n} = \{0, 1, 2, \ldots, n-1 \} $,
    \item for all $x\in V_{n}$, $ \neg_L x = (n-1) - x $,
    \item for all $x, y \in V_{n}$, $x \rightarrow_L y = \min (n-1, (n-1) - x + y) $,
    \item for all $x, y \in V_{n}$, $x \vee y = (x \rightarrow_L y) \rightarrow_L y = \max (x, y) $,
    \item for all $x, y \in V_{n}$, $x \wedge y = \neg_L (\neg_L x \vee \neg_L y) = \min (x, y)
    $,
    \item for a subset $M \subseteq V_{n}$, $\widetilde{\exists}(M) = \max (M)
    $, where $\max (M)$ is a maximal element of $M$,
    \item for a subset $M \subseteq V_{n}$, $\widetilde{\forall}(M) = \min
    (M)$, where $\min (M)$ is a minimal element of $M$,
    \item $ \{n -1\} $ is the set of designated truth values.
    \end {enumerate}

The truth value $0\in V_{n}$ is false, the truth value $n-1\in
V_{n}$ is true, and other truth values $x\in V_{n} \backslash
\{0,n-1\}$ are
neutral.\\

By $L _ {n} $ denote the set of all superpositions of the
functions $\neg_L, \rightarrow_L, \widetilde{\exists},
\widetilde{\forall}$.\\

We can construct various truth tables on the basis of $n
$-valued \textsc{{\L}ukasie\-wicz}'s matrix logic.\\

1. The truth table for \textsc{{\L}ukasiewicz}'s negation in $
\mathfrak {M}_{\L_{n}}$: \vspace{14 pt}

\begin{center}
\begin{tabular}{|c|c|}
  \hline
  $p$ & $\neg_L p$ \\
  \hline
  $n-1$ & 0 \\
  $n - 2$ & 1 \\
  $n - 3$ & 2 \\
  \ldots & \ldots \\
  0 & $n-1$ \\
  \hline
\end{tabular}
\end{center} \vspace{24 pt}

2. The truth table for \textsc{{\L}ukasiewicz}'s implication in $
\mathfrak {M}_{\L_{n}}$: \vspace{14 pt}

\begin{center}
\begin{tabular}{|c|c|c|c|c|c|}
  \hline
  $\rightarrow_L$ & $n-1$ & $n - 2$ & $n - 3$ & \ldots & 0 \\
  \hline
  $n-1$ & $n-1$ & $n - 2$ & $n - 3$ & \ldots & 0 \\
  $n - 2$ & $n-1$ & $n-1$ & $n - 2$ & \ldots & $1$ \\
  $n - 3$ & $n-1$ & $n-1$ & $n-1$ & \ldots & $2$ \\
  \ldots & \ldots & \ldots & \ldots& \ldots & \ldots \\
  0 & $n-1$ & $n-1$ & $n-1$ & \ldots & $n-1$ \\
  \hline
\end{tabular}
\end{center}
\vspace{24 pt}

3. The truth table for the disjunction in $ \mathfrak
{M}_{\L_{n}}$:\vspace{14 pt}

\begin{center}
\begin{tabular}{|c|c|c|c|c|c|}
  \hline
  $\vee$ & $n-1$ & $n - 2$ & $n - 3$ & \ldots & 0 \\
  \hline
  $n-1$ & $n-1$ & $n-1$ & $n-1$ & \ldots & $n-1$ \\
  $n - 2$ & $n-1$ & $n - 2$ & $n - 2$ & \ldots & $n - 2$ \\
  $n - 3$ & $n-1$ & $n - 2$ & $n - 3$ & \ldots & $n - 3$ \\
  \ldots & \ldots & \ldots & \ldots& \ldots & \ldots \\
  0 & $n-1$ & $n - 2$ & $n - 3$ & \ldots & 0 \\
  \hline
\end{tabular}
\end{center}
\vspace{24 pt}

4. The truth table for the conjunction in $ \mathfrak
{M}_{\L_{n}}$:\vspace{14 pt}

\begin{center}
\begin{tabular}{|c|c|c|c|c|c|}
  \hline
  $\wedge$ & $n-1$ & $n - 2$ & $n - 3$ & \ldots & 0 \\
  \hline
  $n-1$ & $n-1$ & $n-2$ & $n-3$ & \ldots & 0 \\
  $n - 2$ & $n-2$ & $n - 2$ & $n - 3$ & \ldots & 0 \\
  $n - 3$ & $n-3$ & $n - 3$ & $n - 3$ & \ldots & 0 \\
  \ldots & \ldots & \ldots & \ldots& \ldots & \ldots \\
  0 & 0 & 0 & 0 & \ldots & 0 \\
  \hline
\end{tabular}
\end{center}
\vspace{24 pt}

We can define different $n$-valued matrix logics according to the
chose of different logical operations as initial ones. As an
example, we considered $n$-valued \textsc{{\L}ukasiewicz}'s matrix
logic\index{$n $-valued \textsc{{\L}ukasiewicz}'s matrix logic} in
which $\neg_L$, $\rightarrow_L$, $\vee$, $\wedge$ are basic
operations.\\

Also, an $n$-valued logic $\mathfrak{M}$ is given by a set of
truth values $V(\mathfrak{M}) = \{0$, $1$, $2$, \ldots, $n-1\}$,
the set of designated truth values $V_+(\mathfrak{M})$, and a set
of truth functions $\widetilde{\square}_i$: $V(\mathfrak{M})^i \to
V(\mathfrak{M})$ for all connectives $\square_i$.\\

We denote the set of tautologies of the matrix logic
$\mathfrak{M}$ by $\mathrm{Taut}(\mathfrak{M})$. We say that an
$n$-valued logic $\mathfrak{M}_1$ is \textit{better} than
$\mathfrak{M}_2$ ($\mathfrak{M}_1 \vartriangleleft\mathfrak{M}_2$)
iff $\mathrm{Taut}(\mathfrak{M}_1) \subset
\mathrm{Taut}(\mathfrak{M}_2)$.\\

It is obvious that all $n$-valued propositional logics for a
language $\mathcal{L}$ can be enumerated mechanically:

\newtheorem{propositions}{Proposition}\begin{propositions} Assume that we have
$r$ propositional connectives $\widetilde{\square}^{m_j}_j$ of
arity $m_j$ ($1\leqslant j \leqslant r$). There are
$\prod\limits^r_{j=1} n^{n^{m_j}}$ many $n$-valued logics.
\end{propositions}

\noindent \textit{Proof}. The number of different truth functions
$V^{m_j} \to V$ equals\\

~~~~~~~~~~~~~~$\left|V^{V^{m_j}}\right| = n^{n^{m_j}}$.\hfill
$\Box $

\begin{propositions} For any truth function $\widetilde{\square}^i$: $V(\mathfrak{M})^i \to
V(\mathfrak{M})$ and for any subset of truth values $W \subseteq
V$ there exists disjunctive clauses $\Phi_j(x_1, \ldots, x_i)$ of
the form $(x_1 \in R_{j,1} \vee \ldots \vee x_i \in R_{j,i})$
where $R_{j,k}$ are subsets of $V$, with $1 \leqslant j \leqslant
n^{i-1}$ and $1 \leqslant k \leqslant i$, such that for all $x_1,
\ldots, x_i \in V$:
$$\widetilde{\square}^i(x_1,\ldots, x_i) \in W \leftrightarrow \mathop  \bigwedge \limits_{j = 1}^{n^{i - 1} }\Phi_j(x_1, \ldots, x_i).$$
\end{propositions}

\noindent \emph{Proof}. See \cite{baaz2}.\hfill $\Box $\\

This proposition allows any truth function to be regarded as a
conjunctive normal form.

\section{Originality of $(p + 1) $-valued {\L}ukasiewicz's logics}
The algebraic system $ \mathfrak {M}_{P_{n+1}} = \langle  V_{n+1},
\neg_P,$ $\vee, \{n \}\rangle  $, where
\begin {enumerate}
    \item $V_{n+1} = \{0, 1, 2, \ldots, n \} $,
    \item $ \neg_P x = x + 1 \mod (n+1) $,
    \item $x \vee y = \max (x, y) $,
    \item $ \{n \} $ is the set of designated truth values
    (verum),
\end {enumerate}
is said to be the $(n+1) $-\textit {valued Postian matrix}\index{$(n+1) $-valued Postian matrix} (it is proposed by \textsc{Post} in \cite{Post}).\\

Let $P_{n+1} $ be the set of all functions of the $(n+1) $-ary
Postian matrix logic. We say that the system $F$ of functions is
\textit {precomplete}\index{precomplete} in $P_{n+1} $ if $F $
isn't a complete set, but the adding to $F $ any function $f$ such
that $f \in P_{n+1} $ and $f \not \in F $ converts $F $ into a
complete set. As an example, take the set of all functions in
$P_{n+1} $ preserving $0 $ and $n$. Denote this set by $T_{n+1} $.
By assumption, $f (x_1, \ldots, x_m) \in T_{n+1} $ iff $f (x_1,
\ldots, x_m) \in \{0, n\} $, where $x_i \in \{0, n\} $, $1
\leqslant i \leqslant m $. The
class $T_{n+1} $ of functions is \textit{precomplete}.\\

It is known that we can specify each $(p+1) $-valued
\textsc{{\L}ukasiewicz} matrix logic\index{$(p+1) $-valued
\textsc{{\L}ukasiewicz} matrix logic} for any prime number $p $
(see \cite {karp2}):

\begin {theorems} $L _ {n+1} = T _ {n+1} $ iff $n $ (for any
$n \geqslant 2 $) is a prime number.\hfill $\Box $
\end {theorems}

This means that the set of logical functions in the logic $
\mathfrak M_{\L _ {p+1} }$, where $p $ is a prime number, forms a
precomplete set.

\newtheorem {Cor 4} {Corollary 4.}
\begin {Cor 4}
Suppose there exists the infinite sequence of $(p_{s}+1)$-valued
\textsc{{\L}ukasiewicz}'s matrix logics \emph {$ \mathfrak
M_{\L_{p_{s}+1} }$} (\emph {$p_{s} $} is $s$-th prime number).
Then for each precomplete sets $T _ {p _ {s} +1}$ of functions we
have that $L _ {p _ {s} +1} = T _ {p _ {s} +1} $ for all $s = 1,
2, \ldots $ \hfill $\Box $
\end {Cor 4}

Take the sequence of finite-valued \textsc{{\L}ukasiewicz} matrix
logics

$$ \mathfrak M_{\L _ {2+1} }, \mathfrak M_{\L _ {3+1} },
\mathfrak M_{\L _ {5+1} },  \mathfrak M_{\L _ {7+1}} , \ldots$$

We can show that it is sufficient if we consider just this
sequence instead of the sequence of all finite-valued
\textsc{{\L}ukasiewicz}'s logics $ \mathfrak M_{\L _ {2+1}} $, $
\mathfrak M_{ \L _ {3+1} }$, $ \mathfrak M_{\L _ {4+1} }$, $
\mathfrak M_{\L _ {5+1} }$, $ \mathfrak M_{\L _ {6+1}} $, \ldots
(it was considered by \textsc{Karpenko} in \cite
{karp1}, \cite{karp2}).\\

Indeed, let $ \varphi (n) $ be \textsc{Euler}'s
function\index{\textsc{Euler}'s function}, i.e., the function
defined for all positive integers $n $ and equal to a number $k$
of integers such that $k\leqslant n$ and $k$ is relatively prime
to $ n$. Now assume that $ \varphi ^\ast (n) = \varphi (n) + 1 $.
It is necessary to notice that if $n = p $, then $ \varphi ^\ast
(n) = (p - 1) + 1 = p $. Therefore we have the following
algorithm, by which any natural number $n $ is assigned to a prime
number $p $ and hence we assign any logic $\mathfrak M_{ \L _
{n+1} }$ to a logic $ \mathfrak M_{\L _ {p+1}
}$:\\

$0 $. Let $n = n_1 $ and $n_1 \neq p_i $.\\

$1 $. Either $ \varphi_1 ^\ast (n_1) = p_i $ or $ \varphi_1 ^\ast
(n_1) = n_2 $, where $n_2 <n_1 $.\\

$2 $. Either $ \varphi_2 ^\ast (n_2) = p_i $ or $ \varphi_2 ^\ast
(n_2) = n_3 $, where $n_3 <n_2 $.\\

\begin {center} $ \vdots $\end {center}

$k $. $ \varphi_k ^\ast (n_k) = p_i $, i.e., by $ \varphi_k ^\ast
(n) $ denote $k $-th application of the function $ \varphi ^\ast
(n) $.\\

Since there exists the above mentioned algorithm, it follows that
the function $ \varphi_k ^\ast (n_k) $ induces a partition from
sets $L _ {n+1} $ into their equivalence classes:\\

$L _ {n _ {1} +1} \cong L _ {n _ {2} +1} $ iff there exist $k $
and $l $ such that $ \varphi_k ^\ast (n_1) = \varphi_l ^\ast (n_2)
$.\\

By $ \mathcal {X} _ {p _ {s} + 1} $ denote equivalence classes.
Every class contains a unique precomplete set $L _ {p+1} $. Using
inverse \textsc{Euler}'s function $ \varphi ^ {-1} (n) $, it is
possible to set the algorithm that takes each precomplete set $L _
{p+1} $ to
an equivalence class $ \mathcal {X} _ {p + 1} $.\\

Further, let us consider the inverse function $ \varphi ^ {\ast -
1} (m) $. We may assume that $m = p $, where $p $ is a prime
number.\\

$0 $. We subtract $1 $ from $p $, i.e., we have $p - 1 $.\\

1. We set the range of values for $ \varphi ^{-1} (p - 1) $. By
assumption, this family consists of two classes $ \{\nu_o \} _1 $
and $ \{\nu_e \}_1 $, where $ \{\nu_o \}_1 $ is the class of odd
values, $p$ isn't contained in $ \{\nu_o \}_1 $, and $ \{\nu_e
\}_1 $ is the class of even values. By construction, we disregard
the class $ \{\nu_e \} _i $ for any $i$, because $ \nu_e - 1 $ is
the odd number and consequently cannot be a value of
\textsc{Euler}'s function $ \varphi (n) $. If the class $ \{\nu_o
\} _1 $ is empty (for example, in the case $ \varphi ^ {\ast-1}
(3) $ or $ \varphi ^{\ast - 1} (5) $), then we get the equivalence
class $ \mathcal {X}_{p + 1} $. Conversely, if $ \{\nu_o \}_1 $
isn't empty, then we obtain the range of values $ \varphi ^{-1}
(\nu_o - 1) $ for any $ \nu_o $ in the class $ \{\nu_o \} _1 $.
Here we have two
subcases.\\

$2 $. (a) Either $ \{\nu_o \}_2 = \emptyset $ or (b) $ \{\nu_o
\}_2 \neq \emptyset $. In the first case the process of
construction $ \mathcal {X}_{p + 1} $ is finished. If $ \{\nu_o
\}_2 \neq \emptyset $, then all is repeated. Here we have also two
subcases.\\

$3 $. (a) Either $ \{\nu_o \}_3 = \emptyset $ or (b) $ \{\nu_o
\}_3 \neq \emptyset $, etc.\\

\begin {center} $ \vdots $\end {center}

Thus, the sequence of finite-valued \textsc{{\L}ukasiewicz}'s
matrix logics\index{sequence of finite-valued
\textsc{{\L}ukasiewicz}'s matrix logics} $ \mathfrak M_{\L _ {2+1}
}$, $ \mathfrak M_{\L _ {3+1} }$, $ \mathfrak M_{\L _ {5+1}} $,
$\mathfrak M_{ \L _ {7+1}} $, \ldots corresponds to the sequence
of equivalence classes $ \mathcal {X} _ {2+1} $, $ \mathcal {X} _
{3+1} $, $ \mathcal {X} _ {5+1} $, $ \mathcal {X} _ {7+1} $,
\ldots of all finite-valued \textsc{{\L}ukasiewicz}'s matrix
logics $ \mathfrak M_{\L _ {2+1} }$, $ \mathfrak M_{\L _ {3+1} }$,
$ \mathfrak M_{\L _ {4+1}} $, $ \mathfrak M_{\L _ {5+1}} $, \ldots

\section{$n$-valued {\L}ukasiewicz's calculi of Hilbert's type}

Consider the $n$-ary \textsc{{\L}ukasiewicz} propositional
calculus $ \L_{n} $ of \textsc{Hilbert}'s type\index{$n$-valued
\textsc{{\L}ukasiewicz}'s calculi of \textsc{Hilbert}'s type}. The
axioms of this calculus are as follows:
    \begin {equation} (p \rightarrow_L q) \rightarrow_L ((q\rightarrow_L r) \rightarrow_L (p \rightarrow_L r)), \end {equation}
 \begin {equation} p \rightarrow_L (q\rightarrow_L p), \end {equation}
  \begin {equation}  ((p \rightarrow_L q) \rightarrow_L q) \rightarrow_L ((q \rightarrow_L p) \rightarrow_L p), \end {equation}
 \begin {equation} (p \rightarrow_L^{n} q) \rightarrow_L (p \rightarrow_L ^ {n-1} q) \end {equation}

\noindent for any $n\geqslant 1 $; notice that $p \rightarrow_L^0
q = q $ and $p \rightarrow_L ^ {k+1} q = p \rightarrow_L (p
\rightarrow_L^k q) $,

\begin {equation}  (p \wedge q) \rightarrow_L p, \end {equation}
\begin {equation}   (p \wedge q) \rightarrow_L q, \end {equation}
\begin {equation}   (p \rightarrow_L q) \rightarrow_L ((p \rightarrow_L r) \rightarrow_L (p \rightarrow_L (q \wedge r))), \end {equation}
\begin {equation}  p \rightarrow_L (p \vee q), \end {equation}
\begin {equation}  q \rightarrow_L (p \vee q), \end {equation}
\begin {equation} ((p \rightarrow_L r) \rightarrow_L ((q \rightarrow_L r) \rightarrow_L (p \vee q) \rightarrow_L r)), \end {equation}
\begin {equation}  (\neg_L p \rightarrow_L \neg_L q) \rightarrow_L (q \rightarrow_L p), \end {equation}
\begin {equation}  (p \leftrightarrow (p \rightarrow_L ^ {s-1} \neg_L p)) \rightarrow_L ^ {n-1} p \end {equation}

    \noindent for any $1 \leqslant s \leqslant n-1$ such that $n-1 $ doesn't divide by $s $
    and we have $p \leftrightarrow q = (p \rightarrow_L q) \wedge (q \rightarrow_L p)
    $.\\

There are two inference rules in the system $\L_{n}$: modus ponens
and substitution rule. This formalization of $\L_{n}$ was
created by \textsc{Tuziak} in \cite {tuz}.\\

Let us show by means of the truth table that the formula $(p
\rightarrow_L^2 q)\rightarrow_L (p \rightarrow_L q)$ isn't
tautology in 3-valued \textsc{{\L}ukasiewicz}'s logic.

\vspace{14 pt}

\begin{center}
\begin{tabular}{|c|c|c|c|c|}
  \hline
  $p$ & $q$ & $p \rightarrow_L q$ & $p \rightarrow_L (p \rightarrow_L q)$ & $(p \rightarrow_L (p \rightarrow_L q)) \rightarrow_L (p \rightarrow_L q)$ \\
  \hline
  2 & 2 & 2 & 2 & 2 \\
  2 & 1 & 1 & 1 & 2 \\
  2 & 0 & 0 & 0 & 2 \\
  1 & 2 & 2 & 2 & 2 \\
  1 & 1 & 2 & 2 & 2 \\
  1 & 0 & 1 & 2 & 1 \\
  0 & 2 & 2 & 2 & 2 \\
  0 & 1 & 2 & 2 & 2 \\
  0 & 0 & 2 & 2 & 2 \\
  \hline
\end{tabular}
\end{center}
\vspace{14 pt}

But the formula $(p \rightarrow_L^3 q)\rightarrow_L (p
\rightarrow_L^{2} q)$ is a tautology:

\vspace{14 pt} \footnotesize
\begin{center}
\begin{tabular}{|c|c|c|c|c|c|}
  \hline
  $p$ & $q$ & $p \rightarrow_L q$ & $p \rightarrow_L (p \rightarrow_L q)$ & $p \rightarrow_L (p \rightarrow_L (p \rightarrow_L q))$ & $(p \rightarrow_L^3 q)\rightarrow_L (p
\rightarrow_L^{2} q)$ \\
  \hline
  2 & 2 & 2 & 2 & 2& 2 \\
  2 & 1 & 1 & 1 & 1& 2 \\
  2 & 0 & 0 & 0 & 0& 2 \\
  1 & 2 & 2 & 2 & 2& 2 \\
  1 & 1 & 2 & 2 & 2& 2 \\
  1 & 0 & 1 & 2 & 2& 2 \\
  0 & 2 & 2 & 2 & 2& 2 \\
  0 & 1 & 2 & 2 & 2& 2 \\
  0 & 0 & 2 & 2 & 2& 2 \\
  \hline
\end{tabular}
\end{center}\normalsize
\vspace{14 pt}

We already know that the formula $(p \rightarrow_L^k
q)\rightarrow_L (p \rightarrow_L^{k-1} q)$ is tautology for any $k
\geqslant n$ in $n $-valued \textsc{{\L}ukasiewicz}'s logic. Thus,
some tautologies of the classical logic are ignored in
finite-valued \textsc{{\L}ukasiewicz}'s logic as well as the other
tautologies of the classical logic are ignored in the other
nonclassical logics.

\section{Sequent calculi for $n$-valued {\L}ukasiewicz's logics}
Sequent calculi and natural deduction systems for $n$-valued logic
$\L_n$\index{sequent calculi for $n$-valued
\textsc{{\L}ukasiewicz}'s logics}, where $n$ is a finite natural
number, are considered by \textsc{Baaz}, \textsc{Ferm\"{u}ller},
and \textsc{Zach} in \cite{Baaz5}, \cite{baaz}, \cite{baaz2}. An
$n$-valued sequent is regarded as an $n$-tuple of finite sets
$\Gamma_i$ ($1\leq i \leq n$) of formulas, denoted by $\Gamma_1
\mid \Gamma_2 \mid\ldots\mid \Gamma_n$. It is defined to be
satisfied by an interpretation iff for some $i\in\{1,\ldots,n\}$
at least one formula in $\Gamma_i$ takes the truth value
$i-1\in\{0,\ldots,n-1\}$, where
$\{0,\ldots,n-1\}$ is the set of truth values for $\L_n$.\\

By this approach, a two-valued sequent with \textsc{Gentzen}'s
standard notation $\Gamma_1\hookrightarrow\Gamma_2$, where
$\Gamma_1$ and $\Gamma_2$ are finite sequences of formulas, is
interpreted truth-functionally: either one of the formulas in
$\Gamma_1$ is false or one of the formulas in $\Gamma_2$ is true.
In other words, we can denote a sequent\index{sequent}
$\Gamma_1\hookrightarrow\Gamma_2$ by $\Gamma_1\mid\Gamma_2$ and we
can define it to be satisfied by an interpretation iff for some
$i\in\{1,2\}$ at least one formula in
$\Gamma_i$ takes the truth value $i-1\in\{0,1\}$.\\

Let $I$ be an interpretation. The $I$ $\mathfrak
p$-\textit{satisfies}\index{$\mathfrak p$-satisfies} ($\mathfrak
n$-\textit{satisfies}\index{$\mathfrak n$-satisfies}) a sequent
$\Gamma_1 \mid \Gamma_2 \mid\ldots\mid \Gamma_n$ iff there is an
$i$ ($1 \leqslant i \leqslant n$) such that, for some formula
$\varphi \in \Gamma_i$, $\mathrm{val}_I(\varphi) = i-1\in V_n$
($\mathrm{val}_I(\varphi) \neq i-1\in V_n$). This $I$ is called a
$\mathfrak p$-model ($\mathfrak n$-model) of $\Gamma$, i.e.\ $I
\models^p \Gamma$ ($I \models^n \Gamma$). A
sequent is called $\mathfrak p$- ($\mathfrak n$)-\textit{valid}\index{$\mathfrak p$-valid}\index{$\mathfrak n$-valid}, if it is $\mathfrak p$- ($\mathfrak n$)-satisfied by every interpretation. \\

Also, a sequent $\Gamma$ is called $\mathfrak p$-satisfiable
($\mathfrak n$-satisfiable) iff there is an interpretation $I$
such that $I \models^p \Gamma$ ($I \models^n \Gamma$), and
$\mathfrak p$-valid ($\mathfrak n$-valid) iff for every
interpretation $I$, $I \models^p \Gamma$ ($I \models^n \Gamma$).
The concept of $\mathfrak p$-satisfiability was
proposed by \textsc{Rousseau} in \cite{Roussea}.\\

Notice that according to $\mathfrak p$-satisfiability a sequent is
understood as a positive disjunction and according to $\mathfrak
n$-satisfiability as a negative disjunction. Therefore the
negation of a $\mathfrak p$-sequent ($\mathfrak n$-sequent) is
equivalent to a
conjunction of $\mathfrak n$-sequents ($\mathfrak p$-sequents).\\

By $[i:\psi]$ denote a sequent in that a formula $\psi$ occurs at
place $i+1$.\\

Consider the sequent containing only one formula $i:\psi$ with the
truth value $i$. Then we obtain the following result:

\begin{propositions} A sequent $[i:\psi]$ is $\mathfrak p$-unsatisfiable
($\mathfrak n$-unsatisfiable) iff it is $\mathfrak n$-valid
($\mathfrak p$-valid).
\end{propositions}

\noindent \textit{Proof}. The negation of the $\mathfrak
p$-sequent $\psi^{i}$ is the
$\mathfrak n$-sequent $\neg \psi^{i}$.\\

On the other hand, the $n$-sequent $\neg \psi^{i}$ can also be
written as a $\mathfrak p$-sequent $\bigvee_{j\neq i} \psi^{j}$ ,
and hence the $\mathfrak p$-unsatisfiability of $[i:\psi]$ can be
established by proving $[V \backslash \{i\}:\psi]$ $\mathfrak
p$-valid.\hfill $\Box $

\begin{propositions} Let $\psi$ be a formula. Then the following are
equivalent:
\begin{itemize}
    \item $\psi$ is a tautology.
    \item The sequent $[V_+: \psi]$ is $\mathfrak p$-valid
    \item The sequents $[j: \psi]$, where $j \in V \backslash V_+$, are all
$\mathfrak n$-valid.\hfill $\Box $
\end{itemize}
\end{propositions}

\begin{propositions} Let $\psi$ be a formula. Then the following are
equivalent:
\begin{itemize}
    \item $\psi$ is a unsatisfiable.
    \item The sequent $[V \backslash V_+: \psi]$ is $\mathfrak p$-valid.
    \item The sequents $[j: \psi]$, where $j \in V_+$, are all
$\mathfrak n$-valid.\hfill $\Box $
\end{itemize}
\end{propositions}

\begin {definitions} An introduction rule for a connective
$\square$\index{introduction rule for a connective} at place $i$
in the $n$-valued \textsc{{\L}ukasiewicz}'s logic $\L_n$ is a
schema of the form:
$$\frac{\langle \Gamma^j_1, \Delta^j_1\mid\ldots\mid\Gamma^j_n, \Delta^j_n\rangle _{j \in N}}{\Gamma_1 \mid \ldots\mid \Gamma_i , \square(\psi_1, \ldots, \psi_m)\mid \ldots\mid \Gamma_n} \square : i$$
where the arity of $\square$ is $m$, $N$ is a finite set,
$\Gamma_l = \bigcup_{j\in N} \Gamma^j_l$ , $\Delta^j_l \subseteq
\{\psi_1,\ldots,\psi_m\}$, and for every interpretation $I$ the
following are equivalent:
\begin{enumerate}
    \item $\square(\psi_1, \ldots, \psi_n)$ takes (resp.\ does not take) the truth
value $i-1$.
    \item For all $j \in N$, an interpretation $I$ $\mathfrak p$- (resp.\ $\mathfrak n$)-satisfies the
sequents $\Delta^j_1\mid\ldots\mid\Delta^j_n$.
\end{enumerate}
\end {definitions}

Note that these introduction rules are generated in a mechanical
way from the truth table $\widetilde{\square}^i$:
$V(\mathfrak{M}_{\L_n})^i \to V(\mathfrak{M}_{\L_n})$ of the
connective $\square^i$ through conjunctive normal forms (see
proposition 2).

\begin {definitions} An introduction rule for a quantifier
$\mathrm{Q}$\index{introduction rule for a quantifier} at place
$i$ in the $n$-valued \textsc{{\L}ukasiewicz}'s logic $\L_n$ is a
schema of the form:
$$\frac{\langle \Gamma^j_1, \Delta^j_1\mid\ldots\mid\Gamma^j_n, \Delta^j_n\rangle _{j \in N}}{\Gamma_1 \mid \ldots\mid \Gamma_i , \mathrm{Q}x ~\psi(x)\mid \ldots\mid \Gamma_n} \mathrm{Q} : i$$
where
\begin{itemize}
    \item $N$ is a finite set,
    \item $\Gamma_l = \bigcup_{j\in N} \Gamma^j_l$ ,
$\Delta^j_l \subseteq \{\psi(a_1),\ldots,\psi(a_p)\}
\cup\{\psi(t_1),\ldots,\psi(t_q)\}$,
    \item the $a_l$ are free variables
satisfying the condition that they do not occur in the lower
sequent,
    \item the $t_k$ are arbitrary terms,
\end{itemize}
and for every interpretation $I$ the following are equivalent:
\begin{enumerate}
    \item $\mathrm{Q}x~\psi(x)$ takes (does not take) the truth value $i-1$
under $I$.
    \item For all $d_1$, \ldots, $d_p \in D$, there are $e_1, \ldots , e_q
\in D$ such that for all $j \in N$, an interpretation $I$
$\mathfrak p$- ($\mathfrak n$)-satisfies
$\Delta'^j_1\mid\ldots\mid\Delta'^j_n$ where $\Delta'^j_l$ is
obtained from $\Delta^j_l$ by instantiating the eigenvariable
$a_k$ (term variable $t_k$) with $d_k$ ($e_k$).
\end{enumerate}
\end {definitions}

A $\mathfrak p$-\textit{sequent calculus}\index{$\mathfrak
p$-sequent calculus} for a logic $\L_n$ is given by:

\begin{itemize}
    \item Axioms of the form: $\varphi \mid \varphi \mid \ldots \mid \varphi$, where $\varphi$
is any formula.
    \item For every connective $\square$ of the logic $\L_n$ and every truth value $i-1$ an appropriate introduction rule
    $\square:i$.
    \item For every quantifier $\mathrm{Q}$ and every truth value $i-1$
an appropriate introduction rule $\mathrm{Q}:i$.
    \item Weakening rules for every place $i$:

$$\frac{\Gamma_1\mid\ldots\mid\Gamma_i\mid\ldots \mid\Gamma_n}{\Gamma_1 \mid \ldots\mid \Gamma_i, \psi\mid \ldots\mid \Gamma_n} w : i$$

    \item Cut rules\index{cut rule} for every pair of truth values $(i-1, j-1)$ such that
$i \neq j $:

$$\frac{\Gamma_1\mid\ldots\mid\Gamma_i, \psi\mid\ldots \mid\Gamma_n \quad \Delta_1\mid\ldots\mid\Delta_j, \psi\mid\ldots \mid\Delta_n}{\Gamma_1, \Delta_1 \mid \ldots\mid \Gamma_n, \Delta_n} cut : ij$$

\end{itemize}

An $\mathfrak n$-\textit{sequent calculus}\index{$\mathfrak
n$-sequent calculus} for a logic $\L_n$ is given by:

\begin{itemize}
    \item Axioms of the form: $\Delta_1 \mid \ldots \mid \Delta_n$,
where $\Delta_i = \Delta_j = \{\psi\}$ for some $i$, $j$ such that
$i \neq j$ and $\Delta_k = \emptyset$ otherwise ($\psi$ is any
formula).
    \item For every connective $\square$ and every truth value $i-1$ an
appropriate introduction rule $\square:i$.
    \item For every quantifier $\mathrm{Q}$ and every truth value $i-1$
an appropriate introduction rule $\mathrm{Q}:i$.
    \item Weakening rules identical to the ones of $\mathfrak p$-sequent
calculi.
    \item The cut rule\index{cut rule}:

$$\frac{\langle \Gamma^j_1\mid\ldots\mid\Gamma^j_i\mid\ldots\Gamma^j_n, \Delta^j_n\rangle ^n_{j =1}}{\Gamma_1 \mid \ldots\mid \Gamma_n} cut : $$

where $\Gamma_l = \bigcup_{1\leqslant j \leqslant n} \Gamma^j_l$.
\end{itemize}

A sequent is {\it $\mathfrak p$- ($\mathfrak
n$)-provable}\index{$\mathfrak p$-provable}\index{$\mathfrak
n$-provable} if there is an upward tree of sequents such that
every topmost sequent is an axiom and every other sequent is
obtained from the ones standing immediately above it by an
application of one of the rules of $\mathfrak p$- ($\mathfrak
n$)-sequent calculus.

\begin{theorems}[Soundness and  Completeness] \index{completeness theorem}\index{soundness theorem}
For every $\mathfrak p$- ($\mathfrak n$)-sequent calculus the
following holds: A sequent is $\mathfrak p$- ($\mathfrak
n$)-provable without cut rule(s) iff it is $\mathfrak p$-
($\mathfrak n$)-valid.
\end{theorems}
\noindent \emph{Proof}. See \cite{baaz2}.\\

\textbf{Example: 3-valued \textsc{{\L}ukasiewicz}'s propositional
logic}\index{3-valued \textsc{{\L}ukasiewicz}'s propositional
logic}. Let us consider here the 3-valued \textsc{{\L}ukasiewicz}
logic with the set of basic connectives $\{\rightarrow_L,
\neg_L\}$. Recall that the 3 values
are denoted by $\{0, 1, 2\}$.\\

1. Introduction rules for $\neg_L$ in $\mathfrak{ p}$-sequent calculus:\\

\begin{align*}
& \frac{\Gamma_1 \mid \Gamma_2\mid \Gamma_3, \psi}{\Gamma_1,\neg_L \psi\mid \Gamma_2 \mid \Gamma_3}\neg_L: 0 & \qquad \frac{\Gamma_1 \mid \Gamma_2,\psi\mid \Gamma_3}{\Gamma_1 \mid \Gamma_2,\neg_L\psi\mid \Gamma_3}\neg_L: 1  \\
&{}&{}\\
& \frac{\Gamma_1,\psi \mid \Gamma_2\mid \Gamma_3}{\Gamma_1 \mid
\Gamma_2\mid \Gamma_3, \neg_L\psi}\neg_L: 2 &
\end{align*}\\

2. Introduction rules for $\rightarrow_L$ in $\mathfrak{
p}$-sequent calculus:

$$
\frac{\Gamma_1 \mid \Gamma_2\mid \Gamma_3,\psi \quad
\Gamma_1,\varphi \mid \Gamma_2\mid \Gamma_3}{\Gamma_1,\psi
\rightarrow_L \varphi\mid \Gamma_2\mid \Gamma_3}\rightarrow_L: 0$$
\\

$$\frac{\Gamma_1\mid \Gamma_2, \psi\mid \Gamma_3,\psi \quad \Gamma_1\mid \Gamma_2,\psi,\varphi\mid \Gamma_3\quad \Gamma_1,\varphi\mid \Gamma_2\mid \Gamma_3,\psi}{\Gamma_1\mid \Gamma_2, \psi \rightarrow_L \varphi\mid \Gamma_3}\rightarrow_L: 1$$
\\

$$\frac{\Gamma_1,\psi\mid \Gamma_2,\varphi\mid \Gamma_3,\varphi \quad \Gamma_1,\psi\mid \Gamma_2,\psi\mid \Gamma_3,\varphi}{\Gamma_1\mid \Gamma_2\mid \Gamma_3,\psi \rightarrow_L
\varphi}\rightarrow_L: 2$$

\section{Hypersequent calculus for $3$-valued {\L}ukasie\-wicz's propositional logic}

The hypersequent formalization of 3-valued
\textsc{{\L}ukasiewicz}'s propositional logic $ \L_{3}
$\index{hypersequent calculus for $3$-valued
\textsc{{\L}ukasiewicz}'s propositional logic} was proposed by
\textsc{Avron} in \cite{Arnon1}. Let us remember what is a
hypersequent.

\begin{definitions}A hypersequent\index{hypersequent} is a structure of the form:

$$\Gamma \hookrightarrow \Delta \mid \Gamma' \hookrightarrow \Delta' \mid \ldots \mid \Gamma'' \hookrightarrow \Delta'',$$

\noindent where $\Gamma \hookrightarrow \Delta$, $ \Gamma'
\hookrightarrow \Delta'$,$ \ldots$, $\Gamma'' \hookrightarrow
\Delta''$ are finite sequences of ordinary sequents in
\textsc{Gentzen}'s sense.
\end{definitions}

We shall use $G$ and $H$ as variables for possibly empty
hypersequents.\\

The standard interpretation of the $\mid$ symbol is usually
disjunctive, i.e.\ a hypersequent is true if and only if one
of its components is true.\\

The only \textit{axiom}\index{axiom} of this calculus:~~~ $\psi\hookrightarrow\psi.$\\

The \textit{inference rules}\index{inference rules} are as follows.\\

\begin{align*}
& \frac{G |\Gamma
\hookrightarrow \Delta}{G |\Gamma  \hookrightarrow \Delta|\Gamma'  \hookrightarrow \Delta'},& \qquad \frac{G |\Gamma  \hookrightarrow \Delta|\Gamma  \hookrightarrow \Delta}{G |\Gamma  \hookrightarrow \Delta}, \\
&{}&{}\\
& \frac{G |\Gamma' \hookrightarrow \Delta'|\Gamma \hookrightarrow
\Delta}{G |\Gamma \hookrightarrow \Delta|\Gamma' \hookrightarrow
\Delta'},
\end{align*}

\begin{align*}
    &
\frac{G | \Gamma\hookrightarrow \Delta | H}{G | \psi , \Gamma
\hookrightarrow \Delta | H},& \qquad \frac{G |
\Gamma \hookrightarrow \Delta | H}{G |  \Gamma \hookrightarrow \Delta, \psi| H},\\
&{}&{}\\
& \frac{G | \psi, \varphi, \Gamma \hookrightarrow \Delta | H}{G |
\varphi, \psi , \Gamma \hookrightarrow \Delta | H},& \qquad
\frac{G | \Gamma \hookrightarrow \Delta, \psi, \varphi | H}{G |
\Gamma \hookrightarrow \Delta, \varphi, \psi | H},
\end{align*}

\begin{align*}
    & \frac{G | \psi, \varphi, \Gamma
\hookrightarrow \Delta | H}{G | \psi \wedge \varphi, \Gamma
\hookrightarrow \Delta | H},& \qquad \frac{G | \Gamma
\hookrightarrow \Delta, \psi, \varphi | H}{G | \Gamma
\hookrightarrow \Delta, \psi \vee \varphi | H},
\end{align*}

\[\frac{G | \Gamma \hookrightarrow \Delta,
\psi | H \qquad G | \Gamma \hookrightarrow \Delta, \varphi | H}{G
| \Gamma \hookrightarrow \Delta, \psi\wedge \varphi | H},\]

\[\frac{ G |
\psi,\Gamma \hookrightarrow \Delta | H \qquad G | \varphi, \Gamma
\hookrightarrow \Delta | H}{G | \psi \vee \varphi, \Gamma
\hookrightarrow \Delta | H},\]

\begin{align*}
    &\frac{G |\psi, \Gamma  \hookrightarrow
\Delta, \varphi }{G |\Gamma \hookrightarrow \Delta, \psi
\rightarrow_L \varphi},& \qquad\frac{G| \Gamma \hookrightarrow
\Delta, \psi\quad G |\varphi, \Gamma \hookrightarrow \Delta }{G
|\psi \rightarrow_L \varphi, \Gamma\hookrightarrow
\Delta},\end{align*}

\begin{align*}
& \frac{G |\psi, \Gamma  \hookrightarrow \Delta }{G |\Gamma
\hookrightarrow \Delta, \neg_L \psi},& \qquad \frac{G| \Gamma
\hookrightarrow
\Delta, \psi }{G | \neg_L \psi, \Gamma\hookrightarrow \Delta}, \\
\end{align*}
$$
\frac{   G | \Gamma_{1}, \Gamma_{2}, \Gamma_{3} \hookrightarrow
\Delta_{1}, \Delta_{2}, \Delta_{3} | H  \qquad
   G | \Gamma'_{1},
\Gamma'_{2}, \Gamma'_{3} \hookrightarrow \Delta'_{1}, \Delta'_{2},
\Delta'_{3} | H  }{G | \Gamma_{1}, \Gamma'_{1} \hookrightarrow
\Delta_{1}, \Delta'_{1} |\Gamma_{2}, \Gamma'_{2} \hookrightarrow
\Delta_{2}, \Delta'_{2} |\Gamma_{3}, \Gamma'_{3} \hookrightarrow
\Delta_{3}, \Delta'_{3} | H}.$$

\chapter{Infinite valued {\L}ukasiewicz's logics}
\section{Preliminaries}The ordered system $\langle  V _ {\mathbf{Q}}$, $\neg_L$, $\rightarrow_L$, $\&_L$,
$\vee$, $\wedge$, $\widetilde{\exists}$, $\widetilde{\forall}, \{1
\} \rangle $ is called \textit {rational valued
\textsc{{\L}uka\-siewicz}'s matrix logic}\index{rational valued
\textsc{{\L}uka\-siewicz}'s matrix logic} $ \mathfrak
{M}_{\mathbf{Q}}$, where

\begin
{enumerate}
    \item $V_{\mathbf{Q}} = \{x\colon x\in \mathbf{Q} \} \cap [0,
    1]$,
    \item for all $x\in V_{\mathbf{Q}}$, $ \neg_L x = 1 - x $,
    \item for all $x, y \in V_{\mathbf{Q}}$, $x \rightarrow_L y = \min (1, 1 - x + y) $,
    \item for all $x, y \in V_{\mathbf{Q}}$, $x\&_L y =
    \neg_L(x \rightarrow_L \neg_L y)$,
    \item for all $x, y \in V_{\mathbf{Q}}$, $x \vee y = (x \rightarrow_L y) \rightarrow_L y = \max (x, y) $,
    \item for all $x, y \in V_{\mathbf{Q}}$, $x \wedge y = \neg_L (\neg_L x \vee \neg_L y) = \min (x, y)
    $,
    \item for a subset $M \subseteq V_{\mathbf{Q}}$, $\widetilde{\exists}(M) = \max (M)
    $, where $\max (M)$ is a maximal element of $M$,
    \item for a subset $M \subseteq V_{\mathbf{Q}}$, $\widetilde{\forall}(M) = \min
    (M)$, where $\min (M)$ is a minimal element of $M$,
    \item $ \{1 \} $ is the set of designated truth values.

    \end {enumerate}

The truth value $0\in V_{\mathbf{Q}}$ is false, the truth value
$1\in V_{\mathbf{Q}}$ is true, and other truth values $x\in
V_{\mathbf{Q}}\backslash \{0,1\}$ are neutral.\\

\textit {Real valued \textsc{{\L}ukasiewicz}'s matrix logic}
\index{real valued \textsc{{\L}ukasiewicz}'s matrix logic}$
\mathfrak {M}_{\mathbf{R}}$ is the ordered system $\langle V _
{\mathbf{R}}$, $\neg_L$, $\rightarrow_L$, $\&_L$, $\vee$,
$\wedge$, $\widetilde{\exists}$, $\widetilde{\forall}$, $\{1 \}
\rangle  $, where
\begin
{enumerate}
    \item $V_{\mathbf{R}} = \{x\colon x\in \mathbf{R}\} \cap [0,
    1]$,
    \item for all $x\in V_{\mathbf{R}}$, $ \neg_L x = 1 - x $,
    \item for all $x, y \in V_{\mathbf{R}}$, $x \rightarrow_L y = \min (1, 1 - x + y) $,
    \item for all $x, y \in V_{\mathbf{R}}$, $x\&_L y =
    \neg_L(x \rightarrow_L \neg_L y)$,
    \item for all $x, y \in V_{\mathbf{R}}$, $x \vee y = (x \rightarrow_L y) \rightarrow_L y = \max (x, y) $,
    \item for all $x, y \in V_{\mathbf{R}}$, $x \wedge y = \neg_L (\neg_L x \vee \neg_L y) = \min (x, y)
    $,
    \item for a subset $M \subseteq V_{\mathbf{R}}$, $\widetilde{\exists}(M) = \max (M)
    $, where $\max (M)$ is a maximal element of $M$,
    \item for a subset $M \subseteq V_{\mathbf{R}}$, $\widetilde{\forall}(M) = \min
    (M)$, where $\min (M)$ is a minimal element of $M$,
    \item $ \{1 \} $ is the set of designated truth values.

    \end {enumerate}

The truth value $0\in V_{\mathbf{R}}$ is false, the truth value
$1\in V_{\mathbf{R}}$ is true, and other truth values $x\in
V_{\mathbf{R}}\backslash \{0,1\}$ are neutral.\\

The logics $ \mathfrak {M}_{\mathbf{Q}}$ and $ \mathfrak
{M}_{\mathbf{R}}$ will be denoted by $ \mathfrak {M}_{[0,1]}$.
They are called \textit{infinite valued \textsc{{\L}ukasiewicz}'s
matrix logic}.

\section{Hilbert's type calculus for infinite valued {\L}u\-kasiewicz's logic}

\textsc{{\L}ukasiewicz}'s infinite valued
logic\index{\textsc{{\L}ukasiewicz}'s infinite valued
logic}\index{infinite valued \textsc{{\L}ukasiewicz}'s
logic}\index{\textsc{Hilbert}'s type calculus for infinite valued
\textsc{{\L}u\-kasiewicz}'s logic} is denoted by $\L_\infty$. The
basic operations of $\L_\infty$ are $\bot$ (truth constant
`falsehood') and \textsc{{\L}ukasiewicz}'s implications
$\rightarrow_L$. Other connectives are derivable:

$$\neg_L\psi=: \psi \rightarrow_L \bot,$$
$$\psi\&_L\varphi=: \neg_L(\psi\rightarrow_L\neg_L\varphi),$$
$$\psi\wedge\varphi=:\psi\&_L(\psi\rightarrow_L\varphi),$$
$$\psi\vee\varphi=:(\psi\rightarrow_L\varphi)\rightarrow_L\varphi,$$
$$\top=:\neg_L\bot.$$

The \textsc{Hilbert}'s type calculus for $\L_\infty$ consists of
the axioms:

\begin{equation}
    \psi\rightarrow_L(\varphi\rightarrow_L\psi)\label{shuma1},
\end{equation}
\begin{equation}
(\psi\rightarrow_L\varphi)\rightarrow_L((\varphi\rightarrow_L\chi)\rightarrow_L(\psi\rightarrow_L\chi)),
\end{equation}
\begin{equation}
((\psi\rightarrow_L\varphi)\rightarrow_L\varphi)\rightarrow_L((\varphi\rightarrow_L\psi)\rightarrow_L\psi),\label{SchuL}
\end{equation}
\begin{equation}
(\neg_L\psi\rightarrow_L\neg_L\varphi)\rightarrow_L(\varphi\rightarrow_L\psi),\label{shuma2}
\end{equation}
\begin {equation} \forall x ~\varphi (x) \rightarrow_L \varphi [x/t],
\end {equation}
\begin {equation} \varphi [x/t] \rightarrow_L \exists x ~\varphi(x),
\end {equation}
\\

\noindent where the formula $\varphi [x/t]$ is the result of
substituting
the term $t$ for all free occurrences of $x$ in $\varphi$,\\

\begin {equation} \forall x (\chi\rightarrow_L\varphi) \rightarrow_L (\chi\rightarrow_L\forall x\varphi ),
\end {equation}
\begin {equation} \forall x (\varphi\rightarrow_L\chi) \rightarrow_L (\exists x \varphi \rightarrow_L\chi),
\end {equation}
\begin {equation} \forall x (\chi\vee\varphi) \rightarrow_L (\chi\vee\forall x\varphi ),
\end {equation}
\\

\noindent where $x$ is not free in $\chi$.\\

In $\L_\infty$ there are the following inference
rules\index{inference rules}:

\begin{enumerate}
    \item \textit {Modus ponens}\index{modus ponens}: from $
\varphi$ and $ \varphi \rightarrow_L \psi$ infer $ \psi$:

\[ \frac {\varphi,\quad \varphi \rightarrow_L \psi} {\psi}.
\]

    \item \textit{Substitution rule}\index{substitution rule}: we can substitute any formulas for propositional variables.

    \item \textit{Generalization}\index{generalization}: from $\varphi$ infer $\forall x \varphi
    (x)$:

\[ \frac {\varphi } {\forall x ~\varphi(x)}.
\]
\end{enumerate}

The \textsc{Hilbert}'s type propositional calculus for $\L_\infty$
can be obtained by extending the axiom system \eqref{sch1} --
\eqref{sch2} by the expression \eqref{SchuL}.

\section{Sequent calculus for infinite valued {\L}u\-ka\-siewicz's propositional logic}

An original interpretation of a sequent\index{sequent} for
infinite valued \textsc{{\L}u\-kasiewicz}'s logic $\L_\infty$ was
proposed by \textsc{Metcalfe}, \textsc{Olivetti}, and
\textsc{Gabbay} in \cite{Gabbay3}. For setting this interpretation
they used the following proposition:

\begin{propositions}
Let $\mathfrak M_{[-1,0]}$ be the structure $\langle
[-1,0],\max,\min, \&_L, \rightarrow_L,0\rangle$ where
\begin{itemize}
    \item $\&_L=:\max(-1, x+y)$ and
    \item $x\rightarrow_L y=:\min(0, y-x)$,
    \item $0$ is the designated truth value,
\end{itemize}
then $\psi$ is logically valid in $\mathfrak M_{[-1,0]}$ iff
$\psi$ is logically valid in \textsc{{\L}u\-kasiewicz}'s matrix
logic $\mathfrak M_{[0,1]}$.
\end{propositions}

\noindent \emph{Proof}. See \cite{Gabbay3}.\hfill $\Box $\\

\begin {definitions}A sequent\index{sequent} written
$\Gamma_1 \hookrightarrow \Gamma_2 $, where
$\Gamma_1=\{\varphi_1,\dots, \varphi_m\}$ and
$\Gamma_2=\{\psi_1,\dots, \psi_n\}$, has the following
interpretation: $\Gamma_1 \hookrightarrow \Gamma_2 $ is logically
valid in $\L_\infty$ iff $(\psi_1 + \ldots + \psi_n) \rightarrow_L
(\varphi_1 + \ldots + \varphi_m)$ is logically valid in
$\L_\infty$ for all valuations $\mathrm{val}_I$ mapped from
formulas of $\mathcal{L}$ to the structure $\mathfrak M_{[-1,0]}$,
where $\chi_1 + \ldots + \chi_k = \top$ if $k = 0$.
\end {definitions}

\begin{enumerate}
    \item Axioms\index{sequent calculus for infinite valued \textsc{{\L}u\-ka\-siewicz}'s propositional logic} of the sequent calculus:

\begin{align*}
& (ID) \qquad \psi \hookrightarrow \psi, & \qquad  (\Lambda) \qquad \hookrightarrow~~,  & \qquad (\bot) \qquad\bot\hookrightarrow \psi.\\
\end{align*}
    \item Structural rules:

\begin{align*}& \frac{\Gamma  \hookrightarrow \Delta}{\Gamma, \psi  \hookrightarrow  \Delta},& \qquad\frac{\Gamma_1 \hookrightarrow \Delta_1  \quad \Gamma_2 \hookrightarrow \Delta_2}
{\Gamma_1, \Gamma_2 \hookrightarrow  \Delta_1 , \Delta_2},\\
\end{align*}

\[\frac{\overbrace{\Gamma, \Gamma, \ldots, \Gamma}^{n} \hookrightarrow \overbrace{\Delta, \Delta, \ldots, \Delta}^{n}}
{\Gamma \hookrightarrow  \Delta}, n > 0.\]
    \item Logical rules:

\begin{align*}
& \frac{\Gamma, \varphi, \varphi \rightarrow_L \psi
\hookrightarrow \Delta, \psi }{\Gamma, \psi\rightarrow_L \varphi
\hookrightarrow  \Delta},& \qquad
\frac{\Gamma\hookrightarrow \Delta \quad \Gamma, \psi\hookrightarrow \varphi, \Delta}{\Gamma \hookrightarrow  \psi\rightarrow_L \varphi, \Delta}, \\
\end{align*}
\begin{align*}
& \frac{\Gamma,\psi,\varphi  \hookrightarrow \Delta \qquad
\Gamma,\bot \hookrightarrow \Delta}{\Gamma, \psi \&_L \varphi
\hookrightarrow \Delta},& \qquad
\frac{\Gamma,(\psi\rightarrow_L\bot)\rightarrow_L\varphi\hookrightarrow\psi,\varphi,\Delta}{\Gamma\hookrightarrow \psi \&_L \varphi,  \Delta}. \\
\end{align*}
\end{enumerate}

\section{Hypersequent calculus for infinite valued {\L}u\-kasiewicz's propositional logic}

The concept of hypersequent calculi\index{hypersequent calculus
for infinite valued \textsc{{\L}u\-kasiewicz}'s propositional
logic} was introduced for non-classical logics by \textsc{Avron}
in \cite{Avro2}. Hypersequents consist of multiple sequents
interpreted disjunctively. Therefore hypersequent rules include,
in addition to single sequent rules, external structural rules
that can operate on more than one
component at a time.\\

\begin {definitions}A hypersequent\index{hypersequent} is a multiset of components written
$$\Gamma_1 \hookrightarrow \Delta_1 | \ldots |\Gamma_n \hookrightarrow \Delta_n$$ with
the following interpretation: $\Gamma_1 \hookrightarrow \Delta_1 |
\ldots |\Gamma_n \hookrightarrow \Delta_n$ is logically valid in
\textsc{{\L}u\-kasiewicz}'s logic $\L_\infty$ iff for all
valuations $\mathrm{val}_I$ mapped from formulas of $\mathcal{L}$
to the structure $\mathfrak M_{[-1,0]}$ there exists $i$ such that
$\sum\limits_{\psi\in \Gamma_i}\mathrm{val}_I(\psi) \leqslant
\sum\limits_{\varphi\in \Delta_i}\mathrm{val}_I(\varphi)$.

\end {definitions}
\begin{enumerate}
    \item The axioms, i.e. the initial sequents\index{initial sequents}\index{axiom}, are as follows:

\begin{align*}
& (ID) \qquad \psi \hookrightarrow \psi, & \qquad  (\Lambda) \qquad \hookrightarrow,  & \qquad (\bot) \qquad\bot\hookrightarrow \psi.\\
\end{align*}
    \item Let $G$ be a variable for possibly empty hypersequents. The
structural rules\index{structural rules}:

$$\frac{G |\Gamma  \hookrightarrow \Delta}{G |\Gamma, \psi \hookrightarrow  \Delta},$$

\begin{align*}
& \frac{G |\Gamma  \hookrightarrow \Delta}{G |\Gamma  \hookrightarrow \Delta|\Gamma'  \hookrightarrow \Delta'},& \qquad \frac{G |\Gamma  \hookrightarrow \Delta|\Gamma  \hookrightarrow \Delta}{G |\Gamma  \hookrightarrow \Delta}, \\
&{}&{}\\
& \frac{G |\Gamma_1, \Gamma_2  \hookrightarrow \Delta_1 ,
\Delta_2}{G |\Gamma_1  \hookrightarrow \Delta_2 |
\Gamma_2\hookrightarrow \Delta_2},& \qquad
\frac{G |\Gamma_1\hookrightarrow \Delta_1 \quad G |\Gamma_2  \hookrightarrow \Delta_2}{G |\Gamma_1, \Gamma_2 \hookrightarrow \Delta_1, \Delta_2}. \\
\end{align*}
    \item Logical rules\index{logical rules}:

\begin{align*}
& \frac{G |\Gamma,\varphi  \hookrightarrow \psi, \Delta }{G
|\Gamma, \psi \rightarrow_L \varphi \hookrightarrow \Delta},&
\qquad \frac{G| \Gamma \hookrightarrow
\Delta\quad G |\Gamma,\psi \hookrightarrow \varphi, \Delta }{G |\Gamma\hookrightarrow \psi \rightarrow_L \varphi,  \Delta}, \\
\end{align*}
\begin{align*}
& \frac{G |\Gamma,\psi,\varphi  \hookrightarrow \Delta \qquad G
|\Gamma,\bot  \hookrightarrow \Delta}{G |\Gamma, \psi \&_L \varphi
\hookrightarrow \Delta},& \qquad \frac{G| \Gamma \hookrightarrow
\psi,\varphi,\Delta |\Gamma \hookrightarrow \bot, \Delta }{G |\Gamma\hookrightarrow \psi \&_L \varphi,  \Delta}. \\
\end{align*}
\end{enumerate}

As an example, consider the proof of \eqref{SchuL} using just
sequents:\\

$$1.~~~\frac{\varphi \hookrightarrow \varphi \qquad \psi \hookrightarrow \psi \qquad
\varphi \hookrightarrow \varphi \qquad \psi \hookrightarrow
\psi}{\varphi,\psi\hookrightarrow \psi,\varphi \qquad
\varphi,\psi\hookrightarrow \psi,\varphi},$$

$$2.~~~\frac{\varphi,\psi\hookrightarrow
\psi,\varphi \qquad \varphi,\psi\hookrightarrow
\psi,\varphi}{\varphi,\varphi\rightarrow_L\psi\hookrightarrow \psi
\qquad \varphi,\varphi\rightarrow_L\psi,\psi\hookrightarrow
\psi,\varphi},$$

$$3.~~~\frac{\varphi,\varphi\rightarrow_L\psi\hookrightarrow \psi \qquad
\varphi,\varphi\rightarrow_L\psi,\psi\hookrightarrow
\psi,\varphi}{\varphi,\varphi\rightarrow_L\psi\hookrightarrow
\psi,\psi \rightarrow_L\varphi},$$

$$4.~~~\frac{\varphi,\varphi\rightarrow_L\psi\hookrightarrow \psi,\psi
\rightarrow_L\varphi}{(\psi
\rightarrow_L\varphi)\rightarrow_L\varphi,\varphi\rightarrow_L\psi\hookrightarrow
\psi},$$

$$5.~~~\frac{(\psi
\rightarrow_L\varphi)\rightarrow_L\varphi,\varphi\rightarrow_L\psi\hookrightarrow
\psi}{(\psi
\rightarrow_L\varphi)\rightarrow_L\varphi\hookrightarrow
(\varphi\rightarrow_L\psi)\rightarrow_L\psi},$$

$$6.~~~\frac{(\psi \rightarrow_L\varphi)\rightarrow_L\varphi\hookrightarrow
(\varphi\rightarrow_L\psi)\rightarrow_L\psi}{\hookrightarrow((\psi
\rightarrow_L\varphi)\rightarrow_L\varphi)\rightarrow_L(
(\varphi\rightarrow_L\psi)\rightarrow_L\psi)}.$$\\

However there exist some tautologies that can be proved only by
means of hypersequents in the framework of this calculus (see
\cite{Gabbay3}).
\chapter{G\"{o}del's logic}
\section{Preliminaries}
\textit{\textsc{G\"{o}del}'s matrix
logic}\index{\textsc{G\"{o}del}'s matrix logic} $G_{[0,1]}$ is
interesting as the logic of linear order. It is the structure
$\langle [0,1]$, $\neg_G$, $\rightarrow_G$, $\vee$, $\wedge$,
$\widetilde{\exists}$, $\widetilde{\forall}, \{1 \} \rangle $,
where

\begin
{enumerate}
    \item for all $x\in [0,1]$, $ \neg_G x = x \rightarrow_G 0 $,
    \item for all $x, y \in [0,1]$, $x \rightarrow_G y = 1$ if $x\leqslant y$ and
$x\rightarrow_G y =y$ otherwise,
    \item for all $x, y \in [0,1]$, $x \vee y =  \max (x, y) $,
    \item for all $x, y \in [0,1]$, $x \wedge y = \min (x, y)
    $,
    \item for a subset $M \subseteq [0,1]$, $\widetilde{\exists}(M) = \max (M)
    $, where $\max (M)$ is a maximal element of $M$,
    \item for a subset $M \subseteq [0,1]$, $\widetilde{\forall}(M) = \min
    (M)$, where $\min (M)$ is a minimal element of $M$,
    \item $ \{1 \} $ is the set of designated truth values.

    \end {enumerate}

The truth value $0\in [0,1]$ is false, the truth value $1\in
[0,1]$ is true, and other truth values $x\in
(0,1)$ are neutral.\\

\section{Hilbert's type calculus for G\"{o}del's logic}

\textsc{G\"{o}del}'s logic denoted by
$\mathbf{G}$\index{\textsc{Hilbert}'s type calculus for
\textsc{G\"{o}del}'s logic} is one of the main intermediate
between intuitionistic and classical logics. It can be obtained by
adding $(\psi\rightarrow_G\varphi)\vee(\varphi\rightarrow_G\psi)$
to any axiomatization of intuitionistic logic (about intuitionism
see \cite{Heyt}, \cite{Krip}). The \textsc{G\"{o}del} logic was
studied by \textsc{Dummett} in \cite{Dum}.\\

The negation of $\mathbf{G}$ is understood as follows $$ \neg_G
\psi =: \psi \rightarrow_G \bot,$$ where $\bot$ is the truth
constant `falsehood'.\\

As an example, the \textsc{Hilbert}'s type calculus for
$\mathbf{G}$ consists of the following axioms:

\begin{equation}
(\psi\rightarrow_G\varphi)\rightarrow_G((\varphi\rightarrow_G\chi)\rightarrow_G(\psi\rightarrow_G\chi)),
\end{equation}
\begin{equation}
    \psi\rightarrow_G(\psi\vee\varphi),
\end{equation}
\begin{equation}
    \varphi\rightarrow_G(\psi\vee\varphi),
\end{equation}
\begin{equation}
(\varphi\rightarrow_G\chi)\rightarrow_G((\psi\rightarrow_G\chi)\rightarrow_G((\varphi\vee\psi)\rightarrow_G\chi)),
\end{equation}
\begin{equation}
(\varphi\wedge\psi)\rightarrow_G\varphi,
\end{equation}
\begin{equation}
(\varphi\wedge\psi)\rightarrow_G\psi,
\end{equation}
\begin{equation}
(\chi\rightarrow_G\varphi)\rightarrow_G((\chi\rightarrow_G\psi)\rightarrow_G(\chi\rightarrow_G(\varphi\wedge\psi))),
\end{equation}
\begin{equation}
(\varphi\rightarrow_G(\psi\rightarrow_G\chi))\rightarrow_G((\varphi\wedge\psi)\rightarrow_G\chi),
\end{equation}
\begin{equation}
((\varphi\wedge\psi)\rightarrow_G\chi)\rightarrow_G(\varphi\rightarrow_G(\psi\rightarrow_G\chi)),
\end{equation}
\begin{equation}
(\varphi\wedge\neg_G\varphi)\rightarrow_G\psi,
\end{equation}
\begin{equation}
(\varphi\rightarrow_G(\varphi\wedge\neg_G\varphi))\rightarrow_G\neg_G\varphi,
\end{equation}
\begin{equation}
(\psi\rightarrow_G\varphi)\vee(\varphi\rightarrow_G\psi),
\end{equation}
\begin {equation} \forall x ~\varphi (x) \rightarrow_G \varphi [x/t],
\end {equation}
\begin {equation} \varphi [x/t] \rightarrow_G \exists x ~\varphi(x),
\end {equation}
\\

\noindent where the formula $\varphi [x/t]$ is the result of
substituting
the term $t$ for all free occurrences of $x$ in $\varphi$,\\

\begin {equation} \forall x (\chi\rightarrow_G\varphi) \rightarrow_G (\chi\rightarrow_G\forall x\varphi ),
\end {equation}
\begin {equation} \forall x (\varphi\rightarrow_G\chi) \rightarrow_G (\exists x \varphi \rightarrow_G\chi),
\end {equation}
\begin {equation} \forall x (\chi\vee\varphi) \rightarrow_G (\chi\vee\forall x\varphi ),
\end {equation}
\\

\noindent where $x$ is not free in $\chi$.\\

In $\mathbf{G}$ there are the following inference rules:

\begin{enumerate}
    \item \textit {Modus ponens}\index{modus ponens}: from $
\varphi$ and $ \varphi \rightarrow_G \psi$ infer $ \psi$:

\[ \frac {\varphi,\quad \varphi \rightarrow_G \psi} {\psi}.
\]

    \item \textit{Substitution rule}\index{substitution rule}: we can substitute any formulas for propositional variables.

    \item \textit{Generalization}\index{generalization}: from $\varphi$ infer $\forall x \varphi
    (x)$:

\[ \frac {\varphi } {\forall x ~\varphi(x)}.
\]
\end{enumerate}

The \textsc{Hilbert}'s type propositional calculus for
\textsc{G\"{o}del}'s logic $\mathbf{G}$ can be obtained by
extending the axiom system \eqref{sch1} -- \eqref{sch2} by the
following axiom:
$$\psi\rightarrow_G(\psi\wedge\psi).$$

\section{Sequent calculus for G\"{o}del's propositional logic}

A \textit{sequent}\index{sequent} for \textsc{G\"{o}del}'s logic
$\mathbf{G}$ is understood in the standard way, i.e. as follows:
$\Gamma_1\hookrightarrow \Gamma_2$, where
$\Gamma_1=\{\varphi_1,\dots,\varphi_j\}$,
$\Gamma_2=\{\psi_1,\dots,\psi_i\}$, is logically valid in
$\mathbf{G}$ iff
$$\bigwedge_j\varphi_j\rightarrow_G\bigvee_i\psi_i$$ is logically
valid in $\mathbf{G}$\index{sequent calculus for
\textsc{G\"{o}del}'s propositional logic}.

\begin{enumerate}
    \item The initial sequents\index{initial sequents} in $\mathbf{G}$:
\begin{align*}
& (ID) \qquad \psi \hookrightarrow \psi, & \qquad  (\bot\hookrightarrow) \qquad \Gamma,\bot\hookrightarrow  \Delta,& \qquad (\hookrightarrow\bot) \qquad\Gamma\hookrightarrow \bot.\\
\end{align*}

    \item The structural rules\index{structural rules} are as follows:
\begin{align*}&\frac{\Gamma  \hookrightarrow \Delta}{\Gamma, \psi \hookrightarrow  \Delta},& \qquad\frac{\Gamma, \psi, \psi  \hookrightarrow \Delta}{\Gamma, \psi \hookrightarrow  \Delta},\\\end{align*}
\begin{align*}
& \frac{\Gamma_1,\psi\hookrightarrow \Delta\qquad
\Gamma_2\hookrightarrow\psi}{\Gamma_1,\Gamma_2\hookrightarrow
\Delta},& \qquad
\frac{\Gamma\hookrightarrow }{\Gamma\hookrightarrow \psi}. \\
\end{align*}

    \item Logical rules\index{logical rules}:
\begin{align*}
& \frac{\Gamma,\varphi \hookrightarrow \psi }{\Gamma
\hookrightarrow \varphi \rightarrow_G \psi},& \qquad
\frac{\Gamma_1 \hookrightarrow
\psi\quad \Gamma_2,  \varphi\hookrightarrow \Delta }{\Gamma_1,\Gamma_2, \psi \rightarrow_G \varphi\hookrightarrow \Delta}, \\
\end{align*}
\begin{align*}
& \frac{\Gamma_1,\psi\hookrightarrow
\Gamma_2}{\Gamma_1,\psi\wedge\chi\hookrightarrow \Gamma_2},&
\qquad \frac{\Gamma_1,\chi\hookrightarrow
\Gamma_2}{\Gamma_1,\psi\wedge\chi\hookrightarrow \Gamma_2}, \\
\end{align*}
\begin{align*}
& \frac{\Gamma_1,\psi\hookrightarrow \Gamma_2\qquad
\Gamma_1,\chi\hookrightarrow
\Gamma_2}{\Gamma_1,\psi\vee\chi\hookrightarrow \Gamma_2},& \qquad
\frac{\Gamma\hookrightarrow\psi\qquad \Gamma\hookrightarrow \chi}{\Gamma\hookrightarrow \psi\wedge\chi}, \\
\end{align*}
\begin{align*}
& \frac{\Gamma_1\hookrightarrow \psi}{\Gamma_1\hookrightarrow
\psi\vee\chi},& \qquad \frac{\Gamma_1\hookrightarrow
\chi}{\Gamma_1\hookrightarrow \psi\vee\chi}.
 \\
\end{align*}
    \end{enumerate}

\section{Hypersequent calculus for G\"{o}del's propositional logic}

A \textit{hypersequent}\index{hypersequent} in $\mathbf{G}$ is
interpreted in the standard way, i.e.,
disjunctively\index{hypersequent calculus for \textsc{G\"{o}del}'s
propositional logic}.

\begin{enumerate}
    \item The initial sequents\index{initial sequents} in $\mathbf{G}$ are as follows:
\begin{align*}
& (ID) \qquad \psi \hookrightarrow \psi, & \qquad  (\bot\hookrightarrow) \qquad \Gamma,\bot\hookrightarrow  \Delta,& \qquad (\hookrightarrow\bot) \qquad\Gamma\hookrightarrow \bot.\\
\end{align*}

    \item Let $G$ be a variable for possibly empty hypersequents. The
structural rules\index{structural rules}:
\begin{align*}&\frac{G |\Gamma  \hookrightarrow \Delta}{G |\Gamma, \psi \hookrightarrow  \Delta},& \qquad\frac{G |\Gamma, \psi, \psi  \hookrightarrow \Delta}{G |\Gamma, \psi \hookrightarrow  \Delta},\\\end{align*}
\begin{align*}
& \frac{G |\Gamma  \hookrightarrow \Delta}{G |\Gamma  \hookrightarrow \Delta|\Gamma'  \hookrightarrow \Delta'},& \qquad \frac{G |\Gamma  \hookrightarrow \Delta|\Gamma  \hookrightarrow \Delta}{G |\Gamma  \hookrightarrow \Delta}, \\
&{}&{}\\
& \frac{G |\Gamma_1, \Pi_2  \hookrightarrow \Delta_1\qquad G
|\Gamma_2, \Pi_2 \hookrightarrow \Delta_2}{G |\Gamma_1,\Gamma_2
\hookrightarrow \Delta_1 | \Pi_1,\Pi_2\hookrightarrow \Delta_2},&
\qquad
\frac{G |\Gamma\hookrightarrow }{G |\Gamma\hookrightarrow \psi}. \\
\end{align*}

    \item Logical rules\index{logical rules}:
\begin{align*}
& \frac{G |\Gamma,\varphi  \hookrightarrow \psi }{G |\Gamma
\hookrightarrow \varphi \rightarrow_G \psi},& \qquad \frac{G|
\Gamma_1 \hookrightarrow
\psi\quad G |\Gamma_2,  \varphi\hookrightarrow \Delta }{G |\Gamma_1,\Gamma_2, \psi \rightarrow_G \varphi\hookrightarrow  \Delta}, \\
\end{align*}
\begin{align*}
& \frac{G| \Gamma_1,\psi\hookrightarrow
\Gamma_2}{G|\Gamma_1,\psi\wedge\chi\hookrightarrow \Gamma_2},&
\qquad \frac{G|\Gamma_1,\chi\hookrightarrow
\Gamma_2}{G| \Gamma_1,\psi\wedge\chi\hookrightarrow \Gamma_2}, \\
\end{align*}
\begin{align*}
& \frac{G|\Gamma_1,\psi\hookrightarrow \Gamma_2\qquad
G|\Gamma_1,\chi\hookrightarrow
\Gamma_2}{G|\Gamma_1,\psi\vee\chi\hookrightarrow \Gamma_2},&
\qquad
\frac{G|\Gamma\hookrightarrow\psi\qquad G| \Gamma\hookrightarrow \chi}{G|\Gamma\hookrightarrow \psi\wedge\chi}, \\
\end{align*}
\begin{align*}
& \frac{G|\Gamma_1\hookrightarrow \psi}{G|\Gamma_1\hookrightarrow
\psi\vee\chi},& \qquad \frac{G|\Gamma_1\hookrightarrow
\chi}{G|\Gamma_1\hookrightarrow \psi\vee\chi}.
 \\
\end{align*}
    \item Cut rule\index{cut rule}:
$$\frac{G|\Gamma_1,\psi\hookrightarrow \Delta\qquad G| \Gamma_2\hookrightarrow\psi}{G|\Gamma_1,\Gamma_2\hookrightarrow \Delta}.$$
    \end{enumerate}

\chapter{Product logic}
\section{Preliminaries}
\textit{Product matrix logic}\index{Product matrix logic}
$\Pi_{[0,1]}$ behaves like $\L_\infty$ on the interval $(0, 1]$
and like $\mathbf{G}$ at 0. It is the structure $\langle [0,1]$,
$\neg_\Pi$, $\rightarrow_\Pi$, $\&_\Pi$, $\wedge$, $\vee$,
$\widetilde{\exists}$, $\widetilde{\forall}, \{1 \} \rangle $,
where

\begin
{enumerate}
    \item for all $x\in [0,1]$, $ \neg_\Pi x = x \rightarrow_\Pi 0 $,
    \item for all $x, y \in [0,1]$, $x \rightarrow_\Pi y = \left\{%
\begin{array}{ll}
    \frac{y}{x}, & \hbox{if $x>y$;} \\
    1, & \hbox{otherwise,} \\
\end{array}%
\right.    $
    \item for all $x, y \in [0,1]$, $x \&_\Pi y =  x \cdot y $,
    \item for all $x, y \in [0,1]$, $x \wedge y =  x \cdot (x\rightarrow_\Pi y) $,
    \item for all $x, y \in [0,1]$, $x \vee y = (( x \rightarrow_\Pi y)\rightarrow_\Pi y) \wedge ((y\rightarrow_\Pi x)\rightarrow_\Pi x)$,
    \item for a subset $M \subseteq [0,1]$, $\widetilde{\exists}(M) = \max (M)
    $, where $\max (M)$ is a maximal element of $M$,
    \item for a subset $M \subseteq [0,1]$, $\widetilde{\forall}(M) = \min
    (M)$, where $\min (M)$ is a minimal element of $M$,
    \item $ \{1 \} $ is the set of designated truth values.

    \end {enumerate}

The truth value $0\in [0,1]$ is false, the truth value $1\in
[0,1]$ is true, and other truth values $x\in
(0,1)$ are neutral.\\

\section{Hilbert's type calculus for Product logic}

In the Product logic denoted by
$\mathbf{\Pi}$\index{\textsc{Hilbert}'s type calculus for Product
logic} there are the following abridged notations:

$$ \neg_\Pi \psi =: \psi \rightarrow_\Pi \bot ,$$

\noindent where $\bot$ is the truth constant `falsehood',

$$\varphi \wedge \psi =:  \varphi \&_\Pi (\varphi\rightarrow_\Pi \psi) ,$$

$$\varphi \vee \psi =: (( \varphi \rightarrow_\Pi \psi)\rightarrow_\Pi \psi) \wedge
((\psi\rightarrow_\Pi \varphi)\rightarrow_\Pi \varphi).$$

The \textsc{Hilbert}'s type calculus for the Product logic
$\mathbf{\Pi}$ consists of the following axioms:

\begin{eqnarray}(\varphi\rightarrow_\Pi\psi)\rightarrow_\Pi ((\psi\rightarrow_\Pi\chi)\rightarrow_\Pi(\varphi\rightarrow_\Pi\chi)),\end{eqnarray}
\begin{eqnarray}(\varphi \&_\Pi \psi)\rightarrow_\Pi \varphi,\end{eqnarray}
\begin{eqnarray}(\varphi \&_\Pi \psi)\rightarrow_\Pi (\psi\&_\Pi\varphi),\end{eqnarray}
\begin{eqnarray}(\varphi \&_\Pi (\varphi \rightarrow_\Pi\psi))\rightarrow_\Pi(\psi\&_\Pi(\psi\rightarrow_\Pi\varphi)),\end{eqnarray}
\begin{eqnarray}(\varphi \rightarrow_\Pi (\psi \rightarrow_\Pi\chi))\rightarrow_\Pi ((\varphi\&_\Pi\psi)\rightarrow_\Pi\chi),\end{eqnarray}
\begin{eqnarray}((\varphi\&_\Pi\psi)\rightarrow_\Pi\chi)\rightarrow_\Pi (\varphi \rightarrow_\Pi (\psi \rightarrow_\Pi\chi)),\end{eqnarray}
\begin{eqnarray}((\varphi \rightarrow_\Pi \psi) \rightarrow_\Pi\chi)\rightarrow_\Pi (((\psi\rightarrow_\Pi\varphi)\rightarrow_\Pi\chi)\rightarrow_\Pi\chi),\end{eqnarray}
\begin{eqnarray}\bot \rightarrow_\Pi \psi,\end{eqnarray}
\begin{eqnarray}\neg_\Pi\neg_\Pi\chi\rightarrow_\Pi(((\varphi\&_\Pi\chi)\rightarrow_\Pi(\psi\&_\Pi\chi))\rightarrow_\Pi(\varphi\rightarrow_\Pi\psi)),\label{prod1}\end{eqnarray}
\begin{eqnarray}(\varphi\wedge\neg_\Pi\varphi) \rightarrow_\Pi \bot,\label{prod2}\end{eqnarray}
\begin {equation} \forall x ~\varphi (x) \rightarrow_\Pi \varphi [x/t],
\end {equation}
\begin {equation} \varphi [x/t] \rightarrow_\Pi \exists x ~\varphi(x),
\end {equation}
\\

\noindent where the formula $\varphi [x/t]$ is the result of
substituting
the term $t$ for all free occurrences of $x$ in $\varphi$,\\

\begin {equation} \forall x (\chi\rightarrow_\Pi\varphi) \rightarrow_\Pi (\chi\rightarrow_\Pi\forall x\varphi ),
\end {equation}
\begin {equation} \forall x (\varphi\rightarrow_\Pi\chi) \rightarrow_\Pi (\exists x \varphi \rightarrow_\Pi\chi),
\end {equation}
\begin {equation} \forall x (\chi\vee\varphi) \rightarrow_\Pi (\chi\vee\forall x\varphi ),
\end {equation}
\\

\noindent where $x$ is not free in $\chi$.\\

In $\mathbf{\Pi}$ there are the following inference
rules\index{inference rules}:

\begin{enumerate}
    \item \textit {Modus ponens}\index{modus ponens}: from $
\varphi$ and $ \varphi \rightarrow_\Pi \psi$ infer $ \psi$:

\[ \frac {\varphi,\quad \varphi \rightarrow_\Pi \psi} {\psi}.
\]

    \item \textit{Substitution rule}\index{substitution rule}: we can substitute any formulas for propositional variables.

    \item \textit{Generalization}\index{generalization}: from $\varphi$ infer $\forall x \varphi
    (x)$:

\[ \frac {\varphi } {\forall x ~\varphi(x)}.
\]
\end{enumerate}

We see that the \textsc{Hilbert}'s type propositional calculus for
Product logic is obtained by extending the axiom system
\eqref{sch1} -- \eqref{sch2} by the new axioms \eqref{prod1},
\eqref{prod2}.

\section{Sequent calculus for Product propositional logic}

A \textit{sequent}\index{sequent} for the Product logic
$\mathbf{\Pi}$\index{sequent calculus for Product propositional
logic} is interpreted as follows: $\Gamma_1\hookrightarrow
\Gamma_2$, where $\Gamma_1=\{\varphi_1,\dots,\varphi_j\}$,
$\Gamma_2=\{\psi_1,\dots,\psi_i\}$, is logically valid in
$\mathbf{\Pi}$ iff
$${\&_\Pi}{}_j\varphi_j\rightarrow_\Pi{\&_\Pi}{}_i\psi_i$$ is logically
valid in $\mathbf{\Pi}$, i.e. for all valuations $\mathrm{val}_I$
in $\Pi_{[0,1]}$ we have
$\prod\limits_j\mathrm{val}_I(\varphi_j)\leqslant\prod\limits_i\mathrm{val}_I(\psi_i)$.

\begin{enumerate}
    \item The initial sequents\index{initial sequents} in $\mathbf{\Pi}$:
\begin{align*}
& (ID) \qquad \psi \hookrightarrow \psi, & \qquad  (\bot\hookrightarrow) \qquad \Gamma,\bot\hookrightarrow  \Delta,& \qquad (\Lambda) \qquad\hookrightarrow ~~.\\
\end{align*}

    \item The structural rules\index{structural rules} are as follows:
\begin{align*}& \frac{\Gamma  \hookrightarrow \Delta}{\Gamma, \psi  \hookrightarrow  \Delta},& \qquad\frac{\Gamma_1 \hookrightarrow \Delta_1  \quad \Gamma_2 \hookrightarrow \Delta_2}
{\Gamma_1, \Gamma_2 \hookrightarrow  \Delta_1 , \Delta_2},\\
\end{align*}
\[\frac{\overbrace{\Gamma, \Gamma, \ldots, \Gamma}^{n} \hookrightarrow \overbrace{\Delta, \Delta, \ldots, \Delta}^{n}}
{\Gamma \hookrightarrow  \Delta}, n > 0.\]

    \item Logical rules\index{logical rules}:
\begin{align*}
& \frac{\Gamma  \hookrightarrow \psi }{\Gamma
\neg_\Pi\psi\hookrightarrow \Gamma_2},& \qquad \frac{\Gamma_1
\hookrightarrow
\Gamma_2\quad \Gamma_1, \psi\hookrightarrow \varphi,\Gamma_2 }{\Gamma_1 \hookrightarrow  \psi \rightarrow_\Pi \varphi,\Gamma_2}, \\
\end{align*}
\begin{align*}\frac{\Gamma_1,\psi,\varphi\hookrightarrow
\Gamma_2}{\Gamma_1,\psi\&_\Pi\varphi\hookrightarrow \Gamma_2},&
\qquad \frac{\Gamma_1\hookrightarrow
\psi,\varphi,\Gamma_2}{\Gamma_1\hookrightarrow \psi\&_\Pi\varphi,\Gamma_2}, \\
\end{align*}
$$\frac{\Gamma_1,\neg_\Pi\psi\hookrightarrow \Gamma_2\qquad \Gamma_1,\varphi,\varphi\rightarrow_\Pi\psi\hookrightarrow \psi,\Gamma_2}{\Gamma_1, \psi\rightarrow_\Pi\varphi\hookrightarrow\Gamma_2}.$$
    \end{enumerate}

\section{Hypersequent calculus for Product propositional logic}

A \textit{hypersequent}\index{hypersequent} in
$\mathbf{\Pi}$\index{hypersequent calculus for Product
propositional logic} is interpreted in the standard way, i.e.,
disjunctively.

\begin{enumerate}
    \item The initial sequents\index{initial sequents} in $\mathbf {\Pi}$:
\begin{align*}
& (ID) \qquad \psi \hookrightarrow \psi, & \qquad  (\bot\hookrightarrow) \qquad \Gamma,\bot\hookrightarrow  \Delta,& \qquad (\Lambda) \qquad\hookrightarrow ~~.\\
\end{align*}

    \item Let $G$ be a variable for possibly empty hypersequents. The
structural rules\index{structural rules}:

$$\frac{G |\Gamma  \hookrightarrow \Delta}{G |\Gamma,
\psi \hookrightarrow  \Delta},$$
\begin{align*}
& \frac{G |\Gamma  \hookrightarrow \Delta}{G |\Gamma  \hookrightarrow \Delta|\Gamma'  \hookrightarrow \Delta'},& \qquad \frac{G |\Gamma  \hookrightarrow \Delta|\Gamma  \hookrightarrow \Delta}{G |\Gamma  \hookrightarrow \Delta}, \\
&{}&{}\\
& \frac{G |\Gamma_1, \Gamma_2  \hookrightarrow \Delta_1 ,
\Delta_2}{G |\Gamma_1  \hookrightarrow \Delta_2 |
\Gamma_2\hookrightarrow \Delta_2},& \qquad
\frac{G |\Gamma_1\hookrightarrow \Delta_1 \quad G |\Gamma_2  \hookrightarrow \Delta_2}{G |\Gamma_1, \Gamma_2 \hookrightarrow \Delta_1, \Delta_2}. \\
\end{align*}

    \item Logical rules\index{logical rules}:
\begin{align*}
& \frac{G |\Gamma  \hookrightarrow \psi }{G |\Gamma
\neg_\Pi\psi\hookrightarrow \Gamma_2},& \qquad \frac{G| \Gamma_1
\hookrightarrow
\Gamma_2\quad G |\Gamma_1, \psi\hookrightarrow \varphi,\Gamma_2 }{G |\Gamma_1 \hookrightarrow  \psi \rightarrow_\Pi \varphi,\Gamma_2}, \\
\end{align*}
\begin{align*}\frac{G| \Gamma_1,\psi,\varphi\hookrightarrow
\Gamma_2}{G|\Gamma_1,\psi\&_\Pi\varphi\hookrightarrow \Gamma_2},&
\qquad \frac{G|\Gamma_1\hookrightarrow
\psi,\varphi,\Gamma_2}{G| \Gamma_1\hookrightarrow \psi\&_\Pi\varphi,\Gamma_2}, \\
\end{align*}
$$\frac{G|\Gamma_1,\neg_\Pi\psi\hookrightarrow \Gamma_2\qquad G|\Gamma_1,\varphi\hookrightarrow \psi,\Gamma_2}{G|\Gamma_1, \psi\rightarrow_\Pi\varphi\hookrightarrow\Gamma_2}.$$
    \end{enumerate}

\chapter{Nonlinear many valued logics}
\textsc{{\L}ukasiewicz}'s infinite-valued logic $\L_\infty$ is
linear in the sense that its truth functions are linear.
\textsc{G\"{o}del}'s infinite-valued logic $\mathbf{G}$ is linear
too, but its truth
function $\neg_G$ is discontinuous at the point 0.\\

Consider some sequences of nonlinear many-valued logics depending
on natural parameter $n$. They can be convergent to linear
many-valued logics (i.e., to $\L_\infty$ and $\mathbf{G}$) as
$n\to\infty$. For the first time these logics were regarded by
Russian logician D. \textsc{Bochvar}.

\section{Hyperbolic logics}
\begin{definitions} A many-valued logic $H_n$ is said to be hyperbolic\index{hyperbolic logic} if the
following truth operations $\neg_{H_n}$ and $\rightarrow_{H_n}$
for an appropriate matrix logic $ \mathfrak {M}_{H_n}$ are
hyperbola functions.
\end{definitions}

Let us consider two cases of hyperbolic matrix logic:
\textsc{{\L}u\-kasiewicz}'s and \textsc{G\"{o}del}'s hyperbolic
matrix logics for any positive integer $n$.\\

The ordered system $\langle  V _ {[0,1]}, \neg_{HL_n},
\rightarrow_{HL_n}, \vee_{HL_n}, \wedge_{HL_n},
\widetilde{\exists}, \widetilde{\forall}, \{1 \} \rangle $ is
called \textit {\textsc{{\L}u\-kasiewicz}'s hyperbolic matrix
logic}\index{\textsc{{\L}u\-kasiewicz}'s hyperbolic matrix logic}
$ \mathfrak {M}_{HL_n}$, where for any positive integer $n$,

\begin
{enumerate}
    \item $V_{[0,1]} = [0,
    1]$,
    \item for all $x\in V_{[0,1]}$, $ \neg_{HL_n} x = n\cdot \frac{1 - x}{1+x} $,
    \item for all $x, y \in V_{[0,1]}$, $x \rightarrow_{HL_n} y = \left\{%
\begin{array}{ll}
    1, & \hbox{if $x\leqslant y$;} \\
    n\cdot \frac{1 -(x-y)}{n+(x-y)}, & \hbox{if $x>y$.} \\
\end{array}%
\right.     $,
    \item for all $x, y \in V_{[0,1]}$, $x \vee_{HL_n} y = (x \rightarrow_{HL_n} y) \rightarrow_{HL_n} y $,
    \item for all $x, y \in V_{[0,1]}$, $x \wedge_{HL_n} y = \neg_{HL_n} (\neg_{HL_n} x \vee \neg_{HL_n} y) $,
    \item for a subset $M \subseteq V_{[0,1]}$, $\widetilde{\exists}(M) = \max (M)
    $, where $\max (M)$ is a maximal element of $M$,
    \item for a subset $M \subseteq V_{[0,1]}$, $\widetilde{\forall}(M) = \min
    (M)$, where $\min (M)$ is a minimal element of $M$,
    \item $ \{1 \} $ is the set of designated truth values.

    \end {enumerate}

The truth value $0\in V_{[0,1]}$ is false, the truth value $1\in
V_{[0,1]}$ is true, and other truth values $x\in
V_{[0,1]}\backslash \{0,1\}$ are neutral.\\

Obviously that if $n\to\infty$, then
\begin{itemize}
    \item $\neg_{HL_n} x =1-x$,
    \item $x \rightarrow_{HL_n} y =\min(1, 1-x +y)$,
    \item $x \vee_{HL_n} y =\max(x,y)$,
    \item $x \wedge_{HL_n} y =\min(x,y)$,
\end{itemize}

in other words, we obtain {\textsc{{\L}u\-kasiewicz}'s infinite
valued logic $\L_\infty$.\\

Some examples of many-valued tautologies for $ \mathfrak
{M}_{HL_n}$ are as follows:
$$\psi\rightarrow_{HL_n}\psi,$$
$$\neg_{HL_n}\neg_{HL_n}\psi\rightarrow_{HL_n}\psi,$$
$$\psi\rightarrow_{HL_n}\neg_{HL_n}\neg_{HL_n}\psi.$$

The ordered system $\langle  V _ {[0,1]}, \neg_{HG_n},
\rightarrow_{HG_n}, \vee, \wedge, \widetilde{\exists},
\widetilde{\forall}, \{1 \} \rangle $ is called \textit
{\textsc{G\"{o}\-del}'s hyperbolic matrix
logic}\index{\textsc{G\"{o}\-del}'s hyperbolic matrix logic} $
\mathfrak {M}_{HG_n}$, where for any positive integer $n$,

\begin
{enumerate}
    \item $V_{[0,1]} = [0,
    1]$,
    \item for all $x\in V_{[0,1]}$, $ \neg_{HG_n} x = (\frac{1 - x}{1+x})^n $,
    \item for all $x, y \in V_{[0,1]}$, $x \rightarrow_{HG_n} y = \left\{%
\begin{array}{ll}
    1, & \hbox{if $x\leqslant y$;} \\
    \frac{(n+1)\cdot y}{n+x}, & \hbox{if $x>y$.} \\
\end{array}%
\right.     $,
    \item for all $x, y \in V_{[0,1]}$, $x \vee y = \max(x, y)$,
    \item for all $x, y \in V_{[0,1]}$, $x \wedge y = \min(x, y)$,
    \item for a subset $M \subseteq V_{[0,1]}$, $\widetilde{\exists}(M) = \max (M)
    $, where $\max (M)$ is a maximal element of $M$,
    \item for a subset $M \subseteq V_{[0,1]}$, $\widetilde{\forall}(M) = \min
    (M)$, where $\min (M)$ is a minimal element of $M$,
    \item $ \{1 \} $ is the set of designated truth values.

    \end {enumerate}

The truth value $0\in V_{[0,1]}$ is false, the truth value $1\in
V_{[0,1]}$ is true, and other truth values $x\in
V_{[0,1]}\backslash \{0,1\}$ are neutral.\\

It is evident that if $n\to\infty$, then
\begin{itemize}
    \item $\neg_{HG_n} x =\left\{%
\begin{array}{ll}
    1, & \hbox{if $x=0$;} \\
    0, & \hbox{if $x\neq y$,} \\
\end{array}%
\right.    $
    \item $x \rightarrow_{HG_n} y =\left\{%
\begin{array}{ll}
    1, & \hbox{if $x\leqslant y$;} \\
    $y$, & \hbox{if $x> y$.} \\
\end{array}%
\right.    $,
\end{itemize}

i.e., we obtain \textsc{G\"{o}\-del}'s infinite valued logic
$\mathbf{G}$.

\section{Parabolic logics}
\begin{definitions} A many-valued logic $P_n$ is said to be parabolic\index{parabolic logic} if the
following truth operations $\neg_{P_n}$ and $\rightarrow_{P_n}$
for an appropriate matrix logic $ \mathfrak {M}_{P_n}$ are
parabola functions.
\end{definitions}

The ordered system $\langle  V _ {[0,1]}, \neg_{P_n},
\rightarrow_{P_n}, \vee_{P_n}, \wedge_{P_n}, \widetilde{\exists},
\widetilde{\forall}, \{1 \} \rangle $ is called \textit {parabolic
matrix logic} $ \mathfrak {M}_{P_n}$, where for any positive
integer $n$,

\begin
{enumerate}
    \item $V_{[0,1]} = [0,
    1]$,
    \item for all $x\in V_{[0,1]}$, $ \neg_{P_n} x = \frac{1-x^2}{n} $,
    \item for all $x, y \in V_{[0,1]}$, $x \rightarrow_{P_n} y = \left\{%
\begin{array}{ll}
    1, & \hbox{if $x^2\leqslant y$;} \\
    \frac{1-x^2}{n}+y, & \hbox{if $x^2>y$.} \\
\end{array}%
\right.     $,
    \item for all $x, y \in V_{[0,1]}$, $x \vee_{P_n} y = (x \rightarrow_{P_n} y) \rightarrow_{P_n} y $,
    \item for all $x, y \in V_{[0,1]}$, $x \wedge_{P_n} y = \neg_{P_n} (\neg_{P_n} x \vee \neg_{P_n} y) $,
    \item for a subset $M \subseteq V_{[0,1]}$, $\widetilde{\exists}(M) = \max (M)
    $, where $\max (M)$ is a maximal element of $M$,
    \item for a subset $M \subseteq V_{[0,1]}$, $\widetilde{\forall}(M) = \min
    (M)$, where $\min (M)$ is a minimal element of $M$,
    \item $ \{1 \} $ is the set of designated truth values.

    \end {enumerate}

The truth value $0\in V_{[0,1]}$ is false, the truth value $1\in
V_{[0,1]}$ is true, and other truth values $x\in
V_{[0,1]}\backslash \{0,1\}$ are neutral.\\

The ordered system $\langle  V _ {[0,1]}, \neg_{P_n},
\rightarrow_{P_n}, \vee_{P_n}, \wedge_{P_n}, \widetilde{\exists},
\widetilde{\forall}, \{1 \} \rangle $ is called \textit
{quasipa\-rabolic matrix logic}\index{quasipa\-rabolic matrix
logic} $ \mathfrak {M}_{P_n}$, where for any positive integer $n$,

\begin
{enumerate}
    \item $V_{[0,1]} = [0,
    1]$,
    \item for all $x\in V_{[0,1]}$, $ \neg_{P_n} x = \frac{1-x^2}{1+\frac{n}{n+1}\cdot x} $,
    \item for all $x, y \in V_{[0,1]}$, $x \rightarrow_{P_n} y = \min(1, \frac{1-x^2}{1+\frac{n}{n+1}\cdot x}+y)$,
    \item for all $x, y \in V_{[0,1]}$, $x \vee_{P_n} y = (x \rightarrow_{P_n} y) \rightarrow_{P_n} y $,
    \item for all $x, y \in V_{[0,1]}$, $x \wedge_{P_n} y = \neg_{P_n} (\neg_{P_n} x \vee \neg_{P_n} y) $,
    \item for a subset $M \subseteq V_{[0,1]}$, $\widetilde{\exists}(M) = \max (M)
    $, where $\max (M)$ is a maximal element of $M$,
    \item for a subset $M \subseteq V_{[0,1]}$, $\widetilde{\forall}(M) = \min
    (M)$, where $\min (M)$ is a minimal element of $M$,
    \item $ \{1 \} $ is the set of designated truth values.

    \end {enumerate}

The truth value $0\in V_{[0,1]}$ is false, the truth value $1\in
V_{[0,1]}$ is true, and other truth values $x\in
V_{[0,1]}\backslash \{0,1\}$ are neutral.\\

If $n\to\infty$, then the last matrix logic is transformed into
{\textsc{{\L}u\-kasiewicz}'s infinite valued logic $\L_\infty$.
This matrix is parabolic in a true sense just in case $n=0$.

\chapter{Non-Archimedean valued logics}

\section{Standard many-valued logics}

\begin{definitions}Let $V$ be a set of truth values. We will say that its
members are exclusive\index{exclusive members} if the powerset
$\mathcal{P}(V)$ is a Boolean algebra. \end{definitions}

It can be easily shown that all the elements of truth value sets
$\{0,1,\dots,n\}$, $[0,1]$ for \textsc{{\L}u\-kasiewicz}'s,
\textsc{G\"{o}\-del}'s, and Product logics are exclusive: for any
their members $x, y$ we have $\{x\} \cap \{y\} = \emptyset$ in
$\mathcal{P}(V)$. Therefore \textsc{{\L}ukasiewicz}'s,
\textsc{G\"{o}\-del}'s, and Product logics are based on the
premise of existence of \textit{\textsc{Shafer}'s
model}\index{\textsc{Shafer}'s model}. In other words, these
logics are built on the families of exclusive elements (see
\cite{Sentz}, \cite{Shafer}).\\

However, for a wide class of fusion problems, ``the intrinsic
nature of hypotheses can be only vague and imprecise in such a way
that precise refinement is just impossible to obtain in reality so
that the exclusive elements $\theta_i$ cannot be properly
identified and precisely separated'' (see \cite{Smar1}). This
means that if some elements $\theta_i$ of a frame $\Theta$ have
non-empty intersection, then sources of evidence don't provide
their beliefs with the same absolute interpretation of elements of
the same frame $\Theta$ and the conflict between sources arises
not only because of the possible unreliability of sources, but
also because of possible different and relative interpretation of
$\Theta$ (see \cite{Dezert2}, \cite{Dezert3}).

\section{Many-valued logics on DSm models\index{DSm models}\index{\textsc{Dezert-Smarandache} model}}
\label{SchumannSec3}

\begin{definitions} Let $V$ be a set of truth values. We will say that its
members are non-exclusive\index{non-exclusive members} if the
powerset $\mathcal{P}(V)$ is not closed under intersection
(consequently, it is not a Boolean algebra).
\end {definitions}

A many-valued logic is said to be a \textit{many-valued logic on
non-exclusive elements}\index{many-valued logic on non-exclusive
elements} if the elements of its set $V$ of truth values are
non-exclusive. These logics are also said to be a many-valued
logic on DSm model\index{DSm
models}\index{\textsc{Dezert-Smarandache} model}
(\textsc{Dezert-Smarandache} model).\\

Recall that a DSm model\index{DSm
models}\index{\textsc{Dezert-Smarandache} model}
(\textit{\textsc{Dezert-Smarandache} model}) is formed as a
hyper-power set\index{hyper-power set}. Let $\Theta = \{\theta_1,
\ldots , \theta_n\}$ be a finite set (called frame) of $n$
non-exclusive elements. The \textit{hyper-power set} $D^\Theta$ is
defined as the set of all composite propositions built from
elements of $\Theta$ with $\cap$ and $\cup$ operators such that:

\begin{enumerate}
    \item $\emptyset, \theta_1, \ldots , \theta_n \in D^\Theta$;
    \item if $A,B \in D^\Theta$, then $A \cap B  \in D^\Theta$ and $A \cup B  \in
    D^\Theta$;
    \item  no other elements belong to $D^\Theta$, except those obtained by using rules 1 or 2.
\end{enumerate}

The cardinality of $D^\Theta$ is majored by $2^{2^n}$ when the
cardinality of $\Theta$ equals $n$, i.e. $|\Theta| = n$. Since for
any given finite set $\Theta$, $|D^\Theta| \geq |2^\Theta|$, we
call $D^\Theta$ the hyper-power set of $\Theta$. Also, $D^\Theta$
constitutes what is called the \textit{DSm model\index{DSm
models}\index{\textsc{Dezert-Smarandache} model}}
$\mathcal{M}^f(\Theta)$. However elements $\theta_i$ can be truly
exclusive. In such case, the hyper-power set $D^\Theta$ reduces
naturally to the classical power set $2^\Theta$ and this
constitutes the most restricted hybrid DSm model\index{DSm
models}\index{\textsc{Dezert-Smarandache} model}, denoted by
$\mathcal{M}^0(\Theta)$, coinciding with \textsc{Shafer}'s model.
As an example, suppose that $\Theta = \{\theta_1, \theta_2\}$ with
$D^\Theta = \{\emptyset, \theta_1\cap\theta_2, \theta_1, \theta_2,
\theta_1\cup\theta_2\}$, where $\theta_1$ and $\theta_2$ are truly
exclusive (i.e., \textsc{Shafer}'s model $\mathcal{M}^0$ holds),
then because $ \theta_1 \cap \theta_2 =_{\mathcal{M}^0}
\emptyset$, one gets $D^\Theta = \{\emptyset, \theta_1 \cap
\theta_2 =_{\mathcal{M}^0} \emptyset, \theta_1, \theta_2, \theta_1
\cup \theta_2\} = \{\emptyset, \theta_1,
\theta_2, \theta_1 \cup \theta_2\} = 2^\Theta$.\\

Now let us show that every non-Archimedean
structure\index{non-Archimedean structure} is an infinite DSm
model\index{DSm models}\index{\textsc{Dezert-Smarandache} model},
but no vice versa. The most easy way of setting non-Archimedean
structures\index{non-Archimedean structure} was proposed by
\textsc{Robinson} in \cite{Robin}. Consider a set $\Theta$
consisting only of exclusive members. Let $I$ be any infinite
index set. Then we can construct an indexed family $\Theta^I$,
i.e., we
can obtain the set of all functions: $f\colon I \to \Theta$ such that $f(\alpha) \in \Theta$ for any $\alpha\in I$.\\

A \textit{filter}\index{filter} $\mathcal{F}$ on the index set $I$
is a family of sets $\mathcal{F} \subset \wp(I)$ for which:

\begin{enumerate}
    \item $A \in \mathcal{F}$, $A \subset B \Rightarrow B \in \mathcal{F}$;
    \item $A_1, \ldots, A_n \in \mathcal{F}\Rightarrow \bigcap\limits_{k = 1}^n A_k \in
    \mathcal{F}$;
    \item $\emptyset \notin \mathcal{F}$.
\end{enumerate}

The set of all complements for finite subsets of $I$ is a filter
and it is called a \textit{\textsc{Frechet}
filter}\index{\textsc{Frechet} filter} and it is denoted by
$\mathcal{U}$. A maximal filter (ultrafilter) that contains a
\textsc{Frechet} filter is called a
\textit{\textsc{Frechet} ultrafilter}\index{\textsc{Frechet} ultrafilter}.\\

Let $\mathcal{U}$ be a \textsc{Frechet} ultrafilter on $I$. Define
a new relation $\backsim$ on the set $\Theta^I$ by

\begin{equation}
f\backsim g \leftrightarrow \{\alpha \in I \colon f(\alpha) =
g(\alpha)\} \in \mathcal{U}. \label{SchumannEq5}
\end{equation}

It is easily proved that the relation $\backsim$ is an
equivalence. Notice that formula \eqref{SchumannEq5}
 means that $f$ and $g$ are
equivalent iff $f$ and $g$ are equal on an infinite index subset.
For each $f \in \Theta^I$ let $[f]$ denote the equivalence class
of $f$ under $\backsim$. The \textit{ultrapower}\index{ultrapower}
$\Theta^I/\mathcal{U}$ is then defined to be the set of all
equivalence classes $[f]$ as $f$ ranges over $\Theta^I$:
$$\Theta^I/\mathcal{U} \triangleq \{[f]
\colon f \in \Theta^I\}.$$

Also, \textsc{Robinson} has proved that each non-empty set
$\Theta$ has an ultrapower with respect to a \textsc{Frechet}
filter/ultrafilter $\mathcal{U}$. This ultrapower
$\Theta^I/\mathcal{U}$ is said to be a \textit{proper nonstandard
extension}\index{proper nonstandard extension} of $\Theta$ and it
is denoted by ${}^*\Theta$. Let us notice that if $\mathcal{U}$ is
not ultrafilter (it is just the \textsc{Frechet} filter), then
${}^*\Theta$ is not well-ordered.

\begin {propositions}Let $X$ be a
non-empty set. A nonstandard extension of $X$ consists of a
mapping that assigns a set ${}^*A$ to each $A \subseteq X^m$ for
all $m \geq 0$, such that ${}^*X$ is non-empty and the following
conditions are satisfied for all $m, n \geq 0$:

\begin{enumerate}
 \item The mapping preserves Boolean operations on subsets of $X^m$:
if $A \subseteq X^m$, then ${}^*A \subseteq ({}^*X)^m$; if $A,B
\subseteq X^m$, then ${}^*(A \cap B) = ({}^*A \cap {}^*B), {}^*(A
\cup B) = ({}^*A \cup {}^*B)$, and ${}^*(A \backslash B) = ({}^*A)
\backslash ({}^*B)$.
 \item The mapping preserves Cartesian products:
if $A \subseteq X^m$ and $B \subseteq X^n$, then ${}^*(A \times B)
= {}^*A \times {}^*B$, where $A \times B\subseteq X^{m+n}$. \hfill
$\Box $

\end{enumerate}
\end {propositions}

This proposition is proved in \cite{Hurd}.\\

Recall that each element of ${}^*\Theta$ is an equivalence class
$[f]$ as $f \colon I \to \Theta$. There exist two groups of
members of ${}^*\Theta$:

\begin{enumerate}
  \item functions that are constant, e.g., $f(\alpha)=m \in \Theta$
  for infinite index subset $\{\alpha \in I\}$. A constant function $[f=m]$ is denoted
  by ${}^*m$,
  \item functions that aren't constant.
\end{enumerate}

The set of all constant functions of ${}^*\Theta$ is called
\textit{standard set}\index{standard set} and it is denoted by
${}^\sigma\Theta$. The members of ${}^\sigma\Theta$ are called
\textit{standard}\index{standard members}. It is readily seen that
${}^\sigma\Theta$ and $\Theta$ are isomorphic:
${}^\sigma\Theta\simeq\Theta$.\\

The following proposition can be easily proved:

\begin {propositions}
For any set $\Theta$ such that $|\Theta|\geq 2$, there exists a
proper nonstandard extension ${}^*\Theta$ for which ${}^*\Theta
\backslash {}^\sigma \Theta \neq \emptyset$.
\end {propositions}

\noindent \emph{Proof.} Let $I_1 = \{\alpha_1, \alpha_2, \ldots,
\alpha_n, \ldots\} \subset I$ be an infinite set and let
$\mathcal{U}$ be a \textsc{Frechet} filter. Suppose that $\Theta_1
= \{m_1,\ldots, m_n\}$ such that $|\Theta_1| \geq 1$ is the subset
of $\Theta$ and there is a mapping:

\[f(\alpha) = \left\{ \begin{array}{ll}
    m_k & \hbox{if $\alpha = \alpha_k$;} \\
    m_0 \in \Theta & \hbox{if $\alpha\in I \backslash I_1$} \\
\end{array}
\right.
\]

and $f(\alpha) \neq m_k$ if $\alpha = \alpha_k\mod (n+1)$, $k=1,
\ldots,n$ and $\alpha \neq \alpha_k$.\\

Show that $[f] \in {}^*\Theta \backslash {}^\sigma\Theta$. The
proof is by reductio ad absurdum. Suppose there is $m \in \Theta$
such that $m \in [f(\alpha)]$. Consider the set:

\[I_2 = \{\alpha \in I \colon f(\alpha) = m\} = \left\{ \begin{array}{ll}
    \{\alpha_k\} & \hbox{if $m = m_k$, $k = 1, \ldots,n$;} \\
    I \backslash I_1 & \hbox{if $m = m_0$.} \\
    \emptyset & \hbox{if $m \notin \{m_0, m_1, \ldots, m_n\}$.} \\
\end{array}
\right.
\]

In any case $I_2\notin \mathcal{U}$, because we have
$\{\alpha_k\}\notin \mathcal{U}$, $\emptyset\notin \mathcal{U}$,
$I \backslash I_1 \notin \mathcal{U}$. Thus, $[f] \in {}^*\Theta
\backslash {}^\sigma
\Theta$.\hfill$\Box$\\

The standard members of ${}^*\Theta $ are
exclusive\index{exclusive members}, because their intersections
are empty. Indeed, the members of $\Theta $ were exclusive,
therefore the members of ${}^\sigma\Theta $ are exclusive too.
However the members of ${}^*\Theta \backslash {}^\sigma \Theta $
are non-exclusive. By definition, if a member $a \in {}^*\Theta$
is nonstandard, then there exists a standard member $b\in
{}^*\Theta $ such that $a \cap b \neq \emptyset$ (for example, see
the proof of proposition 2). We can also prove that there exist
non-exclusive members\index{non-exclusive members} $a\in
{}^*\Theta$, $b\in {}^*\Theta$ such that $a \cap b \neq
\emptyset$.

\begin {propositions}
There exist two inconstant functions $f_1, f_2$ such that the
intersection of $f_1, f_2$ isn't empty.
\end {propositions}
\noindent \emph{Proof}. Let $f_1 \colon I \to \Theta$ and $f_2
\colon I \to \Theta$. Suppose that $[f_i\neq n]$, $\forall n \in
\Theta$, $i = 1, 2$, i.e., $f_1, f_2$ aren't constant. By
proposition 2, these functions are nonstandard members of
${}^*\Theta$. Further consider an indexed family $F(\alpha)$ for
all $\alpha \in I$ such that $\{\alpha \in I \colon f_i(\alpha)
\in F(\alpha)\} \in \mathcal{U} \leftrightarrow [f_i] \in B$ as $i
= 1, 2$. Consequently it is possible that, for some $\alpha_j \in
I$, $f_1(\alpha_j) \cap f_2(\alpha_j) = n_j$
and $n_j \in F(\alpha_j)$. \hfill $\Box $\\

\begin {propositions}
Define the structure $\langle \mathcal{P}({}^*\Theta), \cap, \cup,
\neg, {}^*\Theta\rangle$ as follows

\begin{itemize}
    \item for any $A,B \in\mathcal{P}({}^*\Theta)$, $A\cap B=$
$\{f(\alpha)\colon f(\alpha)\in A \wedge f(\alpha)\in B\}$,
    \item for any $A,B \in\mathcal{P}({}^*\Theta)$, $A\cup B=$
$\{f(\alpha)\colon f(\alpha)\in A \vee f(\alpha)\in B\}$,
    \item for any $A \in\mathcal{P}({}^*\Theta)$, $\neg A=$
$\{f(\alpha)\colon f(\alpha)\in {}^*\Theta\backslash A \}$.
\end{itemize}

The structure $\langle \mathcal{P}({}^*\Theta), \cap, \cup, \neg,
{}^*\Theta\rangle$ is not a Boolean algebra if $|\Theta|\geq 2$.
\end {propositions}

\noindent \emph{Proof}. The set $\mathcal{P}({}^*\Theta)$ is not
closed under intersection. Indeed, some elements of ${}^*\Theta$
have a non-empty intersection (see the previous proposition) that
doesn't belong to $\mathcal{P}({}^*\Theta)$, i.e. they aren't
atoms of
$\mathcal{P}({}^*\Theta)$ of the form $[f]$.\hfill $\Box $\\

However, the structure $\langle \mathcal{P}({}^\sigma\Theta),
\cap, \cup, \neg, {}^\sigma\Theta\rangle$ is a Boolean algebra. In
the meantime
$\mathcal{P}({}^\sigma\Theta)\subset\mathcal{P}({}^*\Theta)$ if
$|\Theta|\geq 2$.\\

Thus, \textbf{non-Archimedean structures\index{non-Archimedean
structure} are infinite DSm-models\index{DSm
models}\index{\textsc{Dezert-Smarandache} model}, because
these contain non-exclusive members}.\\

In the next sections, we shall consider the following
non-Archimedean structures\index{non-Archimedean structure}:

\begin{enumerate}
  \item the nonstandard extension ${}^*\mathbf{Q}$ (called the field
  of hyperrational numbers\index{hyperrational numbers}),
  \item the nonstandard extension ${}^*\mathbf{R}$ (called the field
  of hyperreal numbers)\index{hyperreal numbers},
  \item the nonstandard extension $\mathbf{Z}_p$ (called the ring of $p$-adic integers\index{$p$-adic integers}) that we
  obtain as follows. Let the set $\mathbf{N}$ of natural numbers be the index set and let $\Theta = \{0,\ldots,
  p-1\}$. Then the nonstandard extension $\Theta^\mathbf{N}\backslash
  \mathcal{U}=\mathbf{Z}_p$.
\end{enumerate}

Further, we shall set the following logics on non-Archimedean
structures:

\begin{itemize}
    \item hyperrational valued \textsc{{\L}u\-kasiewicz}'s,
\textsc{G\"{o}\-del}'s, and Product logics\index{hyperrational
valued \textsc{{\L}u\-kasiewicz}'s logic}\index{hyperrational
valued \textsc{G\"{o}\-del}'s logic}\index{hyperrational valued
Product logic},
    \item hyperreal valued
\textsc{{\L}u\-kasiewicz}'s\index{hyperreal valued
\textsc{{\L}u\-kasiewicz}'s logic},
\textsc{G\"{o}\-del}'s\index{hyperreal valued
\textsc{G\"{o}\-del}'s logic}, and Product logics\index{hyperreal
valued Product logic},
    \item $p$-adic valued \textsc{{\L}u\-kasie\-wicz}'s\index{$p$-adic
valued  \textsc{{\L}u\-kasiewicz}'s logic},
\textsc{G\"{o}\-del}'s\index{$p$-adic valued
\textsc{G\"{o}\-del}'s logic}, and \textsc{Post}'s
logics\index{$p$-adic valued \textsc{Post}'s logic}.
\end{itemize}

Note that these many-valued logics are the particular cases of
logics on DSm models\index{DSm models}\index{\textsc{Dezert-Smarandache} model}.\\

Recall that non-Archimedean logical multiple-validities were
considered by \textsc{Schumann} in \cite{schu1}, \cite{schu2},
\cite{schu3}, \cite{schu4}, \cite{schu7}, \cite{schu8}.

\section{Hyper-valued partial order structure}

Assume that
${}^*\mathbf{Q}_{[0,1]}=\mathbf{Q}_{[0,1]}^\mathbf{N}/\mathcal{U}$
is a nonstandard extension of the subset
$\mathbf{Q}_{[0,1]}=\mathbf{Q} \cap [0,1]$ of rational numbers,
where $\mathcal{U}$ is the \textsc{Frechet} filter that may be no
ultrafilter, and
${}^\sigma\mathbf{Q}_{[0,1]}\subset{}^*\mathbf{Q}_{[0,1]}$ is the
subset of standard members. We can extend the usual order
structure on $\mathbf{Q}_{[0,1]}$ to a partial order structure on
${}^*\mathbf{Q}_{[0,1]}$:

\begin{enumerate}
    \item for rational
numbers $x$, $y \in \mathbf{Q}_{[0,1]}$ we have $x \leq y$ in
$\mathbf{Q}_{[0,1]}$ iff $[f] \leq [g]$ in
${}^*\mathbf{Q}_{[0,1]}$, where $\{\alpha \in \mathbf{N}\colon
f(\alpha) = x\} \in \mathcal{U}$ and $\{\alpha \in
\mathbf{N}\colon g(\alpha) = y\} \in \mathcal{U}$,

i.e., $f$ and $g$ are constant functions such that $[f] = {}^*x$
and $[g] = {}^*y$,
    \item each positive rational number ${}^*x\in {}^\sigma\mathbf{Q}_{[0,1]}$ is greater than any number $[f]\in {}^*\mathbf{Q}_{[0,1]}\backslash{}^\sigma\mathbf{Q}_{[0,1]}$,

    i.e., ${}^*x
    > [f]$ for any positive $x \in \mathbf{Q}_{[0,1]}$ and $[f] \in
{}^*\mathbf{Q}_{[0,1]}$, where $[f]$ isn't constant function.
\end{enumerate}

\noindent These conditions have the following informal sense:

\begin{enumerate}
    \item The sets ${}^\sigma\mathbf{Q}_{[0,1]}$ and $\mathbf{Q}_{[0,1]}$
    have isomorphic order structure.
    \item The set ${}^*\mathbf{Q}_{[0,1]}$ contains actual
    infinities that are less than any positive rational number of
    ${}^\sigma\mathbf{Q}_{[0,1]}$.
\end{enumerate}

Define this partial order structure on
${}^*\mathbf{Q}_{[0,1]}$\index{hyper-valued partial order
structure} as follows:

\begin{description}
\item [$\mathcal{O}_{{}^*\mathbf{Q}}$]
\begin{enumerate}
\item For any hyperrational numbers\index{hyperrational numbers}
$[f], [g] \in {}^*\mathbf{Q}_{[0,1]}$, we set $[f] \leq [g]$ if
$$\{\alpha \in \mathbf{N} \colon f(\alpha) \leq g(\alpha) \}
\in\mathcal{U}.$$ \item For any hyperrational
numbers\index{hyperrational numbers} $[f], [g] \in
{}^*\mathbf{Q}_{[0,1]}$, we set $[f] < [g]$ if $$\{\alpha \in
\mathbf{N} \colon f(\alpha) \leq g(\alpha) \} \in \mathcal{U}$$
and $[f] \neq [g]$, i.e., $\{\alpha \in \mathbf{N}\colon f(\alpha)
\neq g(\alpha) \} \in \mathcal{U}$. \item For any hyperrational
numbers\index{hyperrational numbers} $[f], [g] \in
{}^*\mathbf{Q}_{[0,1]}$, we set $[f] = [g]$ if $f \in[ g]$.
\end{enumerate}
\end{description}

This ordering relation is not linear, but partial, because there
exist elements $[f], [g] \in {}^*\mathbf{Q}_{[0,1]}$, which are
incompatible.\\

Introduce two operations $\max$, $\min$ in the partial order
structure $\mathcal{O}_{{}^*\mathbf{Q}}$:

\begin{enumerate}
\item for all hyperrational numbers\index{hyperrational numbers}
$[f], [g] \in {}^*\mathbf{Q}_{[0,1]}$, $\min([f], [g]) = [f]$ if
and only if $[f] \leq [g]$ under condition
$\mathcal{O}_{{}^*\mathbf{Q}}$, \item for all hyperrational
numbers\index{hyperrational numbers} $[f], [g] \in
{}^*\mathbf{Q}_{[0,1]}$, $\max([f], [g]) = [g]$ if and only if
$[f] \leq [g]$ under condition $\mathcal{O}_{{}^*\mathbf{Q}}$,
\item  for all hyperrational numbers\index{hyperrational numbers}
$[f], [g] \in {}^*\mathbf{Q}_{[0,1]}$, $\min([f], [g])=\max([f],
[g])= [f]=[g]$ if and only if $[f] = [g]$ under condition
$\mathcal{O}_{{}^*\mathbf{Q}}$, \item for all hyperrational
numbers\index{hyperrational numbers} $[f], [g] \in
{}^*\mathbf{Q}_{[0,1]}$, if $[f], [g]$ are incompatible under
condition $\mathcal{O}_{{}^*\mathbf{Q}}$, then $\min([f], [g]) =
[h]$ iff there exists $[h] \in {}^*\mathbf{Q}_{[0,1]}$ such that
$$\{\alpha \in \mathbf{N} \colon \min(f(\alpha),
g(\alpha))=h(\alpha) \} \in \mathcal{U}.$$ \item for all
hyperrational numbers\index{hyperrational numbers} $[f], [g] \in
{}^*\mathbf{Q}_{[0,1]}$, if $[f], [g]$ are incompatible under
condition $\mathcal{O}_{{}^*\mathbf{Q}}$, then $\max([f], [g]) =
[h]$ iff there exists $[h] \in {}^*\mathbf{Q}_{[0,1]}$ such that
$$\{\alpha \in \mathbf{N} \colon \max(f(\alpha),
g(\alpha))=h(\alpha) \} \in \mathcal{U}.$$
\end{enumerate}

\noindent It is easily seen that conditions 1 -- 3 are corollaries
of conditions 4, 5.

Note there exist the maximal number ${}^*1 \in
{}^*\mathbf{Q}_{[0,1]}$ and the minimal number ${}^*0 \in
{}^*\mathbf{Q}_{[0,1]}$ under condition
$\mathcal{O}_{{}^*\mathbf{Q}}$. Therefore, for any $[f]\in
{}^*\mathbf{Q}_{[0,1]}$, we have: $\max({}^*1, [f])= {}^*1$, $\max({}^*0, [f])= [f]$, $\min({}^*1,[f])= [f]$ and $\min({}^*0, [f])= {}^*0$.\\

Let us consider a nonstandard extension
${}^*\mathbf{R}_{[0,1]}=\mathbf{R}_{[0,1]}^\mathbf{N}/\mathcal{U}$
for the subset $\mathbf{R}_{[0,1]}=\mathbf{R} \cap [0,1]$ of real
numbers. Let
${}^\sigma\mathbf{R}_{[0,1]}\subset{}^*\mathbf{R}_{[0,1]}$ be the
subset of standard members. We can extend the usual order
structure on $\mathbf{R}_{[0,1]}$ to a partial order structure on
${}^*\mathbf{R}_{[0,1]}$:

\begin{enumerate}
    \item for real
numbers $x$, $y \in \mathbf{R}_{[0,1]}$ we have $x \leq y$ in
$\mathbf{R}_{[0,1]}$ iff $[f] \leq [g]$ in
${}^*\mathbf{R}_{[0,1]}$, where $\{\alpha \in \mathbf{N}\colon
f(\alpha) = x\} \in \mathcal{U}$ and $\{\alpha \in
\mathbf{N}\colon g(\alpha) = y\} \in \mathcal{U}$,

    \item each positive real number ${}^*x\in {}^\sigma\mathbf{R}_{[0,1]}$ is greater than any number $[f]\in {}^*\mathbf{R}_{[0,1]}\backslash{}^\sigma\mathbf{R}_{[0,1]}$,

\end{enumerate}

\noindent As before, these conditions have the following informal
sense:

\begin{enumerate}
    \item The sets ${}^\sigma\mathbf{R}_{[0,1]}$ and $\mathbf{R}_{[0,1]}$
    have isomorphic order structure.
    \item The set ${}^*\mathbf{R}_{[0,1]}$ contains actual
    infinities\index{actual infinities} that are less than any positive real number of
    ${}^\sigma\mathbf{R}_{[0,1]}$.
\end{enumerate}

\noindent Define this partial order structure on
${}^*\mathbf{R}_{[0,1]}$ as follows:

\begin{description}
\item [$\mathcal{O}_{{}^*\mathbf{R}}$] \begin{enumerate} \item For
any hyperreal numbers $[f], [g] \in {}^*\mathbf{R}_{[0,1]}$, we
set $[f] \leq [g]$ if $$\{\alpha \in \mathbf{N} \colon f(\alpha)
\leq g(\alpha) \} \in \mathcal{U}.$$ \item For any hyperreal
numbers $[f], [g] \in {}^*\mathbf{R}_{[0,1]}$, we set $[f] < [g]$
if $\{\alpha \in \mathbf{N} \colon f(\alpha) \leq g(\alpha) \} \in
\mathcal{U}$ and $[f] \neq [g]$, i.e., $\{\alpha \in
\mathbf{N}\colon f(\alpha) \neq g(\alpha) \} \in \mathcal{U}$.
\item For any hyperreal numbers $[f], [g] \in
{}^*\mathbf{R}_{[0,1]}$, we set $[f] =[g]$ if $f \in[ g]$.
\end{enumerate}
\end{description}

\noindent Introduce two operations $\max$, $\min$ in the partial
order structure $\mathcal{O}_{{}^*\mathbf{R}}$:

\begin{enumerate}
\item for all hyperreal numbers $[f], [g] \in
{}^*\mathbf{R}_{[0,1]}$, $\min([f], [g]) = [f]$ if and only if
$[f] \leq [g]$ under condition $\mathcal{O}_{{}^*\mathbf{R}}$,
\item for all hyperreal numbers $[f], [g] \in
{}^*\mathbf{R}_{[0,1]}$, $\max([f], [g]) = [g]$ if and only if
$[f] \leq [g]$ under condition $\mathcal{O}_{{}^*\mathbf{R}}$,
\item  for all hyperreal numbers $[f], [g] \in
{}^*\mathbf{R}_{[0,1]}$, $\min([f], [g])=\max([f], [g])= [f]=[g]$
if and only if $[f] = [g]$ under condition
$\mathcal{O}_{{}^*\mathbf{R}}$, \item for all hyperreal numbers
$[f], [g] \in {}^*\mathbf{R}_{[0,1]}$, if $[f], [g]$ are
incompatible under condition $\mathcal{O}_{{}^*\mathbf{R}}$, then
$\min([f], [g]) = [h]$ iff there exists $[h] \in
{}^*\mathbf{R}_{[0,1]}$ such that $$\{\alpha \in \mathbf{N} \colon
\min(f(\alpha), g(\alpha))=h(\alpha) \} \in \mathcal{U}.$$ \item
for all hyperreal numbers $[f], [g] \in {}^*\mathbf{R}_{[0,1]}$,
if $[f], [g]$ are incompatible under condition
$\mathcal{O}_{{}^*\mathbf{R}}$, then $\max([f], [g]) = [h]$ iff
there exists $[h] \in {}^*\mathbf{R}_{[0,1]}$ such that $$\{\alpha
\in \mathbf{N} \colon \max(f(\alpha), g(\alpha))=h(\alpha) \} \in
\mathcal{U}.$$
\end{enumerate}

Note there exist the maximal number ${}^*1 \in
{}^*\mathbf{R}_{[0,1]}$ and the minimal number ${}^*0 \in
{}^*\mathbf{R}_{[0,1]}$ under condition
$\mathcal{O}_{{}^*\mathbf{R}}$.\\

\section{Hyper-valued matrix logics}

\noindent Now define \textit{hyperrational-valued
\textsc{{\L}u\-kasiewicz}'s logic}\index{hyperrational-valued
\textsc{{\L}u\-kasiewicz}'s logic} $ \mathfrak
{M}_{{}^*\mathbf{Q}}$:

\begin{definitions}
The ordered system $\langle  V_{{}^*\mathbf{Q}}, \neg_L,
\rightarrow_L, \vee,\wedge, \widetilde{\exists},
\widetilde{\forall}, \{{}^*1 \} \rangle  $ is called
hyper\-rational valued \textsc{{\L}u\-kasiewicz}'s matrix logic $
\mathfrak {M}_{{}^*\mathbf{Q}}$, where

\begin
{enumerate}
    \item $V_{{}^*\mathbf{Q}}= {}^*\mathbf{Q}_{[0,1]}$ is the subset of hyperrational numbers\index{hyperrational numbers},
    \item for all $[x]\in V_{{}^*\mathbf{Q}}$, $ \neg_L [x] = {}^*1 - [x] $,
    \item for all $[x], [y] \in V_{{}^*\mathbf{Q}}$, $[x] \rightarrow_L [y] = \min ({}^*1, {}^*1 - [x] + [y]) $,
    \item for all $[x], [y] \in V_{{}^*\mathbf{Q}}$, $[x] \vee [y] = ([x] \rightarrow_L [y]) \rightarrow_L [y] = \max ([x], [y]) $,
    \item for all $[x], [y] \in V_{{}^*\mathbf{Q}}$, $[x] \wedge [y] = \neg_L (\neg_L [x] \vee \neg_L [y]) = \min ([x], [y])
    $,
    \item for a subset $M \subseteq V_{{}^*\mathbf{Q}}$, $\widetilde{\exists}(M) = \max (M)
    $, where $\max (M)$ is a maximal element of $M$,
    \item for a subset $M \subseteq V_{{}^*\mathbf{Q}}$, $\widetilde{\forall}(M) = \min
    (M)$, where $\min (M)$ is a minimal element of $M$,
    \item $ \{{}^*1\} $ is the set of designated truth values.
    \end {enumerate}
\end{definitions}

The truth value ${}^*0\in V_{{}^*\mathbf{Q}}$ is false, the truth
value ${}^*1\in V_{{}^*\mathbf{Q}}$ is true, and other truth
values
$x\in V_{{}^*\mathbf{Q}}\backslash \{{}^*0,{}^*1\}$ are neutral.\\

Continuing in the same way, define \textit{hyperreal valued
\textsc{{\L}u\-kasiewicz}'s matrix logic}\index{hyperreal valued
\textsc{{\L}u\-kasiewicz}'s matrix logic} $ \mathfrak
{M}_{{}^*\mathbf{R}}$:

\begin{definitions}
The ordered system $\langle  V_{{}^*\mathbf{R}}, \neg_L,
\rightarrow_L, \vee,\wedge, \widetilde{\exists},
\widetilde{\forall}, \{{}^*1 \} \rangle  $ is called hyper\-real
valued \textsc{{\L}u\-kasiewicz}'s matrix logic $ \mathfrak
{M}_{{}^*\mathbf{R}}$, where

\begin
{enumerate}
    \item $V_{{}^*\mathbf{R}}= {}^*\mathbf{R}_{[0,1]}$ is the subset of hyperreal numbers,
    \item for all $[x]\in V_{{}^*\mathbf{R}}$, $ \neg_L [x] = {}^*1 - [x] $,
    \item for all $[x], [y] \in V_{{}^*\mathbf{R}}$, $[x] \rightarrow_L [y] = \min ({}^*1, {}^*1 - [x] + [y]) $,
    \item for all $[x], [y] \in V_{{}^*\mathbf{R}}$, $[x] \vee [y] = ([x] \rightarrow_L [y]) \rightarrow_L [y] = \max ([x], [y]) $,
    \item for all $[x], [y] \in V_{{}^*\mathbf{R}}$, $[x] \wedge [y] = \neg_L (\neg_L [x] \vee \neg_L [y]) = \min ([x], [y])
    $,
    \item for a subset $M \subseteq V_{{}^*\mathbf{R}}$, $\widetilde{\exists}(M) = \max (M)
    $, where $\max (M)$ is a maximal element of $M$,
    \item for a subset $M \subseteq V_{{}^*\mathbf{R}}$, $\widetilde{\forall}(M) = \min
    (M)$, where $\min (M)$ is a minimal element of $M$,
    \item $ \{{}^*1\} $ is the set of designated truth values.
    \end {enumerate}
\end{definitions}

The truth value ${}^*0\in V_{{}^*\mathbf{R}}$ is false, the truth
value ${}^*1\in V_{{}^*\mathbf{R}}$ is true, and other truth
values $x\in V_{{}^*\mathbf{R}}\backslash \{{}^*0,{}^*1\}$ are
neutral.

\begin{definitions}
Hyper-valued \textit{\textsc{G\"{o}del}'s matrix
logic}\index{\textsc{G\"{o}del}'s matrix logic} $G_{{}^*[0,1]}$ is
the structure $\langle {}^*[0,1]$, $\neg_G$, $\rightarrow_G$,
$\vee$, $\wedge$, $\widetilde{\exists}$, $\widetilde{\forall},
\{{}^*1 \} \rangle $, where

\begin
{enumerate}
    \item for all $[x]\in {}^*[0,1]$, $ \neg_G [x] = [x] \rightarrow_G {}^*0 $,
    \item for all $[x], [y] \in {}^*[0,1]$, $[x] \rightarrow_G [y] = {}^*1$ if $[x]\leqslant [y]$ and
$[x]\rightarrow_G [y] =[y]$ otherwise,
    \item for all $[x], [y] \in {}^*[0,1]$, $[x] \vee [y] =  \max ([x], [y]) $,
    \item for all $[x], [y] \in {}^*[0,1]$, $[x] \wedge [y] = \min ([x], [y])
    $,
    \item for a subset $M \subseteq {}^*[0,1]$, $\widetilde{\exists}(M) = \max (M)
    $, where $\max (M)$ is a maximal element of $M$,
    \item for a subset $M \subseteq {}^*[0,1]$, $\widetilde{\forall}(M) = \min
    (M)$, where $\min (M)$ is a minimal element of $M$,
    \item $ \{{}^*1 \} $ is the set of designated truth values.

    \end {enumerate}
\end{definitions}

The truth value ${}^*0\in {}^*[0,1]$ is false, the truth value
${}^*1\in {}^*[0,1]$ is true, and other truth values $[x]\in
{}^*(0,1)$ are neutral.\\

\begin{definitions} Hyper-valued Product matrix logic\index{hyper-valued Product matrix logic} $\Pi_{{}^*[0,1]}$ is the structure $\langle {}^*[0,1]$, $\neg_\Pi$,
$\rightarrow_\Pi$, $\&_\Pi$, $\wedge$, $\vee$,
$\widetilde{\exists}$, $\widetilde{\forall}, \{{}^*1 \} \rangle $,
where

\begin
{enumerate}
    \item for all $[x]\in {}^*[0,1]$, $ \neg_\Pi [x] = [x] \rightarrow_\Pi {}^*0 $,
    \item for all $[x], [y] \in {}^*[0,1]$, $[x] \rightarrow_\Pi [y] = \left\{%
\begin{array}{ll}
    {}^*1, & \hbox{if $[x]\leqslant[y]$,} \\
    \min({}^*1, \frac{[y]}{[x]}), & \hbox{otherwise;}\\
\end{array}%
\right.    $
    \item for all $[x], [y] \in {}^*[0,1]$, $[x]\&_\Pi [y] =  [x ]\cdot [y] $,
    \item for all $[x], [y] \in {}^*[0,1]$, $[x] \wedge [y] =  [x] \cdot ([x]\rightarrow_\Pi [y]) $,
    \item for all $[x], [y] \in {}^*[0,1]$, $[x] \vee [y] = (( [x] \rightarrow_\Pi [y])\rightarrow_\Pi [y]) \wedge (([y]\rightarrow_\Pi [x])\rightarrow_\Pi [x])$,
    \item for a subset $M \subseteq {}^*[0,1]$, $\widetilde{\exists}(M) = \max (M)
    $, where $\max (M)$ is a maximal element of $M$,
    \item for a subset $M \subseteq {}^*[0,1]$, $\widetilde{\forall}(M) = \min
    (M)$, where $\min (M)$ is a minimal element of $M$,
    \item $ \{{}^*1 \} $ is the set of designated truth values.

    \end {enumerate}
\end{definitions}

The truth value ${}^*0\in {}^*[0,1]$ is false, the truth value
${}^*1\in {}^*[0,1]$ is true, and other truth values $[x]\in
{}^*(0,1)$ are neutral.\\

\section{Hyper-valued probability theory and hyper-valued fuzzy logic}

Let $X$ be an arbitrary set and let $\mathcal{A}$ be an
\textit{algebra of subsets}\index{algebra of subsets} $A\subseteq
X$, i.e.\
\begin{enumerate}
    \item union, intersection, and difference of two subsets of $X$ also belong
to $\mathcal{A}$;
    \item $\emptyset, X$ belong to $ \mathcal{A}$.
\end{enumerate}

Recall that a \textit{finitely additive probability
measure}\index{finitely additive probability measure} is a
nonnegative set function $\mathbf{P} (\cdot)$ defined for sets $A
\in \mathcal{A}$ that satisfies the following properties:
\begin{enumerate}
    \item $\mathbf{P} (A) \geq 0$ for all $A \in \mathcal{A},$
    \item $\mathbf{P} (X) = 1$ and $\mathbf{P} (\emptyset) = 0$,
    \item if $A \in \mathcal{A}$ and $B \in \mathcal{A}$ are disjoint, then
$\mathbf{P}(A \cup B) = \mathbf{P}(A) + \mathbf{P}(B)$. In
particular $\mathbf{P}(\neg A) = 1 - \mathbf{P}(A)$ for all $A \in
\mathcal{A}$.
\end{enumerate}

The algebra $\mathcal{A}$ is called a
$\sigma$-\textit{algebra}\index{$\sigma$-algebra} if it is assumed
to be closed under countable union (or equivalently, countable
intersection), i.e.\ if for every $n$, $A_n \in \mathcal{A}$
causes $A = \bigcup\limits_n A_n \in\mathcal{A}$.

A set function $\mathbf{P}(\cdot)$ defined on a $\sigma$-algebra
is called a \textit{countable additive probability
measure}\index{countable additive probability measure} (or a
$\sigma$-\textit{additive probability
measure}\index{$\sigma$-additive probability measure}) if in
addition to satisfying equations of the definition of finitely
additive probability measure, it satisfies the following countable
additivity property: for any sequence of pairwise disjoint sets
$A_n$, $\mathbf{P} (A) = \sum\limits_n \mathbf{P}(A_n)$. The
ordered system $\langle X, \mathcal{A}, \mathbf{P}\rangle$ is
called a \textit{probability space}\index{probability space}.

Now consider hyper-valued probabilities. Let $I$ be an arbitrary
set, let $\mathcal{A}$ be an algebra of subsets $A \subseteq I$,
and let $\mathcal{U}$ be a \textsc{Frechet}
ultrafilter\index{\textsc{Frechet} ultrafilter} on $I$. Set for $A
\in \mathcal{A}$:

\[\mu_{\mathcal{U}}(A) = \left\{ \begin{array}{ll}
    1, & \hbox{$A \in \mathcal{U}$;} \\
    0, & \hbox{$A \notin \mathcal{U}$.} \\
\end{array}
\right.
\]

\noindent Hence, there is a mapping $\mu_{\mathcal{U}} \colon
\mathcal{A} \to \{0, 1\}$ satisfying the following properties:

\begin{enumerate}
    \item $\mu_{\mathcal{U}}(\emptyset) = 0$, $\mu_{\mathcal{U}}(I) =
    1$;
    \item if $\mu_{\mathcal{U}}(A_1) = \mu_{\mathcal{U}}(A_2) =
    0$, then $\mu_{\mathcal{U}}(A_1 \cup A_2) = 0$;
    \item if $A_1\cap A_2 = \emptyset$, then $\mu_{\mathcal{U}}(A_1\cup A_2) = \mu_{\mathcal{U}}(A_1)+ \mu_{\mathcal{U}}(A_2)$.
\end{enumerate}

This implies that $\mu_{\mathcal{U}}$ is a probability measure.
Notice that $\mu_{\mathcal{U}}$ isn't $\sigma$-additive. As an
example, if $A$ is the set of even numbers and $B$ is the set of
odd numbers, then $A \in \mathcal{U}$ implies $B \notin
\mathcal{U}$, because the filter $\mathcal{U}$ is maximal. Thus,
$\mu_{\mathcal{U}}(A) = 1$ and $\mu_{\mathcal{U}}(B) = 0$,
although the cardinalities of $A$ and $B$ are equal.\\

The ordered system $\langle I, \mathcal{A},
\mu_{\mathcal{U}}\rangle$ is called a probability space.\\

Let's consider a mapping: $f \colon I \ni \alpha \mapsto f(\alpha)
\in M$. Two mappings $f$, $g$ are equivalent: $f\backsim g$ if
$\mu_{\mathcal{U}}(\{\alpha \in I \colon f(\alpha) = g(\alpha)\})
= 1$. An equivalence class of $f$ is called a probabilistic events
and is denoted by $[f]$. The set ${}^*M$ is the \textit{set of all
probabilistic events} of $M$. This ${}^*M$ is a proper nonstandard
extension defined above.\\

Under condition 1 of proposition 7, we can obtain a nonstandard
extension of an algebra $\mathcal{A}$ denoted by
${}^*\mathcal{A}$. Let ${}^*X$ be an arbitrary nonstandard
extension. Then the nonstandard algebra\index{nonstandard algebra}
${}^*\mathcal{A}$ is an algebra of subsets $A \subseteq {}^*X$ if
the following conditions hold:

\begin{enumerate}
    \item union, intersection, and difference of two subsets of ${}^*X$ also belong
to ${}^*\mathcal{A}$;
    \item $\emptyset, {}^*X$ belong to $ {}^*\mathcal{A}$.
\end{enumerate}

\begin {definitions} A hyperrational (respectively hyperreal) valued finitely additive probability
measure\index{hyperrational valued finitely additive probability
measure}\index{hyperreal valued finitely additive probability
measure} is a nonnegative set function ${}^*\mathbf{P} \colon
{}^*\mathcal{A} \to V_{{}^*\mathbf{Q}}$ (respectively
${}^*\mathbf{P} \colon {}^*\mathcal{A} \to V_{{}^*\mathbf{R}}$)
that satisfies the following properties:
\begin{enumerate}
    \item ${}^*\mathbf{P} (A) \geq {}^*0$ for all $A \in {}^*\mathcal{A},$
    \item ${}^*\mathbf{P} ({}^*X) = {}^*1$ and ${}^*\mathbf{P} (\emptyset) = {}^*0$,
    \item if $A \in {}^*\mathcal{A}$ and $B \in {}^*\mathcal{A}$ are disjoint, then
${}^*\mathbf{P}(A \cup B) = {}^*\mathbf{P}(A) +
{}^*\mathbf{P}(B)$. In particular ${}^*\mathbf{P}(\neg A) = {}^*1
- {}^*\mathbf{P}(A)$ for all $A \in {}^*\mathcal{A}$.
\end{enumerate}
\end {definitions}

\noindent Now consider hyper-valued fuzzy logic\index{hyper-valued
fuzzy logic}.

\begin {definitions}
Suppose ${}^*X $ is a nonstandard extension. Then a hyperrational
(respectively hyperreal) valued fuzzy set $A $ in ${}^*X $ is a
set defined by means of the membership function $ {}^*\mu _ {A} $:
${}^*X \to V_{{}^*\mathbf{Q}}$ (respectively by means of the
membership function $ {}^*\mu _ {A} $: ${}^*X \to
V_{{}^*\mathbf{R}}$).
\end {definitions}

A set $A \subseteq {}^*X$ is called \textit{crisp}\index{crisp
set} if $ {}^*\mu _ {A} (u)= {}^*1 $ or $ {}^*\mu _ {A} (u) =
{}^*0 $ for any $u \in
{}^*X$.\\

The logical operations on hyper-valued fuzzy sets are defined as
follows:
\begin {enumerate}
    \item $ {}^*\mu _ {A\cap B} (x) = \min({}^*\mu _ {A} (x), {}^*\mu _ {B} (x)) $;
    \item $ {}^*\mu _ {A\cup B} (x) = \max({}^*\mu _ {A} (x), {}^*\mu _ {B} (x)) $;
    \item $ {}^*\mu _ {A + B} (x) = {}^*\mu _ {A} (x) + {}^*\mu _ {B} (x) - {}^*\mu _ {A} (x)\cdot{}^* \mu _ {B} (x) $;
    \item $ {}^*\mu _ {\neg A} (x)  = \neg{}^*\mu _ {A} (x)  = {}^*1- {}^*\mu _ {A} (x)$.
\end {enumerate}

\chapter{$p$-Adic valued logics}
\section{Preliminaries}
Let us remember that the expansion \\

$n =\alpha_{-N}\cdot p^{-N} + \alpha_{-N+1}\cdot p^{-N+1}
+\ldots+\alpha_{-1}\cdot p^{-1} + \alpha_0 + \alpha_1\cdot p +
\ldots + \alpha_k\cdot p^k + \ldots= \sum\limits_{k=-N}^{+\infty}
\alpha_k\cdot p^k,$\\

\noindent where $\alpha_k \in \{0, 1,\ldots,p -1\} $, $\forall k
\in \mathbf{Z}$, and $\alpha_{-N} \neq 0$, is called the
\textit{canonical expansion of $p$-adic number}\index{canonical
expansion of $p$-adic number}\index{$p$-adic numbers} $n$ (or
$p$-adic expansion for $n$). The number $n$ is called $p$-adic.
This number can be identified with sequences of digits: $ n
=\ldots \alpha_2 \alpha_1 \alpha_0, \alpha_{-1} \alpha_{-2} \ldots
\alpha_{-N}$. We denote the set of such numbers by
$\mathbf{Q}_p$.\\

The expansion $n =\alpha_0 + \alpha_1\cdot p + \ldots +
\alpha_k\cdot p^k + \ldots=\sum\limits_{k=0}^\infty \alpha_k\cdot
p^k,$ where $\alpha_k \in \{0, 1,\ldots,p -1\} $, $\forall k \in
\mathbf{N}\cup\{0\}$, is called the \textit{expansion of $p$-adic
integer}\index{expansion of $p$-adic integer}\index{$p$-adic
integers} $n$. The integer $n$ is called $p$-adic. This number
sometimes has the following notation: $ n =\ldots \alpha_3
\alpha_2 \alpha_1 \alpha_0$. We denote the set of such numbers by
$\mathbf{Z}_p$.\\

If $n \in \mathbf{Z}_p$, $n \neq 0$, and its canonical expansion
contains only a finite number of nonzero digits $\alpha_j$, then
$n$ is natural number (and vice versa). But if $n \in
\mathbf{Z}_p$ and its expansion contains an infinite number of
nonzero digits $\alpha_j$, then $n$ is an infinitely large natural
number. Thus the set of $p$-adic integers contains actual
infinities $n \in \mathbf{Z}_p \backslash \mathbf{N}$, $n \neq 0$.
This is one of the most important features of non-Archimedean
number systems, therefore it is natural to compare $\mathbf{Z}_p$
with the set of nonstandard numbers ${}^*\mathbf{Z}$. Also, the
set $\mathbf{Z}_p$ contains {\it non-exclusive elements}.\\

It is evident that
$\mathbf{Q}_p=\{0,1,\dots,p-1\}^\mathbf{Z}\backslash
  \mathcal{U}$ and
$\mathbf{Z}_p=\{0,1,\dots,p-1\}^\mathbf{N}\backslash
  \mathcal{U}$.

\section{$p$-Adic valued partial order structure}

Extend the standard order structure on $\{0, \ldots, p-1\}$ to a
partial order structure on $\mathbf{Z}_p$. Define this partial
order structure on $\mathbf{Z}_p$ as follows:

\begin{description}
\item [$\mathcal{O}_{\mathbf{Z}_p}$] Let\index{$p$-adic valued
partial order structure} $x = \ldots x_n \ldots x_1 x_0$ and $y =
\ldots y_n \ldots y_1 y_0$ be the canonical expansions of two
$p$-adic integers $x, y \in \mathbf{Z}_p$.
\begin{enumerate}
\item We set $x \leq y$ if we have $x_n \leq y_n$ for each
$n=0,1,\ldots$ \item We set $x < y$ if we have $x_n \leq y_n$ for
each $n=0,1,\ldots$ and there exists $n_0$ such that $x_{n_0} <
y_{n_0}$. \item We set $x = y$ if $x_n = y_n$ for each
$n=0,1,\ldots$
\end{enumerate}
\end{description}

\noindent Now introduce two operations $\max$, $\min$ in the
partial order structure on $\mathbf{Z}_p$:

\begin{description}
\item[1]  for all $p$-adic integers $x, y \in \mathbf{Z}_p$,
$\min(x, y) = x$ if and only if $x \leq y$ under condition
$\mathcal{O}_{\mathbf{Z}_p}$, \item [2]  for all $p$-adic integers
$x, y \in \mathbf{Z}_p$, $\max(x, y) = y$ if and only if $x \leq
y$ under condition $\mathcal{O}_{\mathbf{Z}_p}$, \item [3]  for
all $p$-adic integers $x, y \in \mathbf{Z}_p$, $\max(x, y) =
\min(x, y) =x =y$ if and only if $x = y$ under condition
$\mathcal{O}_{\mathbf{Z}_p}$.
\end{description}

The ordering relation $\mathcal{O}_{\mathbf{Z}_p}$ is not linear,
but partial, because there exist elements $x, z \in \mathbf{Z}_p$,
which are incompatible. As an example, let $p = 2$ and let $x =
-\frac{1}{3} = \ldots 10101\ldots 101$, $ z = -\frac{2}{3} =
\ldots
01010\ldots 010$. Then the numbers $x$ and $z$ are incompatible.\\

\noindent Thus,
\begin{description}
\item[4]  Let $x = \ldots x_n \ldots x_1 x_0$ and $y = \ldots y_n
\ldots y_1 y_0$ be the canonical expansions of two $p$-adic
integers $x, y \in \mathbf{Z}_p$ and $x$, $y$ are incompatible
under condition $\mathcal{O}_{\mathbf{Z}_p}$. We get $\min(x, y) =
z= \ldots z_n \ldots z_1 z_0$, where, for each $n=0,1,\ldots$, we
set
\begin{enumerate}
\item  $z_n = y_n$ if $x_n \geq y_n$, \item  $z_n = x_n$ if $x_n
\leq y_n$, \item  $z_n = x_n = y_n$ if $x_n = y_n$.
\end{enumerate}
We get $\max(x, y) = z= \ldots z_n \ldots z_1 z_0$, where, for
each $n=0,1,\ldots$, we set
\begin{enumerate}
\item  $z_n = y_n$ if $x_n \leq y_n$, \item  $z_n = x_n$ if $x_n
\geq y_n$, \item  $z_n = x_n = y_n$ if $x_n = y_n$.
\end{enumerate}
\end{description}

It is important to remark that there exists the maximal number
$N_{max} \in \mathbf{Z}_p$ under condition
$\mathcal{O}_{\mathbf{Z}_p}$. It is easy to see:

$$N_{max} = - 1 = (p-1) + (p-1)\cdot p + \ldots + (p-1)\cdot p^k + \ldots=
\sum\limits_{k=0}^\infty (p-1)\cdot p^k$$

\noindent Therefore

\begin{description}
    \item[5]  $\min (x, N_{max}) = x$ and $\max (x, N_{max}) = N_{max}$ for any $x \in \mathbf{Z}_p$.
\end{description}

\section{$p$-Adic valued matrix logics}

Now consider $p$-\textit{adic valued \textsc{{\L}u\-kasiewicz}'s
matrix logic}\index{$p$-adic valued \textsc{{\L}u\-kasiewicz}'s
matrix logic} $ \mathfrak {M}_{\mathbf{Z}_p}$.

\begin{definitions} The ordered system $\langle  V _ {\mathbf{Z}_p},
\neg_L, \rightarrow_L, \vee,\wedge,\widetilde{\exists},
\widetilde{\forall},\{N_{max} \} \rangle  $ is called $p$-adic
valued \textsc{{\L}u\-kasiewicz}'s matrix logic $ \mathfrak
{M}_{\mathbf{Z}_p}$, where

\begin
{enumerate}
    \item $V_{\mathbf{Z}_p} = \{0, \ldots, N_{max} \} = \mathbf{Z}_p $,
    \item for all $x\in V_{\mathbf{Z}_p}$, $ \neg_L x = N_{max} - x $,
    \item for all $x, y \in V_{\mathbf{Z}_p}$, $x \rightarrow_L y = (N_{max} - \max(x, y)+ y) $,
    \item for all $x, y \in V_{\mathbf{Z}_p}$, $x \vee y = (x \rightarrow_L y) \rightarrow_L y = \max (x, y) $,
    \item for all $x, y \in V_{\mathbf{Z}_p}$, $x \wedge y = \neg_L (\neg_L x \vee \neg_L y) = \min (x, y)
    $,
    \item for a subset $M \subseteq V_{\mathbf{Z}_p}$, $\widetilde{\exists}(M) = \max (M)
    $, where $\max (M)$ is a maximal element of $M$,
    \item for a subset $M \subseteq V_{\mathbf{Z}_p}$, $\widetilde{\forall}(M) =
    \min
    (M)$, where $\min (M)$ is a minimal element of $M$,
    \item $ \{N_{max} \} $ is the set of designated truth values.
    \end {enumerate}
\end{definitions}

The truth value $0\in \mathbf{Z}_p$ is false, the truth value
$N_{max}\in \mathbf{Z}_p$ is true, and other truth values $x\in
\mathbf{Z}_p\{0,N_{max}\}$ are neutral.

\begin{definitions}
$p$-Adic valued  \textsc{G\"{o}del}'s matrix logic\index{$p$-adic
valued  \textsc{G\"{o}del}'s matrix logic} $G_{\mathbf{Z}_p}$ is
the structure $\langle V_{\mathbf{Z}_p}$, $\neg_G$,
$\rightarrow_G$, $\vee$, $\wedge$, $\widetilde{\exists}$,
$\widetilde{\forall}, \{N_{max} \} \rangle $, where

\begin
{enumerate}
    \item $V_{\mathbf{Z}_p} = \{0, \ldots, N_{max} \} = \mathbf{Z}_p $,
    \item for all $x\in V_{\mathbf{Z}_p}$, $ \neg_G x = x \rightarrow_G 0 $,
    \item for all $x, y \in V_{\mathbf{Z}_p}$, $x \rightarrow_G y = N_{max}$ if $x\leqslant y$ and
$x\rightarrow_G y =y$ otherwise,
    \item for all $x, y \in V_{\mathbf{Z}_p}$, $x \vee y =  \max (x, y) $,
    \item for all $x, y \in V_{\mathbf{Z}_p}$, $x \wedge y = \min (x, y)
    $,
    \item for a subset $M \subseteq V_{\mathbf{Z}_p}$, $\widetilde{\exists}(M) = \max (M)
    $, where $\max (M)$ is a maximal element of $M$,
    \item for a subset $M \subseteq V_{\mathbf{Z}_p}$, $\widetilde{\forall}(M) = \min
    (M)$, where $\min (M)$ is a minimal element of $M$,
    \item $ \{N_{max} \} $ is the set of designated truth values.

    \end {enumerate}
\end{definitions}

\begin{definitions}
$p$-Adic valued \textsc{Post}'s matrix logic\index{$p$-adic valued
\textsc{Post}'s matrix logic} $P_{\mathbf{Z}_p}$ is the structure
$\langle V_{\mathbf{Z}_p}, \neg_P,$ $\vee, \{N_{max} \}\rangle  $,
where
\begin {enumerate}
    \item $V_{\mathbf{Z}_p} = \mathbf{Z}_p $,
    \item for all $\ldots x_n \ldots
x_1 x_0\in V_{\mathbf{Z}_p}$, $ \neg_P \ldots x_n \ldots x_1 x_0 =
\ldots y_n \ldots y_1 y_0$, where $$ y_j= x_j + 1 \mod p $$

\noindent for each $j=0,1,2,\dots$,
    \item for all $x, y \in V_{\mathbf{Z}_p}$, $x \vee y = \max (x, y) $,
    \item $ \{N_{max} \} $ is the set of designated truth values.
\end {enumerate}
\end{definitions}

\begin {propositions}
The 2-adic valued logic\index{2-adic valued logic} $ \mathfrak
{M}_{\mathbf{Z}_2} = \langle V _ {\mathbf{Z}_2}$, $\neg_L$,
$\rightarrow_L$, $\vee$, $\wedge$, $\widetilde{\exists}$,
$\widetilde{\forall}$, $\{N_{max} \} \rangle $ is a Boolean
algebra.
\end {propositions}
\noindent \emph {Proof}. Indeed, the operation $\neg_L$ in $
\mathfrak {M}_{\mathbf{Z}_2}$ is the Boolean complement:

\begin{enumerate}
    \item $\max(x, \neg_L x) = N_{max}$,
    \item $\min(x, \neg_L x) = 0$.\hfill $\Box $
\end{enumerate}

\section{$p$-Adic probability theory and $p$-adic fuzzy logic}

Let us remember that the frequency theory of
probability\index{frequency theory of probability} was created by
von \textsc{Mises} in \cite{Mises}. This theory is based on the
notion of a collective: ``We will say that a collective is a mass
phenomenon or a repetitive event, or simply a long sequence of
observations for which there are sufficient reasons to believe
that the relative frequency of the observed attribute would tend
to a fixed limit if the observations were infinitely continued.
This limit will be called the probability of the attribute
considered
within the given collective'' \cite{Mises}.\\

As an example, consider a random experiment $\mathcal{S}$ and by
$L = \{s_1, \ldots, s_m\}$ denote the set of all possible results
of this experiment. The set $\mathcal{S}$ is called the label set,
or the set of attributes. Suppose there are $N$ realizations of
$\mathcal{S}$ and write a result $x_j$ after each realization.
Then we obtain the finite sample: $x = (x_1,\ldots, x_N), x_j \in
L$. A collective is an infinite idealization of this finite
sample: $x = (x_1,\ldots, x_N, \ldots), x_j \in L$. Let us compute
frequencies $\nu_N(\alpha; x) = n_N(\alpha ;x)/N$, where
$n_N(\alpha ;x)$ is the number of realizations of the attribute
$\alpha$ in the first $N$ tests. There exists the statistical
stabilization of relative frequencies: the frequency
$\nu_N(\alpha; x)$ approaches a limit as $N$ approaches infinity
for every label $\alpha \in L$. This limit $\mathbf{P}(\alpha) =
\lim \nu_N(\alpha; x)$ is said to be the probability of the label
$\alpha$ in the frequency theory of probability. Sometimes this
probability is denoted by $\mathbf{P}_x(\alpha)$ to show a
dependence on the collective $x$. Notice that the limits of
relative frequencies have to be stable with respect to a place
selection (a choice of a subsequence) in the collective.
\textsc{Khrennikov} developed von \textsc{Mises}' idea and
proposed the
frequency theory of $p$-adic probability\index{$p$-adic probability} in \cite{Khren2a}, \cite{Khren4}. We consider here some basic definitions of \textsc{Khrennikov}'s theory.\\

We shall study some ensembles $S = S_N$, which have a $p$-adic
volume\index{$p$-adic volume} $N$, where $N$ is the $p$-adic
integer. If $N$ is finite, then $S$ is the ordinary finite
ensemble. If $N$ is infinite, then $S$ has essentially $p$-adic
structure. Consider a sequence of ensembles $M_j$ having volumes
$l_j\cdot p^j$, $j = 0, 1, \ldots$ Get $S =
\bigcup\limits_{j=0}^\infty M_j$. Then the cardinality $|S| = N$.
We may imagine an ensemble $S$ as Get $S = \cup_{j=0}^\infty M_j$.
Then the cardinality $|S| = N$. We may imagine an ensemble $S$ as
being the population of a tower\index{tower} $T = T_S$, which has
an infinite number of floors with the following distribution of
population through floors: population of $j$-th
floor\index{population of $j$-th floor} is $M_j$. Set $T_k =
\cup_{j=0}^k M_j $. This is population of the first $k
+1$ floors. Let $A \subset S$ and let there exists: $n(A) = \lim \limits_{k\to \infty} n_k(A)$, where $n_k(A) = |A \cap T_k|$. The quantity $n(A)$ is said to be a $p$-adic volume of the set $A$. \\

\noindent We define the probability of $A$ by the standard
proportional relation:
\begin{equation}
\mathbf{P}(A) \triangleq \mathbf{P}_S(A) = \frac{n(A)}{N},
\end{equation}
\noindent
where $|S|= N$, $n(A) = |A\cap S|$.\\

We denote the family of all $A \subset S$, for which
$\mathbf{P}(A)$ exists, by $\mathcal{G}_S$. The sets $A \in
\mathcal{G}_S$ are said to be events. The ordered system $\langle
S, \mathcal{G}_S, \mathbf{P}_S\rangle$ is called a
$p$-\textit{adic ensemble probability space for the
ensemble}\index{$p$-adic ensemble probability space for the
ensemble} $S$.

\begin {propositions} Let $F$ be the set algebra which
consists of all finite subsets and their complements. Then $F
\subset \mathcal{G}_S$. \end {propositions}

\noindent \emph{Proof}. Let $A$ be a finite set. Then $n(A) = |A|$
and the probability of $A$ has the form: $$\mathbf{P}(A) =
\frac{|A|}{|S|}$$

Now let $B = \neg A$. Then $|B\cap T_k| = |T_k|- |A\cap T_k|$.
Hence there exists $\lim\limits_{k\to \infty} |B\cap T_k| = N -
|A|$. This equality implies the standard formula:
$$\mathbf{P}(\neg A) = 1 - \mathbf{P}(A)$$

In particular, we have: $\mathbf{P}(S) = 1$.\hfill $\Box $\\

\noindent The next propositions are proved in \cite{Khren2a}:

\begin {propositions} Let $A_1,A_2 \in \mathcal{G}_S$ and $A_1
\cap A_2 = \emptyset$. Then $A_1 \cup A_2 \in \mathcal{G}_S$ and
$$\mathbf{P}(A_1 \cup A_2) = \mathbf{P}(A_1) + \mathbf{P}(A_2).$$\hfill $\Box $
\end {propositions}

\begin {propositions}  Let $A_1,A_2 \in \mathcal{G}_S$. The
following conditions are equivalent:

\begin{enumerate}
    \item $A_1\cup A_2 \in \mathcal{G}_S$,
    \item $A_1\cap A_2 \in \mathcal{G}_S$,
    \item $A_1\backslash A_2 \in \mathcal{G}_S$,
    \item $A_2\backslash A_1 \in \mathcal{G}_S$.\hfill $\Box $
\end{enumerate}
\end {propositions}

But it is possible to find sets $A_1, A_2 \in \mathcal{G}_S$ such
that, for example, $A_1\cup A_2 \notin \mathcal{G}_S$. Thus, the
family $\mathcal{G}_S$ is not an algebra, but a semi-algebra (it
is closed only with respect to a finite unions of sets, which have
empty intersections). $\mathcal{G}_S$ is not closed with respect
to countable unions of such sets.

\begin {propositions} Let $A\in
\mathcal{G}_S$, $\mathbf{P}(A) \neq 0$ and $B\in \mathcal{G}_A$.
Then $B\in \mathcal{G}_S$ and the following \textsc{Bayes} formula
holds:
\begin{equation}
\mathbf{P}_A(B) = \frac{\mathbf{P}_S(B)}{\mathbf{P}_S(A)}
\end{equation}
\end {propositions}

\noindent \emph{Proof}. The tower $T_A$ of the $A$ has the
following population structure: there are $M_{A_j}$ elements on
the $j$-th floor. In particular, $T_{A_k} = T_k \cap A$. Thus
$$n_{A_k}(B) = |B \cap T_{A_k}| = |B \cap T_k| = n_k(B)$$
\noindent for each $B \subset A$. Hence the existence of $n_A(B) =
\lim\limits_{k\to \infty} n_{A_k}(B)$ implies the existence of
$n_S(B)$ with $n_S(B) = \lim\limits_{k\to \infty}  n_k(B)$.
Moreover, $n_S(B) = n_A(B)$. Therefore,

$$\mathbf{P}_A(B) =\frac{n_A(B)}{n_S(A)}=\frac{n_A(B)/|S|}{n_S(A)/|S|}.$$
\hfill $\Box $

\begin {propositions}  Let $N \in \mathbf{Z}_p$, $N \neq 0$ and
let the ensemble $S_{-1}$ have the $p$-adic volume $-1 = N_{max}$
(it is the largest ensemble).

\begin{enumerate}
    \item Then $S_N \in \mathcal{G}_{S_{-1}}$ and
$$\mathbf{P}_{S_{-1}}(S_N) = \frac{|S_N|}{|S_{-1}|}
= -N$$
    \item Then
$\mathcal{G}_{S_N} \subset \mathcal{G}_{S_{-1}}$ and probabilities
$\mathbf{P}_{S_N} (A)$ are calculated as conditional probabilities
with respect to the subensemble $S_N$ of ensemble $S_{-1}$:

$$\mathbf{P}_{S_N} (A) = \mathbf{P}_{S_{-1}}(\frac{A}{S_N}) =\frac{\mathbf{P}_{S_{-1}}(A)}{\mathbf{P}_{S_{-1}}(S_N)},
A \in \mathcal{G}_{S_N}$$
\end{enumerate}    \hfill $\Box $
\end {propositions}

Transform the \textsc{{\L}u\-kasiewicz} matrix logic $\mathfrak
M_{\mathbf{Z}_p}$ into a $p$-adic probability theory. Let us
remember that a formula $\varphi $ has the truth value $0 \in
\mathbf{Z}_p$ in $\mathfrak M_{\mathbf{Z}_p}$ if $\varphi$ is
false, a formula $\varphi$ has the truth value $N_{max} \in
\mathbf{Z}_p$ in $\mathfrak M_{\mathbf{Z}_p}$ if $\varphi$ is
true, and a formula $\varphi$ has other truth values $\alpha\in
\mathbf{Z}_p\backslash \{0, N_{max}\}$ in $\mathfrak
M_{\mathbf{Z}_p}$ if $\varphi$ is neutral.

\begin {definitions}
A function $\mathbf{P}(\varphi)$ is said to be a probability
measure of a formula $\varphi$ in $\mathfrak M_{\mathbf{Z}_p}$ if
$\mathbf{P}(\varphi)$ ranges over numbers of $\mathbf{Q}_p$ and
satisfies the following axioms:

\begin{enumerate}
    \item $\mathbf{P}(\varphi) = \frac{\mathrm{val}_I(\varphi)}{N_{max}}$, where $\mathrm{val}_I(\varphi)$ is a truth value of $\varphi$;
    \item if a conjunction $\varphi \wedge \psi$ has the truth value $0$,
    then $\mathbf{P}(\varphi \vee \psi) = \mathbf{P}(\varphi)
    + \mathbf{P}(\psi)$,
\item $\mathbf{P}(\varphi \wedge \psi) = \min
(\mathbf{P}(\varphi), \mathbf{P}(\psi))$.
\end{enumerate}
\end {definitions}

\noindent Notice that:
\begin{enumerate}
\item taking into account condition 1 of our definition, if
$\varphi$ has the truth value $N_{max}$ for any its
interpretations, i.e.\ $\varphi$ is a tautology, then
$\mathbf{P}(\varphi) = 1$ in all possible worlds, and if $\varphi$
has the truth value $0$ for any its interpretations, i.e.\
$\varphi$ is a contradiction, then $\mathbf{P}(\varphi) = 0$ in
all possible worlds; \item under condition 1, we obtain also
$\mathbf{P}(\neg \varphi) = 1 - \mathbf{P}(\varphi)$.
\end{enumerate}

Since $\mathbf{P}(N_{max}) = 1$, we have
$$\mathbf{P}(\max\{x\in V_{\mathbf{Z}_p}\})=\sum\limits_{x\in V_{\mathbf{Z}_p}} \mathbf{P}(x) = 1$$

All events have a conditional plausibility in the logical theory
of $p$-adic probability\index{$p$-adic probability}:
\begin{equation}
\mathbf{P}(\varphi) \leftrightarrow \mathbf{P}(\varphi/N_{max}),
\end{equation}

\noindent i.e., for any $\varphi$, we consider the conditional
plausibility that there is an event of $\varphi$, given an event
$N_{max}$,
\begin{equation}
\mathbf{P}(\varphi/\psi)=
\frac{\mathbf{P}(\varphi\wedge\psi)}{\mathbf{P}(\psi)}.
\end{equation}

The probability interpretation of the \textsc{{\L}u\-kasiewicz}
logic $\mathfrak M_{\mathbf{Z}_p}$ shows that this logic is a
special system of fuzzy logic\index{$p$-adic valued fuzzy logic}.
Indeed, we can consider the membership function $ \mu _ {A} $ as a
$p$-adic valued predicate.

\begin {definitions}
Suppose $X $ is a non-empty set. Then a $p $-adic valued fuzzy
set\index{$p $-adic valued fuzzy set} $A $ in $X $ is a set
defined by means of the membership function $ \mu _ {A} $: $X \to
\mathbf {Z} _ {p} $, where $ \mathbf {Z} _ {p} $ is the set of all
$p $-adic integers.
\end {definitions}

It is obvious that the set $A $ is completely determined by the
set of tuples ${\{} \langle  u, \mu _ {A} (u) \rangle
 \colon u \in X {\}}$. We define a norm\index{$p$-adic norm} $|\cdot|_p \colon \mathbf{Q}_p \to \mathbf{R}$ on
$\mathbf{Q}_p$ as follows:
$$|n = \sum\limits_{k=-N}^{+\infty} \alpha_k\cdot p^k|_p \triangleq p^{-L},$$

\noindent where $L = \max\{k\colon n \equiv 0 \mod {p^k}\}\geq 0$,
i.e. $L$ is an index of the first number distinct from zero in
$p$-adic expansion of $n$. Note that $|0|_p \triangleq 0$. The
function $|\cdot|_p$ has values $0$ and $\{p^\gamma\}_{\gamma \in
\mathbf{Z}}$ on $\mathbf{Q}_p$. Finally, $|x|_p \geq 0$ and $|x|_p
= 0 \leftrightarrow x = 0$. A set $A \subset X$ is called
\textit{crisp} if $ | \mu _ {A} (u) | _p = 1 $ or $ | \mu _ {A}
(u) | _p = 0 $ for any $u \in X$. Notice that $ | \mu _ {A} (u) =
1 | _p = 1 $ and $ | \mu _ {A} (u) = 0 | _p = 0 $. Therefore our
membership function is an extension of the classical
characteristic function. Thus, $A = B $ causes $ \mu _ {A} (u) =
\mu _ {B} (u) $ for all $u \in X $ and $A \subseteq B $ causes $ |
\mu _ {A} (u) | _p \leqslant
| \mu _ {B} (u) | _p $ for all $u \in X $.\\

In $p $-adic fuzzy logic, there always exists a non-empty
intersection of two crisp sets. In fact, suppose the sets $A $, $B
$ have empty intersection and $A $, $B $ are crisp\index{crisp
set}. Consider two cases under condition $ \mu _ {A} (u) \neq \mu
_ {B} (u) $ for any $u $. First, $ | \mu _ {A} (u) | _p = 0 $ or $
| \mu _ {A} (u) | _p = 1 $ for all $u $ and secondly $ | \mu _ {B}
(u) | _p = 0 $ or $ | \mu _ {B} (u) | _p = 1 $ for all $u $.
Assume we have $ \mu _ {A} (u_0) = N_{max} $ for some $u_0 $,
i.e., $ | \mu _ {A} (u_0) | _p = 1 $. Then $ \mu _ {B} (u_0) \neq
N_{max} $, but this doesn't mean that $ \mu _ {B} (u_0) = 0$. It
is possible that $ | \mu _ {A} (u_0) | _p =
1 $ and $ | \mu _ {B} (u_0) | _p = 1 $ for $u_0$.\\

Now we set logical operations on $p $-adic fuzzy sets:
\begin {enumerate}
\item $ \mu _ {A\cap B} (x) = \min(\mu _ {A} (x), \mu _ {B} (x))
$; \item $ \mu _ {A\cup B} (x) = \max(\mu _ {A} (x), \mu _ {B}
(x)) $; \item $ \mu _ {A + B} (x) = \mu _ {A} (x) + \mu _ {B} (x)
- \min(\mu _ {A} (x), \mu _ {B} (x)) $; \item $ \mu _ {\neg A} (x)
= \neg\mu _ {A} (x)  = N_{max}- \mu _ {A} (x)= -1 - \mu _ {A} (x)
$.
\end {enumerate}


\chapter{Fuzzy logics}

\section{Preliminaries} In this section we cover the main
essentials of the t-norm based approach, defining fuzzy logics as
logics based on t-norms and their residua. Recall that t-{\it
norm} is an operation $\ast\colon [0,1]^2\to [0,1]$ which is
commutative and associative, non-decreasing in both arguments and
have 1 as unit element and 0 as zero element, i.e.\\

$$x\ast y= y\ast x,$$
$$(x\ast y)\ast z=x\ast (y\ast z),$$
$$x \leq x' \mbox{ and }y \leq y' \mbox{ implies }x\ast y\leq x'\ast y',$$
$$1\ast x= x,\, 0\ast x=0.$$
Each t-norm\index{t-norm} determines uniquely its corresponding
implication $\Rightarrow$ ({\it residuum})\index{t-norm residuum}
satisfying for all $x,y,z\in [0,1]$,
$$z\leq x\Rightarrow y\mbox{ iff } x\ast z\leq y$$

The following are important examples of continuous t-norms and
their residua:
\begin{enumerate}
    \item {\it \textsc{{\L}ukasiewicz}'s logic}\index{\textsc{{\L}ukasiewicz}'s logic}:
\begin{itemize}
    \item $x\ast y= \max(x+y-1,0)$,
    \item $x\Rightarrow y=1$ for $x\leq y$ and $x \Rightarrow y=1-x+y$
otherwise.
\end{itemize}
In this logic $*$ and $\Rightarrow$ are denoted by $\&_L$ and
$\rightarrow_L$ respectively.
    \item {\it \textsc{G\"{o}del}'s logic}\index{\textsc{G\"{o}del}'s logic}: \begin{itemize}
    \item $x\ast y= \min(x,y)$,
    \item $x\Rightarrow y=1$ for $x\leq y$ and $x \Rightarrow y=y$
otherwise.
\end{itemize}
In this logic $*$ and $\Rightarrow$ are denoted by $\&_G$ and
$\rightarrow_G$ respectively.
    \item {\it Product logic}\index{Product logic}:
\begin{itemize}
    \item $x\ast y= x\cdot y$
    \item $x\Rightarrow y=1$ for $x\leq y$ and $x \Rightarrow y=y/x$
otherwise.
\end{itemize}

In this logic $*$ and $\Rightarrow$ are denoted by $\&_\Pi$ and
$\rightarrow_\Pi$ respectively.\end{enumerate}

A {\it regular residuated lattice}\index{regular residuated
lattice} (or a $BL$-{\it algebra})\index{$BL$-algebra} is an
algebra $\mathbf{L}_L=\langle L$, $\wedge$, $\vee$, $\ast$,
$\Rightarrow$, $0$, $1\rangle$ such that (1) $\langle
L,\wedge,\vee,0,1\rangle$ is a lattice with the largest element 1
and the least element 0, (2) $\langle L, *, 1\rangle$ is a
commutative semigroup with the unit element 1, i.e. $\ast$ is
commutative, associative, and $1\ast x =x$ for all $x$, (3) the
following conditions hold

$$z\leq (x\Rightarrow y)\mbox{ iff }x\ast z\leq y \mbox{ for all }x,y,z;$$
$$x\wedge y = x\ast (x\Rightarrow y);$$
$$x\vee y = ((x\Rightarrow y) \Rightarrow y)\wedge ((y\Rightarrow x) \Rightarrow x),$$
$$(x\Rightarrow y) \vee (y\Rightarrow x) =1.$$

It is known the following result. Let $*$ be a continuous t-norm
with residuum $\Rightarrow$, then $\mathcal{A}=\langle[0,1],
\min,\max$, $*$, $\Rightarrow$, $ 0$, $1\rangle$ is a
$BL$-algebra, called a standard $BL$-algebra; if $*$ is the
\textsc{{\L}ukasiewicz}, \textsc{G\"{o}del} or Product t-norm then
$\mathcal{A}$ is an $\L$-algebra, $\mathbf{G}$-algebra or
$\mathbf{\Pi}$-algebra respectively, called the standard
$\L$-algebra, $\mathbf{G}$-algebra or $\mathbf{\Pi}$-algebra.\\

Let us remember that every relation $R$ of $D^n$ can be
represented by its characteristic function $c_R$ : $D^n \to
\{0,1\}$ defined by setting $c_R(x_1,\ldots, x_n) = 1$ if $\langle
x_1,\ldots, x_n\rangle\in R$ and $c_R(x_1,\ldots,$ $ x_n) = 0$ if
$\langle x_1,\ldots, x_n\rangle\notin R$. We can also represent
the extension of a vague predicate by a generalized characteristic
function assuming values in a set $V$. Assume that
$V=\{0,1,\dots,n-1\}$ or $V=[0,1]$. A {\it fuzzy
relation}\index{fuzzy relation} or {\it fuzzy subset}\index{fuzzy
subset} $P$ of $D^n$ is any map $P$ : $D^n\to V$. For instance,
the subset $\emptyset\subset D^n$ (resp.\ $D^n\subseteq D^n$) can
be represented by a fuzzy relation $P$ that is considered as the
constant map $P$ : $D^n\ni x\mapsto 0$ (resp.\ $P$ : $D^n\ni
x\mapsto 1$), where $0$ is the least element of $V$ and $1$ is the
largest element of $V$. We denote by $\mathcal{F}(D^n)$ the class
of fuzzy subsets\index{class of fuzzy subsets} of $D^n$. It is
obvious that $\mathcal{P}(D^n)\subset\mathcal{F}(D^n)$, where
$\mathcal{P}(D^n)$ is the powerset.\\

Let $1$ be the top element of $V$. Observe that, if $P$ and $P'$
are fuzzy subsets of $D^n$, then we can define,\\

\begin{itemize}
    \item the {\it inclusion}\index{inclusion} by setting $P\subseteq P'$ iff
$P(x_1,\ldots, x_n) \leq P'(x_1,\ldots, x_n)$ for every $\langle
x_1,\ldots, x_n\rangle\in D^n$,
    \item {\it \textsc{{\L}ukasiewicz}'s relative complement}\index{\textsc{{\L}ukasiewicz}'s relative complement} $P\rightarrow_{L} P'$ by
    setting $(P\rightarrow_{L} P')$ $(x_1$, \dots, $x_n)= 1 -\max(P(x_1$, \dots, $x_n),
P'(x_1$, \dots, $x_n))+P'(x_1$, \dots, $x_n)$ for every $\langle
x_1,\ldots, x_n\rangle\in D^n$,
    \item {\it \textsc{{\L}ukasiewicz}'s complement}\index{\textsc{{\L}ukasiewicz}'s complement} $\neg_L P$
of $P$ by setting $(\neg_L P)( x_1,\ldots, x_n) = 1 - P(x_1$,
\dots, $x_n)$ for every $\langle x_1,\ldots, x_n\rangle\in D^n$,
         \item {\it \textsc{{\L}ukasiewicz}'s intersection}\index{\textsc{{\L}ukasiewicz}'s intersection} $P\&_{L} P'$ as $(P\&_{L}
P')(x_1$, \dots, $x_n)= \max(0$, $P(x_1$, \dots, $x_n)+P'(x_1$,
\dots, $x_n)-1)$ for every $\langle x_1,\ldots, x_n\rangle\in
D^n$,
        \item the {\it Product relative complement}\index{Product relative complement}
$P\rightarrow_{\Pi} P'$ by setting $(P\rightarrow_{\Pi} P')(x_1$,
\dots, $x_n)= 1$ if $P(x_1$, \dots, $x_n)\leq P'(x_1$, \dots,
$x_n)$  for every $\langle x_1,\ldots, x_n\rangle\in D^n$ and
$(P\rightarrow_{\Pi} P')(x_1$, \dots, $x_n)= \frac {P(x_1, \dots,
x_n)}{P'(x_1, \dots, x_n)}$ otherwise,
     \item the {\it Product complement}\index{Product complement} $\neg_\Pi P$ as $\neg_\Pi P(x_1$,
\dots, $x_n):= P(x_1$, \dots, $x_n)\rightarrow_{\Pi} 0$ for every
$\langle x_1,\ldots, x_n\rangle\in D^n$,
    \item the {\it Product intersection}\index{Product intersection} $P\&_{\Pi} P'$ as $(P\&_{\Pi}
P')(x_1$, \dots, $x_n)= P(x_1$, \dots, $x_n)\cdot P'(x_1$, \dots,
$x_n)$ for every $\langle x_1,\ldots, x_n\rangle\in D^n$,
    \item the {\it union}\index{union} $P\vee P'$
as $(P\vee P')(x_1$, \dots, $x_n)= \max(P(x_1$, \dots, $x_n),
P'(x_1$, \dots, $x_n))$ for every $\langle x_1,\ldots,
x_n\rangle\in D^n$,
    \item the {\it intersection}\index{intersection} $P\wedge P'$ by setting
$(P\wedge P')(x_1$, \dots, $x_n)= \min(P(x_1$, \dots, $x_n),
P'(x_1$, \dots, $x_n))$ for every $\langle x_1,\ldots,
x_n\rangle\in D^n$,
    \item the {\it union} $\bigvee\limits_{i\in
I}P_i$, given a family $(P_i)_{i\in I}$ of fuzzy subsets of $D^n$,
as $$\bigvee\limits_{i\in I}P_i(x_1, \dots, x_n)= \max\{P_i(x_1,
\dots, x_n)\colon i \in I\}$$ for every $\langle x_1,\ldots,
x_n\rangle\in D^n$,
    \item the {\it intersection}
$\bigwedge\limits_{i\in I}P_i$, given a family $(P_i)_{i\in I}$ of
fuzzy subsets of $D^n$, as $$\bigwedge\limits_{i\in I}P_i(x_1,
\dots, x_n)= \min\{P_i(x_1, \dots, x_n)\colon i \in I\}$$ for
every $\langle x_1,\ldots, x_n\rangle\in D^n$.
\end{itemize}

The structure $\mathbf{L}_{V} = \langle \mathcal{F}(D^n)$,
$\wedge$, $\vee$, $\ast$, $\Rightarrow$, $\bot(D^n)$,
$\top(D^n)\rangle $ is a $BL$-algebra\index{$BL$-algebra}, where
$\bot(D^n)$ and $\top(D^n)$ are the constant maps $\bot(D^n)$ : $
D^n \ni x \mapsto 0\in V$ and $\top(D^n)$ : $ D^n \ni x \mapsto
1\in V$ respectively. It is the direct power, with index set
$D^n$, of the structure $\langle V$, $\wedge$, $\vee$, $\ast$,
$\Rightarrow$, $0$, $1\rangle $.

\section{Basic fuzzy logic $BL\forall$}

The basic fuzzy logic\index{basic fuzzy logic} denoted by
$BL\forall$ has just two initial propositional operations: $\&$,
$\rightarrow$, which are understood as t-norm and its residuum
respectively. The negation is derivable:

$$ \neg \psi =: \psi \rightarrow \bot ,$$

\noindent where $\bot$ is the truth constant `falsehood'.\\

In $BL\forall$ we can define the following new operations:

\begin{itemize}
    \item $\varphi\wedge\psi:=\varphi\&(\varphi\rightarrow\psi)$,
    \item $\varphi\vee\psi:=((\varphi\rightarrow\psi)\rightarrow\psi)\wedge((\psi\rightarrow\varphi)\rightarrow\varphi)$,
    \item $\varphi\leftrightarrow\psi:=(\varphi\rightarrow\psi)\&(\psi\rightarrow\varphi)$,
    \item $\varphi\oplus\psi:=\neg\varphi\rightarrow\psi$,
    \item $\varphi\ominus\psi:=\varphi\&\neg\psi$.\end{itemize}

The \textsc{Hilbert}'s type calculus for
$BL\forall$\index{\textsc{Hilbert}'s type calculus for
$BL\forall$} consists of the following axioms:

\begin{eqnarray}(\varphi\rightarrow\psi)\rightarrow ((\psi\rightarrow\chi)\rightarrow(\varphi\rightarrow\chi)),\label{sch1}\end{eqnarray}
\begin{eqnarray}(\varphi \& \psi)\rightarrow \varphi,\end{eqnarray}
\begin{eqnarray}(\varphi \& \psi)\rightarrow (\psi\&\varphi),\end{eqnarray}
\begin{eqnarray}(\varphi \& (\varphi \rightarrow\psi))\rightarrow(\psi\&(\psi\rightarrow\varphi)),\end{eqnarray}
\begin{eqnarray}(\varphi \rightarrow (\psi \rightarrow\chi))\rightarrow ((\varphi\&\psi)\rightarrow\chi),\end{eqnarray}
\begin{eqnarray}((\varphi\&\psi)\rightarrow\chi)\rightarrow (\varphi \rightarrow (\psi \rightarrow\chi)),\end{eqnarray}
\begin{eqnarray}((\varphi \rightarrow \psi) \rightarrow\chi)\rightarrow (((\psi\rightarrow\varphi)\rightarrow\chi)\rightarrow\chi),\end{eqnarray}
\begin{eqnarray}\bot \rightarrow \psi,\label{sch2}\end{eqnarray}
\begin {equation} \forall x ~\varphi (x) \rightarrow \varphi [x/t],
\end {equation}
\begin {equation} \varphi [x/t] \rightarrow \exists x ~\varphi(x),
\end {equation}
\\

\noindent where the formula $\varphi [x/t]$ is the result of
substituting
the term $t$ for all free occurrences of $x$ in $\varphi$,\\

\begin {equation} \forall x (\chi\rightarrow\varphi) \rightarrow (\chi\rightarrow\forall x\varphi ),
\end {equation}
\begin {equation} \forall x (\varphi\rightarrow\chi) \rightarrow (\exists x \varphi \rightarrow\chi),
\end {equation}
\begin {equation} \forall x (\chi\vee\varphi) \rightarrow (\chi\vee\forall x\varphi ),
\end {equation}
\\

where $x$ is not free in $\chi$.\\

In $BL\forall$ there are the following inference
rules\index{inference rules}:

\begin{enumerate}
    \item \textit {Modus ponens}\index{modus ponens}: from $
\varphi$ and $ \varphi \rightarrow \psi$ infer $ \psi$:

\[ \frac {\varphi,\quad \varphi \rightarrow \psi} {\psi}.
\]

    \item \textit{Substitution rule}\index{substitution rule}: we can substitute any formulas for propositional variables.

    \item \textit{Generalization}\index{generalization}: from $\varphi$ infer $\forall x ~\varphi
    (x)$:

\[ \frac {\varphi } {\forall x ~\varphi(x)}.
\]
\end{enumerate}

\begin{theorems}[Soundness and Completeness]  \index{completeness theorem}\index{soundness theorem} Let $\varphi$ be a formula of
$BL\forall$, $T$ a $BL\forall$-theory. Then the following
conditions are equivalent:
\begin{itemize}
    \item $T \vdash\varphi$;
    \item $\mathrm{val}_I(\varphi)=1$ for each $BL$-algebra (with infinite intersection and infinite union) that is model
    for $T$.
\end{itemize}
\end{theorems}

\noindent {\it Proof}. See \cite{Haj}. \hfill $\Box $

\section{Non-Archimedean valued $BL$-algebras}
Now introduce the following new operations defined for all $[x],
[y] \in {}^*\mathbf{Q}$ in the partial order structure
$\mathcal{O}_{{}^*\mathbf{Q}}$:

\begin{itemize}
\item $[x]\rightarrow_{L} [y] = {}^*1 - \max([x],[y]) + [y]$,
\item $[x]\rightarrow_{\Pi} [y] ={}^*1$ if $[x]\leq [y]$ and
$[x]\rightarrow_{\Pi} [y] = \min({}^*1, \frac {[y]}{[x]})$
otherwise,\\

notice that we have $\min({}^*1, \frac {[y]}{[x]} ) = [h]$ iff
there exists $[h] \in {}^*\mathbf{Q}_{[0,1]}$ such that $\{\alpha
\in \mathbf{N} \colon$ $\min(1,\frac
{y(\alpha)}{x(\alpha)})=h(\alpha) \} \in \mathcal{U}$, let us also
remember that the members $[x]$, $[y]$ can be incompatible under
$\mathcal{O}_{{}^*\mathbf{Q}}$, \item  $ \neg_L [x] = {}^*1 - [x]
$, i.e.\ $[x]\rightarrow_L {}^*0$, \item  $ \neg_\Pi [x] = {}^*1$
if $[x]={}^*0$ and $\neg_\Pi [x] ={}^*0$ otherwise, i.e.\
$\neg_\Pi [x] =[x]\rightarrow_\Pi {}^*0$, \item $\Delta [x]
={}^*1$ if $[x]={}^*1$ and $\Delta [x] ={}^*0$ otherwise, i.e.\
$\Delta [x] =\neg_{\Pi} \neg_L [x]$, \item $[x]\&_{L} [y] =
\max([x],{}^*1-[y]) + [y]-{}^*1$, i.e.\ $[x]\&_{L} [y] =
\neg_L([x]\rightarrow_{L} \neg_L [y])$, \item $[x]\&_{\Pi} [y] =
[x] \cdot [y]$, \item $[x]\oplus [y] := \neg_L [x]\rightarrow_{L}
[y]$, \item $[x]\ominus [y] : = [x]\&_{L} \neg_L [y]$, \item
$[x]\wedge [y] = \min([x], [y])$, i.e.\ $[x]\wedge [y]=[x]\&_{L}
([x]\rightarrow_L [y])$, \item $[x]\vee [y] = \max([x], [y])$,
i.e.\ $[x]\vee [y]=([x]\rightarrow_{L} [y])\rightarrow_L [y]$,
\item $[x]\rightarrow_G [y] = {}^*1$ if $[x]\leq [y]$ and
$[x]\rightarrow_G [y] =[y]$ otherwise, i.e.\ $[x]\rightarrow_G
[y]=\Delta([x]\rightarrow_{L} [y])\vee [y]$, \item $\neg_G
[x]:=[x]\rightarrow_{G} {}^*0$.
  \end{itemize}

A {\it hyperrational valued} $BL$-{\it matrix}\index{hyperrational
valued $BL$-matrix} is a structure $\mathbf{L}_{{}^*\mathbf{Q}} =
\langle {}^*\mathbf{Q}_{[0,1]}$, $\wedge$, $\vee$, $\ast$,
$\Rightarrow$, ${}^*0$, ${}^*1\rangle$ such that (1) $\langle
{}^*\mathbf{Q}_{[0,1]},\wedge,\vee,{}^*0,{}^*1\rangle$ is a
lattice with the largest element ${}^*1$ and the least element
${}^*0$, (2) $\langle{}^*\mathbf{Q}_{[0,1]}, *,{}^*1\rangle$ is a
commutative semigroup with the unit element ${}^*1$, i.e. $\ast$
is commutative, associative, and ${}^*1\ast [x] =[x]$ for all
$[x]\in {}^*\mathbf{Q}_{[0,1]}$, (3) the following conditions hold

$$[z]\leq ([x]\Rightarrow [y])\mbox{ iff }[x]\ast [z]\leq [y] \mbox{ for all }[x],[y],[z];$$
$$[x]\wedge [y]= [x]\ast ([x]\Rightarrow [y]);$$
$$[x]\vee [y]= (([x]\Rightarrow [y]) \Rightarrow [y])\wedge (([y]\Rightarrow [x]) \Rightarrow [x]),$$
$$([x]\Rightarrow [y]) \vee ([y]\Rightarrow [x]) ={}^*1.$$

If we replace the set $\mathbf{Q}_{[0,1]}$ by $\mathbf{R}_{[0,1]}$
and the set ${}^*\mathbf{Q}_{[0,1]}$ by ${}^*\mathbf{R}_{[0,1]}$
in all above definitions, then we obtain {\it hyperreal valued}
$BL$-{\it matrix} $\mathbf {L}_{{}^*\mathbf{R}}$\index{hyperreal
valued $BL$-matrix}. Matrices $\mathbf {L}_{{}^*\mathbf{Q}}$,
$\mathbf {L}_{{}^*\mathbf{R}}$ are different versions of a
non-Archimedean valued $BL$-algebra. Continuing in the same way,
we can build non-Archimedean valued $\L$-algebra,
$\mathbf{G}$-algebra, and
$\mathbf{\Pi}$-algebra\index{non-Archimedean valued
$\L$-algebra}\index{non-Archimedean valued
$\mathbf{G}$-algebra}\index{non-Archimedean valued
$\mathbf{\Pi}$-algebra}.\\

Further consider the following new operations defined for all $x,
y \in \mathbf{Z}_p$ in the partial order structure
$\mathcal{O}_{\mathbf{Z}_p}$:

\begin{itemize}
\item $x\rightarrow_{L} y = N_{max} - \max(x,y) + y$, \item
$x\rightarrow_{\Pi} y = N_{max}$ if $x\leq y$ and
$x\rightarrow_{\Pi} y =$ integral part of $\frac {y}{x}$
otherwise, \item  $ \neg_L x = N_{max} - x $, i.e.\
$x\rightarrow_L 0$, \item  $ \neg_\Pi x = N_{max}$ if $x=0$ and $
\neg_\Pi x =0$ otherwise, i.e.\ $\neg_\Pi x =x\rightarrow_\Pi 0$,
\item $\Delta x = N_{max}$ if $x=N_{max}$ and $\Delta x = 0$
otherwise, i.e.\ $\Delta x =\neg_{\Pi} \neg_L x$, \item $x\&_{L} y
= \max(x,N_{max}-y) + y-N_{max}$, i.e.\ $x\&_{L} y =
\neg_L(x\rightarrow_{L} \neg_L y)$, \item $x\&_{\Pi} y = x \cdot
y$, \item $x\oplus y := \neg_L x\rightarrow_{L} y$, \item
$x\ominus y : = x\&_{L} \neg_L y$, \item $x\wedge y = \min(x, y)$,
i.e.\ $x\wedge y=x\&_{L} (x\rightarrow_L y)$, \item $x\vee y =
\max(x, y)$, i.e.\ $x\vee y=(x\rightarrow_{L} y)\rightarrow_L y$,
\item $x\rightarrow_G y = N_{max}$ if $x\leq y$ and
$x\rightarrow_G y =y$ otherwise, i.e.\ $x\rightarrow_G
y=\Delta(x\rightarrow_{L} y)\vee y$, \item $\neg_G
x:=x\rightarrow_{G} 0$.
  \end{itemize}

A $p$-{\it adic valued} $BL$-{\it matrix}\index{$p$-adic valued
$BL$-matrix} is a structure $\mathbf{L}_{\mathbf{Z}_p} = \langle
\mathbf{Z}_p$, $\wedge$, $\vee$, $\ast$, $\Rightarrow$, $0$,
$N_{max}\rangle$.

\section{Non-Archimedean valued predicate logical language}

Recall that for each $i \in [0,1]$, ${}^*i = [f= i]$, i.e.\ it is
a constant function. Every element of ${}^*[0,1]$ has the form of
infinite tuple $[f] =\langle y_0, y_1,\ldots\rangle$, where
$y_i\in [0,1]$ for each $i=0,1,2,\dots$\\

Let $\mathcal{L}$ be a standard first-order language associated
with $p$-valued (resp.\ infinite-valued) semantics. Then we can
get an extension $\mathcal{L}'$ of first-order language
$\mathcal{L}$\index{non-Archimedean valued predicate logical
language} to set later a language of $p$-adic valued (resp.\
hyper-valued) logic.\\

In $\mathcal{L}'$ we build infinite sequences of well-formed
formulas of $\mathcal{L}$:

$$\psi^\infty = \langle\psi_1, \dots, \psi_N, \dots\rangle,$$
$$\psi^i = \langle\psi_1, \dots, \psi_i \rangle,$$

\noindent where $\psi_j \in \mathcal{L}$.\\

A formula $\psi^\infty$ (resp.\ $\psi^i$) is called a
\textit{formula of infinite length}\index{formula of infinite
length} (resp.\ a \textit{formula of $i$-th length})\index{formula
of $i$-th length}.

\begin{definitions} Logical connectives in hyper-valued logic are defined as
follows:
\begin{enumerate}
    \item $\psi^\infty \star\varphi^\infty = \langle\psi_1\star\varphi_1,\ldots,
\psi_N\star\varphi_N, \ldots\rangle$, where
$\star\in\{\&,\rightarrow\}$;
    \item $\neg\psi^\infty = \langle\neg\psi_1,\ldots, \neg\psi_N,
\ldots\rangle$;
    \item $\mathrm{Q} x\,\psi^\infty = \langle\mathrm{Q}
x\,\psi_1,\ldots, \mathrm{Q} x\,\psi_N, \ldots\rangle$,
$\mathrm{Q} \in\{\forall,\exists\}$;
    \item $\psi^\infty \star\varphi^1 = \langle\psi_1\star\varphi,\psi_2\star\varphi,\ldots,
\psi_N\star\varphi, \ldots\rangle$, where
$\star\in\{\&,\rightarrow\}$.
\end{enumerate}
\end{definitions}

\begin{definitions} Logical connectives in $p$-adic valued logic are defined
as follows:
\begin{enumerate}
    \item $\psi^\infty \star\varphi^\infty = \langle\psi_1\star\varphi_1,\ldots,
\psi_N\star\varphi_N, \ldots\rangle$, where
$\star\in\{\&,\rightarrow\}$;
    \item $\neg\psi^\infty = \langle\neg\psi_1,\ldots, \neg\psi_N,
\ldots\rangle$;
    \item $\mathrm{Q} x\,\psi^\infty = \langle\mathrm{Q}
x\,\psi_1,\ldots, \mathrm{Q} x\,\psi_N, \ldots\rangle$,
$\mathrm{Q} \in\{\forall,\exists\}$.
    \item $\psi^\infty \star\varphi^i = \langle\psi_1\star\varphi_1$, \dots,
$\psi_i\star\varphi_i$, $\psi_{i+1}\star\bot$,
$\psi_{i+2}\star\bot$, \dots, $\psi_N\star\bot, \dots\rangle$,
where $\star\in\{\&,\rightarrow\}$.
    \item suppose $i<j$, then $\psi^i
\star\varphi^j = \langle\psi_1\star\varphi_1$, \dots,
$\psi_i\star\varphi_i$, $\psi_{i+1}\star\bot$,
$\psi_{i+2}\star\bot$, \dots, $\psi_{j}\star\bot\rangle$, where
$\star\in\{\&,\rightarrow\}$.
\end{enumerate}
\end{definitions}

An interpretation for a language $\mathcal{L}'$ is defined in the
standard way. Extend the valuation of $\mathcal{L}$ to one of
$\mathcal{L}'$ as follows.

\begin {definitions} Given an interpretation $I = \langle
\mathcal{M}, s\rangle $ and a valuation $\mathrm{val}_I$ of
$\mathcal{L}$, we define the non-Archimedean
$i$-valuation\index{non-Archimedean $i$-valuation}
$\mathrm{val}_I^i$ (resp. $\infty$-valuation\index{non-Archimedean
$\infty$-valuation} $\mathrm{val}_I^\infty$) to be a mapping from
formulas $\varphi^i$ (resp.\ $\varphi^\infty$) of $\mathcal{L}'$
to truth value set $V^i$ (resp.\ ${}^*V$) as follows:
\begin{enumerate}
    \item $\mathrm{val}_I^i(\varphi^i) = \langle
\mathrm{val}_I(\varphi_1),\dots,
\mathrm{val}_I(\varphi_i)\rangle$.
    \item $\mathrm{val}_I^\infty(\varphi^\infty) = \langle
\mathrm{val}_I(\varphi_1),\dots,
\mathrm{val}_I(\varphi_N),\dots\rangle$.
\end{enumerate}\end {definitions}

For example, in $p$-adic valued case
$\mathrm{val}_I^\infty(\psi^\infty \star\varphi^i) =
\langle\mathrm{val}_I(\psi_1\star\varphi_1)$, \dots,
$\mathrm{val}_I(\psi_i\star\varphi_i)$,
$\mathrm{val}_I(\psi_{i+1}\star\bot)$,
$\mathrm{val}_I(\psi_{i+2}\star\bot)$, \dots,
$\mathrm{val}_I(\psi_N\star\bot), \dots\rangle$, where
$\star\in\{\&,\rightarrow\}$.\\

Let $\mathbf{L}_{{}^*V}$ be a non-Archimedean valued
$BL$-matrix\index{non-Archimedean valued $BL$-matrix}. Then the
valuations $\mathrm{val}_I^i$ and $\mathrm{val}_I^\infty$ of
$\mathcal{L}'$ to non-Archimedean valued $BL$-matrix gives the
basic fuzzy logic with the non-Archimedean valued semantics.\\

We say that an $\mathbf{L}_{{}^*V}$-structure $\mathcal{M}$ is an
$i$-{\it model} (resp.\ an $\infty$-{\it model}) of an
$\mathcal{L}'$-theory $T$ iff $\mathrm{val}_I^i(\varphi^i) =
\langle 1, \dots, 1\rangle$ (resp.\ $\mathrm{val}_I^\infty(\varphi
^\infty) = {}^* 1$) on $\mathcal{M}$ for each $\varphi^i\in T$
(resp.\ $\varphi^\infty\in T$).

\section{Non-Archimedean valued basic fuzzy propositional logic $BL_\infty$}

Let us construct a non-Archimedean extension of basic fuzzy
propositional logic $BL$ denoted by
$BL_\infty$\index{non-Archimedean valued basic fuzzy propositional
logic $BL_\infty$}. This logic is built in the language
$\mathcal{L}'$ and it has a non-Archimedean valued $BL$-matrix as
its semantics.\\

Remember that the logic $BL$ has just two propositional
operations: $\&$, $\rightarrow$, which are understood as t-norm
and its residuum respectively.\\

The logic $BL_\infty$ is given by the following axioms:

\begin{eqnarray}(\varphi^i\rightarrow\psi^i)\rightarrow ((\psi^i\rightarrow\chi^i)\rightarrow(\varphi^i\rightarrow\chi^i)),\end{eqnarray}
\begin{eqnarray}(\varphi^i \& \psi^i)\rightarrow \varphi^i,\end{eqnarray}
\begin{eqnarray}(\varphi^i \& \psi^i)\rightarrow (\psi^i\&\varphi^i),\end{eqnarray}
\begin{eqnarray}(\varphi^i \& (\varphi^i \rightarrow\psi^i))\rightarrow(\psi^i\&(\psi^i\rightarrow\varphi^i)),\end{eqnarray}
\begin{eqnarray}(\varphi^i \rightarrow (\psi^i \rightarrow\chi^i))\rightarrow ((\varphi^i\&\psi^i)\rightarrow\chi^i),\end{eqnarray}
\begin{eqnarray}((\varphi^i\&\psi^i)\rightarrow\chi^i)\rightarrow (\varphi^i \rightarrow (\psi^i \rightarrow\chi^i)),\end{eqnarray}
\begin{eqnarray}((\varphi^i \rightarrow \psi^i) \rightarrow\chi^i)\rightarrow (((\psi^i\rightarrow\varphi^i)\rightarrow\chi^i)\rightarrow\chi^i),\end{eqnarray}
\begin{eqnarray}\bot^i \rightarrow \psi^i,\end{eqnarray}
\begin{eqnarray}(\varphi^\infty\rightarrow\psi^\infty)\rightarrow ((\psi^\infty\rightarrow\chi^\infty)\rightarrow(\varphi^\infty\rightarrow\chi^\infty)),\end{eqnarray}
\begin{eqnarray}(\varphi^\infty \& \psi^\infty)\rightarrow \varphi^\infty,\end{eqnarray}
\begin{eqnarray}(\varphi^\infty \& \psi^\infty)\rightarrow (\psi^\infty\&\varphi^\infty),\end{eqnarray}
\begin{eqnarray}(\varphi^\infty \& (\varphi^\infty \rightarrow\psi^\infty))\rightarrow(\psi^\infty\&(\psi^\infty\rightarrow\varphi^\infty)),\end{eqnarray}
\begin{eqnarray}(\varphi^\infty \rightarrow (\psi^\infty \rightarrow\chi^\infty))\rightarrow ((\varphi^\infty\&\psi^\infty)\rightarrow\chi^\infty),\end{eqnarray}
\begin{eqnarray}((\varphi^\infty\&\psi^\infty)\rightarrow\chi^\infty)\rightarrow (\varphi^\infty \rightarrow (\psi^\infty \rightarrow\chi^\infty)),\end{eqnarray}
\begin{eqnarray}((\varphi^\infty \rightarrow \psi^\infty) \rightarrow\chi^\infty)\rightarrow (((\psi^\infty\rightarrow\varphi^\infty)\rightarrow\chi^\infty)\rightarrow\chi^\infty),\end{eqnarray}
\begin{eqnarray}\bot^\infty \rightarrow \psi^\infty.\end{eqnarray}

These axioms are said to be {\it horizontal}. Introduce also some
new axioms that show basic properties of non-Archimedean ordered
structures. These express a connection between formulas of various
length.
\begin{enumerate}
\item \textbf{Non-Archimedean multiple-validity}. It is well known
that there exist infinitesimals that are less than any positive
number of $[0,1]$. This property can be expressed by means of the
following logical axiom:

\begin {eqnarray}(\neg(\psi ^{1} \leftrightarrow
\psi^\infty)\&\neg(\varphi^{1}\leftrightarrow
\bot^\infty))\rightarrow (\psi^\infty\rightarrow \varphi
^{1}),\label{schulast1}\end {eqnarray}

\noindent where $\psi^{1}=\psi_{1}$, i.e.\ it is the first member
of an infinite tuple $\psi^\infty$.

\item \textbf{$p$-adic multiple-validity}. There is a well known
theorem according to that every equivalence class $a$ for which $|
a |_p \leq 1 $ (this means that $a$ is a $p$-adic integer) has
exactly one representative \textsc{Cauchy} sequence
$\{a_i\}_{i\in\omega}$ for which:

\begin{enumerate}
  \item $0 \leq a_i <p^i $ for $i = 1, 2, 3, \ldots $;
  \item $a_i \equiv a _ {i + 1} \mod ~~p^i $ for $i = 1, 2, 3$,
  \ldots
 \end{enumerate}

This property can be expressed by means of the following logical
axioms:

\begin {eqnarray}
((\overline{p^{i+1}-1} \ominus \overline{p^i -1}) \rightarrow_L
\psi^{i+1}) \rightarrow_L {}\nonumber\\(\psi^{i+1} \leftrightarrow
(\underbrace {\overline{p-1} \oplus \dots \oplus
\overline{p-1}}_{p^i} \oplus \psi^{i}),
\end {eqnarray}
\begin {eqnarray}
(\psi^{i+1} \leftrightarrow (\overbrace {\overline{k} \oplus \dots
\oplus \overline{k}}^{p^i}) \oplus \psi
^{i})\rightarrow_L{}\nonumber\\((((\ldots(\overline{p^{i+1}-1}
\ominus \underbrace {\overline{p^i-1}) \ominus \dots) \ominus
\overline{p^i-1} )}_{0< p-k \leq p}\nonumber\\\ominus \neg_L
\overline{k}) \rightarrow_L \psi^{i+1}),
\end {eqnarray}
\begin {eqnarray}
(\psi^{i+1} \leftrightarrow (\overbrace {\overline{k} \oplus \dots
\oplus \overline{k}}^{p^i}) \oplus \psi
^{i})\rightarrow_L{}\nonumber\\(\psi
^{i+1}\rightarrow_L(((\dots(\overline{p^{i+1}-1} \ominus\nonumber\\
\underbrace {\overline{p^i-1}) \ominus \dots) \ominus
\overline{p^i-1} )}_{0\leq (p-1)-k \leq p-1}\ominus \neg_L
\overline{k})),
\end {eqnarray}
\begin {eqnarray}
(\psi^{i+1} \rightarrow_L \overline{p^i-1}) \rightarrow_L (\psi
^{i+1} \leftrightarrow \psi^{i}),
\end {eqnarray}
\begin {eqnarray}
(\psi^{i+1} \leftrightarrow \psi^{i}) \vee(\psi^{i+1}
\leftrightarrow (\psi^{i} \oplus p^i\cdot \overline{1}))
\vee\dots\nonumber\\\vee(\psi^{i+1} \leftrightarrow (\psi^{i}
\oplus p^i\cdot \overline{p-1}))\label{schulast2},
\end {eqnarray}

\noindent where $\overline{p-1}$ is a tautology at the first-order
level and $\overline{p^i -1}$ (respectively
$\overline{p^{i+1}-1}$) a tautology for formulas of $i$-th length
(respectively of $(i+1)$-th length); $\psi^{1}=\psi_{1}$, i.e.\ it
is the first member of an infinite tuple $\psi^\infty$; $\neg_L
\overline{k}$ is a first-order formula that has the truth value
$((p-1) - k)\in \{0,\ldots,p-1\}$ for any its interpretations and
$\overline{k}$ is a first-order formula that has the truth value
$k\in \{0,\ldots,p-1\}$ for any its interpretations;
$\overline{1}$ is a first-order formula that has the truth value 1
for any its interpretations, etc. The denoting $p^i\cdot
\overline{k}$ means
$\underbrace{\overline{k}\oplus\dots\oplus\overline{k}}_{p^i}$.
\end{enumerate}

Axioms \eqref{schulast1} -- \eqref{schulast2} are said to be {\it
vertical}.\\

The deduction rules of $BL_\infty$ is modus ponens: from $\psi$,
$\psi\rightarrow\varphi$ infer $\varphi$.\\

The notions of proof, derivability $\vdash$, theorem, and theory
over $BL_\infty$ is defined as usual.

\begin{theorems}[Soundness and Completeness]  \index{completeness theorem}\index{soundness theorem}Let $\Phi$ be a formula of $\mathcal{L}'$, $T$ an
$\mathcal{L}'$-theory. Then the following conditions are
equivalent:
\begin{itemize}
    \item $T \vdash\Phi$;
    \item $\mathrm{val}_I^i(\Phi)=\langle\underbrace{1,\dots,1}_i\rangle$ (resp.\ $\mathrm{val}_I^\infty(\Phi)={}^*1$) for each $\mathbf{L}_{{}^*V}$-model $\mathcal{M}$ of $T$;
\end{itemize}
\end{theorems}

\noindent {\it Proof}. This follows from theorem 4 and semantic
rules of $BL_\infty$. \hfill $\Box $

\chapter{Neutrosophic sets}

\section{Vague sets}
Let $U$ be the universe of discourse, $U = \{u_1, u_2, \dots ,
u_n\}$, with a generic element of $U$ denoted by $u_i$. A vague
set $A$ in $U$ is characterized by a truth-membership
function\index{truth-membership function} $t_A$ and a
false-membership function\index{false-membership function} $f_A$,

$$t_A \colon U \to [0, 1],$$
$$f_A \colon U \to [0, 1],$$

\noindent where $t_A(u_i)$ is a lower bound on the grade of
membership of $u_i$ derived from the evidence for $u_i$,
$f_A(u_i)$ is a lower bound on the negation of $u_i$ derived from
the evidence against $u_i$, and $t_A(u_i) + f_A(u_i) \leqslant 1$.
The grade of membership of $u_i$ in the vague set $A$ is bounded
to a subinterval $[t_A(u_i), 1-f_A(u_i)]$ of $[0,1]$. The vague
value $[t_A(u_i),  1-f_A(u_i)]$ indicates that the exact grade of
membership $\mu_A(u_i)$ of $u_i$ may be unknown. But it is bounded
by $t_A(u_i)  \leqslant \mu_A(u_i) \leqslant 1- f_A(u_i)$, where
$t_A(u_i) + f_A(u_i) \leqslant 1$. When the universe of discourse
$U$ is continuous, a vague set $A$ can be written as

$$A =  \int_U [t_A(u_i), 1 - f_A(u_i)]/ u_i,~~ u_i \in U.$$

When $U$ is discrete, then

$$A =  \sum_{i=1}^n~ [t_A(u_i), 1 - f_A(u_i)]/ u_i,~~ u_i \in U.$$

Logical operations in vague set\index{vague set} theory are defined as follows:\\

Let $x$ and $y$ be two vague values, $x = [t_x, 1 - f_x]$, $y =
[t_y, 1 - f_y]$, where $t_x \in [0, 1]$, $f_x \in [0, 1]$, $t_y
\in [0, 1]$, $f_y \in [0, 1]$, $t_x+f_x \leqslant 1$ and $t_y +f_y
\leqslant 1$. Then

$$\neg_L x = [1- t_x, f_x],$$
$$x\wedge y = [\min (t_x,t_y), \min(1 - f_x, 1 - f_y)],$$
$$x\vee y = [\max (t_x,t_y), \max(1 - f_x, 1 - f_y)].$$

\section{Neutrosophic set operations}
\begin{definitions} Let $U$ be the universe of discourse, $U = \{u_1, u_2, \dots ,
u_n\}$. A hyper-valued neutrosophic set\index{hyper-valued
neutrosophic set} $A$ in $U$ is characterized by a
truth-membership function\index{truth-membership function} $t_A$,
an indeterminacy-membership
function\index{indeterminacy-membership function} $i_A$, and a
false-membership function\index{false-membership function} $f_A$

$$t_A\ni f \colon U \to {}^*[0, 1],$$
$$i_A\ni f \colon U \to {}^*[0, 1],$$
$$f_A\ni f \colon U \to {}^*[0, 1],$$

\noindent where $t_A$ is the degree of truth-membership function,
$i_A$ is the degree of indeter\-minacy-membership function, and
$f_A$ is the degree of falsity-membership function. There is no
restriction on the sum of $t_A$, $i_A$, and $f_A$, i.e.\
$${}^*0 \leqslant \max t_A(u_i)+
\max i_A(u_i)+ \max f_A(u_i)\leqslant {}^*3.$$ \end{definitions}

\begin{definitions} Let $U$ be the universe of discourse, $U = \{u_1, u_2, \dots ,
u_n\}$. A $p$-adic valued neutrosophic set\index{$p$-adic valued
neutrosophic set} $A$ in $U$ is characterized by a
truth-membership function $t_A$, an indeterminacy-membership
function $i_A$, and a false-membership function $f_A$

$$t_A\ni f \colon U \to \mathbf{Z}_p,$$
$$i_A\ni f \colon U \to \mathbf{Z}_p,$$
$$f_A\ni f \colon U \to \mathbf{Z}_p,$$

\noindent where $t_A$ is the degree of truth-membership function,
$i_A$ is the degree of indeter\-minacy-membership function, and
$f_A$ is the degree of falsity-membership function. There is no
restriction on the sum of $t_A$, $i_A$, and $f_A$, i.e.\
$$0 \leqslant \max t_A(u_i)+
\max i_A(u_i)+ \max f_A(u_i)\leqslant
N_{max}+N_{max}+N_{max}=-3.$$
\end{definitions}

Also, a neutrosophic set\index{neutrosophic set} $A$ is understood
as a triple $ \langle t_A, i_A, f_A\rangle$ and it can be regarded
as consisting of hyper-valued or
$p$-adic valued degrees.\\

As we see, in neutrosophic sets, indeterminacy is quantified
explicitly and truth-member\-ship, indeterminacy-membership and
falsity-membership are independent. This assumption is very
important in many applications such as information fusion in which
we try to combine the data from different sensors. Neutrosophic
sets are proposed for the first time in the framework of
neutrosophy that was introduced by \textsc{Smarandache} in 1980:
``It is a branch of philosophy which studies the origin, nature
and scope of neutralities, as well as their interactions with
different
ideational spectra'' \cite{Smaran2}.\\

Neutrosophic set is a powerful general formal framework which
generalizes the concept of the fuzzy set\index{fuzzy set}
\cite{Zadeh1}, interval valued fuzzy set\index{interval valued
fuzzy set} \cite{Turks}, intuitionistic fuzzy
set\index{intuitionistic fuzzy set} \cite{Atanass1}, and interval
valued
intuitionistic fuzzy set \cite{Atanass2}.\\

Suppose that $t_A, i_A, f_A$ are subintervals of ${}^*[0,1]$. Then
a neutrosophic set\index{interval neutrosophic set} $A$ is called \textit{interval} one.\\

When the universe of discourse $U$ is continuous, an interval
neutrosophic set $A$ can be written as

$$A =  \int_U \langle t_A(u_i), i_A(u_i), f_A(u_i)\rangle / u_i,~~ u_i \in U.$$

When $U$ is discrete, then

$$A =  \sum_{i=1}^n~ \langle t_A(u_i), i_A(u_i), f_A(u_i)\rangle / u_i,~~ u_i \in U.$$

The interval neutrosophic set can represent uncertain, imprecise,
incomplete and inconsistent information which exist in real world.
It can be readily seen that the interval neutrosophic set
generalizes the following sets:

\begin{itemize}
    \item  the classical set, $i_A=\emptyset$, $\min t_A = \max t_A = 0$ or 1, $\min f_A = \max f_A
= 0$ or 1 and $\max t_A + \max f_A = 1$.
    \item the fuzzy set, $i_A=\emptyset$, $\min t_A = \max t_A\in [0, 1]$, $\min f_A = \max f_A
    \in [0, 1]$ and $\max t_A + \max f_A = 1$.
    \item the interval valued fuzzy set, $i_A=\emptyset$, $\min t_A$, $\max t_A$, $\min f_A$, $\max f_A
    \in [0, 1]$, $\max t_A + \min f_A = 1$ and $\min t_A + \max f_A  = 1$.
    \item the intuitionistic fuzzy set, $i_A=\emptyset$, $\min t_A = \max t_A\in [0, 1]$, $\min f_A = \max f_A
    \in [0, 1]$ and $\max t_A + \max f_A \leqslant 1$.
    \item the interval valued intuitionistic fuzzy set, $i_A=\emptyset$, $\min t_A$, $\max t_A$, $\min f_A$, $\max f_A
    \in [0, 1]$, $\max t_A + \min f_A \leqslant 1$.
    \item the paraconsistent set, $i_A=\emptyset$, $\min t_A = \max t_A\in [0, 1]$, $\min f_A = \max f_A
    \in [0, 1]$ and $\max t_A + \max f_A > 1$.
    \item the interval valued paraconsistent set,  $i_A=\emptyset$, $\min t_A$, $\max t_A$, $\min f_A$, $\max f_A
    \in [0, 1]$, $\max t_A + \min f_A > 1$.
\end{itemize}

Let $S_1$ and $S_2$ be two real standard or non-standard subsets
of ${}^*[0,1]$, then $S_1 + S_2 = \{x\colon x = s_1 + s_2, s_1 \in
S_1$ and $s_2 \in S_2\}$, ${}^*a+S_2 = \{x\colon x = {}^*a+ s_2,
s_2 \in S_2\}$, $S_1 - S_2 = \{x\colon x = s_1 - s_2, s_1 \in S_1$
and $s_2 \in S_2\}$, $S_1 \cdot S_2 = \{x\colon x = s_1 \cdot s_2,
s_1 \in S_1$ and $s_2 \in S_2\}$, $\max (S_1, S_2) = \{x\colon x =
\max (s_1, s_2), s_1 \in S_1$ and $s_2 \in S_2\}$, $\min (S_1,
S_2) =
\{x\colon x = \min (s_1, s_2), s_1 \in S_1$ and $s_2 \in S_2\}$.\\

\begin{enumerate}
    \item \textbf{The complement of a neutrosophic set $A$\index{complement of a neutrosophic set} is defined as follows}

\begin{itemize}
    \item the \textsc{{\L}ukasiewicz} complement:\\

$\neg_L A = \langle {}^*1- t_A, {}^*1-i_A, {}^*1-
f_A\rangle,~~t_A, i_A, f_A \subseteq ({}^*[0,1])^U,$\\

$\neg_L A = \langle N_{max}- t_A, N_{max}-i_A, N_{max}-
f_A\rangle,~~t_A, i_A, f_A \subseteq (\mathbf{Z}_p)^U,$\\

    \item the \textsc{G\"{o}del} complement:\\

$\neg_G A = \langle \neg_G t_A, \neg_G i_A, \neg_G
f_A\rangle,~~t_A, i_A, f_A \subseteq ({}^*[0,1])^U,$\\

$\neg_G A = \langle \neg_G t_A, \neg_G i_A, \neg_G
f_A\rangle,~~t_A, i_A, f_A \subseteq (\mathbf{Z}_p)^U,$\\

\noindent where $\neg_G t_A =\{\neg_G x$ : $x\in t_A\}$, $\neg_G
i_A$ = $\{\neg_G x$ : $x\in i_A\}$, $\neg_G f_A$ = $\{\neg_G
x\colon x\in f_A\}$,

    \item the Product complement:\\

$\neg_\Pi A = \langle \neg_\Pi t_A, \neg_\Pi i_A, \neg_\Pi
f_A\rangle,~~t_A, i_A, f_A \subseteq ({}^*[0,1])^U,$\\

$\neg_\Pi A = \langle \neg_\Pi t_A, \neg_\Pi i_A, \neg_\Pi
f_A\rangle,~~t_A, i_A, f_A \subseteq (\mathbf{Z}_p)^U,$\\

\noindent where $\neg_\Pi t_A$ = $\{\neg_\Pi x\colon x\in t_A\}$,
$\neg_\Pi i_A$ = $\{\neg_\Pi x\colon x\in i_A\}$, $\neg_\Pi f_A
=\{\neg_\Pi x\colon x\in f_A\}$.

\end{itemize}
    \item \textbf{The implication of two neutrosophic sets\index{implication of two neutrosophic sets} $A$ and $B$ is defined as follows}

\begin{itemize}
    \item the \textsc{{\L}ukasiewicz} implication:\\

$A \rightarrow_L B = \langle {}^*1- \max (t_A, t_B) + t_B, {}^*1-
\max (i_A, i_B) + i_B, {}^*1- \max (f_A, f_B) + f_B\rangle,~~t_A$,
$i_A$, $f_A$, $t_B$, $i_B$, $f_B \subseteq ({}^*[0,1])^U,$\\

$A \rightarrow_L B= \langle N_{max}- \max (t_A, t_B) + t_B,
N_{max}- \max (i_A, i_B) + i_B, N_{max}- \max (f_A, f_B) +
f_B\rangle,~~t_A$, $i_A$, $f_A$, $t_B$, $i_B$, $f_B \subseteq
(\mathbf{Z}_p)^U,$\\

    \item the \textsc{G\"{o}del} implication:\\

$A \rightarrow_G B = \langle t_A\rightarrow_G t_B,
i_A\rightarrow_G i_B, f_A\rightarrow_G f_B\rangle,~~t_A$,
$i_A$, $f_A$, $t_B$, $i_B$, $f_B \subseteq ({}^*[0,1])^U,$\\

$A \rightarrow_G B= \langle t_A\rightarrow_G t_B, i_A\rightarrow_G
i_B, f_A\rightarrow_G f_B\rangle,~~t_A$, $i_A$, $f_A$, $t_B$,
$i_B$, $f_B \subseteq
(\mathbf{Z}_p)^U,$\\

\noindent where $t_A\rightarrow_G t_B =\{x\colon x=
s_1\rightarrow_G  s_2, s_1 \in t_A$ and $s_2 \in t_B\}$,
$i_A\rightarrow_G i_B =\{x\colon x= s_1\rightarrow_G  s_2, s_1 \in
i_A$ and $s_2 \in i_B\}$, $f_A\rightarrow_G f_B =\{x\colon x=
s_1\rightarrow_G  s_2, s_1 \in f_A$ and $s_2 \in f_B\}$,\\

    \item the Product implication:\\

$A \rightarrow_\Pi B = \langle t_A\rightarrow_\Pi t_B,
i_A\rightarrow_\Pi i_B, f_A\rightarrow_\Pi f_B\rangle,~~t_A,
i_A, f_A,t_B, i_B, f_B \subseteq ({}^*[0,1])^U,$\\

$A \rightarrow_\Pi B= \langle t_A\rightarrow_\Pi t_B,
i_A\rightarrow_\Pi i_B, f_A\rightarrow_\Pi f_B\rangle,~~t_A, i_A,
f_A,t_B, i_B, f_B \subseteq
(\mathbf{Z}_p)^U,$\\

\noindent where $t_A\rightarrow_\Pi t_B =\{x\colon x=
s_1\rightarrow_\Pi  s_2, s_1 \in t_A$ and $s_2 \in t_B\}$,
$i_A\rightarrow_\Pi i_B =\{x\colon x= s_1\rightarrow_\Pi  s_2, s_1
\in i_A$ and $s_2 \in i_B\}$, $f_A\rightarrow_\Pi f_B =\{x\colon
x= s_1\rightarrow_\Pi  s_2, s_1 \in f_A$ and $s_2 \in f_B\}$,\\
\end{itemize}

    \item \textbf{The intersection of two neutrosophic sets\index{intersection of two neutrosophic sets} $A$ and $B$ is defined as follows}

\begin{itemize}
    \item the \textsc{{\L}ukasiewicz} intersection:\\

$A \&_L B = \langle \max (t_A, {}^*1 - t_B) + t_B - {}^*1$, $\max
(i_A, {}^*1 - i_B) + i_B - {}^*1$, $\max (f_A, {}^*1 - f_B) + f_B
- {}^*1\rangle,~~t_A$,
$i_A$, $f_A$, $t_B$, $i_B$, $f_B \subseteq ({}^*[0,1])^U,$\\

$A \&_L B= \langle \max (t_A, N_{max} - t_B) + t_B - N_{max}, \max
(i_A, N_{max} - i_B) + i_B - N_{max}, \max (f_A, N_{max} - f_B) +
f_B - N_{max}\rangle,~~t_A$, $i_A$, $f_A$, $t_B$, $i_B$, $f_B
\subseteq
(\mathbf{Z}_p)^U,$\\

    \item the \textsc{G\"{o}del} intersection:\\

$A \&_G B = \langle \min (t_A, t_B), \min (i_A, i_B), \min (f_A,
f_B)\rangle,~~t_A$, $i_A$, $f_A$, $t_B$, $i_B$, $f_B \subseteq ({}^*[0,1])^U,$\\

$A \&_G B= \langle \min (t_A, t_B), \min (i_A, i_B), \min (f_A,
f_B)\rangle,~~t_A$, $i_A$, $f_A$, $t_B$, $i_B$, $f_B \subseteq
(\mathbf{Z}_p)^U,$\\

    \item the Product intersection:\\

$A \&_\Pi B = \langle t_A\cdot t_B, i_A\cdot i_B, f_A\cdot
f_B\rangle,~~t_A,
i_A, f_A,t_B, i_B, f_B \subseteq ({}^*[0,1])^U,$\\

$A \&_\Pi B= \langle t_A\cdot t_B, i_A\cdot i_B, f_A\cdot
f_B\rangle,~~t_A, i_A, f_A,t_B, i_B, f_B \subseteq
(\mathbf{Z}_p)^U,$\\
\end{itemize}
\end{enumerate}

Thus, we can extend the logical operations of fuzzy logic to the
case of neutrosophic sets.

\chapter{Interval neutrosophic logic}

\section{Interval neutrosophic matrix logic}

Interval neutrosophic logic\index{interval neutrosophic logic}
proposed in \cite{Smaran2}, \cite{Smaran3}, \cite{Wang}
generalizes the interval valued fuzzy logic, the non-Archimedean
valued fuzzy logic, and paraconsistent logics. In the interval
neutrosophic logic, we consider not only truth-degree and
falsity-degree, but also indeterminacy-degree
which can reliably capture more information under uncertainty.\\

Now consider \textit{hyper-valued interval neutrosophic matrix
logic}\index{hyper-valued interval neutrosophic matrix logic}
$INL$ defined as the ordered system $\langle ({}^*[0,1])^3$,
$\neg_{INL}$, $\rightarrow_{INL}$, $\vee_{INL}$, $\wedge_{INL}$,
$\widetilde{\exists}_{INL}$, $\widetilde{\forall}_{INL}$,
$\{\langle {}^*1$, ${}^*0$, ${}^*0\rangle\} \rangle $ where
\begin
{enumerate}
    \item for all $\langle t, i, f\rangle\in ({}^*[0,1])^3$, $ \neg_{INL} \langle t, i, f\rangle = \langle f, 1-i, t\rangle$,
    \item for all $\langle t_1, i_1, f_1\rangle, \langle t_2, i_2, f_2\rangle\in ({}^*[0,1])^3$, $\langle t_1, i_1, f_1\rangle \rightarrow_{INL} \langle t_2, i_2, f_2\rangle=\langle \min ({}^*1, {}^*1- t_1+t_2), \max ({}^*0, i_2 - i_1), \max ({}^*0, f_2 - f_1)\rangle$,
    \item for all $\langle t_1$, $i_1$, $f_1\rangle$, $\langle t_2$, $i_2$, $f_2\rangle\in ({}^*[0,1])^3$, $\langle t_1$, $i_1$, $f_1\rangle \wedge_{INL} \langle t_2$, $i_2$, $f_2\rangle$ = $\langle \min (t_1, t_2), \max (i_1,i_2), \max (f_1,f_2)\rangle$,
    \item for all $\langle t_1$, $i_1$, $f_1\rangle$, $\langle t_2$, $i_2$, $f_2\rangle\in ({}^*[0,1])^3$, $\langle t_1$, $i_1$, $f_1\rangle \vee_{INL} \langle t_2$, $i_2$, $f_2\rangle$ = $\langle \max (t_1, t_2), \min (i_1,i_2), \min (f_1,f_2)\rangle$,
    \item for a subset $\langle M_1$, $M_2$, $M_3 \rangle\subseteq ({}^*[0,1])^3$, $\widetilde{\exists}(\langle M_1$, $M_2$, $M_3 \rangle) = \langle \max (M_1)$, $\min (M_2), \min (M_3) \rangle$,
    \item for a subset $\langle M_1$, $M_2$, $M_3 \rangle \subseteq ({}^*[0,1])^3$, $\widetilde{\forall}(\langle M_1$, $M_2$, $M_3 \rangle) = \langle \min (M_1)$, $\max (M_2), \max (M_3) \rangle$,
    \item $ \{\langle {}^*1$, ${}^*0$,
${}^*0\rangle\} $ is the set of designated truth values.
    \end {enumerate}

Now consider \textit{$p$-adic valued interval neutrosophic matrix
logic}\index{$p$-adic valued interval neutrosophic matrix logic}
$INL$ defined as the ordered system $\langle (\mathbf{Z}_p)^3$,
$\neg_{INL}$, $\rightarrow_{INL}$, $\vee_{INL}$, $\wedge_{INL}$,
$\widetilde{\exists}_{INL}$, $\widetilde{\forall}_{INL}$,
$\{\langle N_{max}$, $0$, $0\rangle\} \rangle $ where
\begin
{enumerate}
    \item for all $\langle t, i, f\rangle\in (\mathbf{Z}_p)^3$, $ \neg_{INL} \langle t, i, f\rangle = \langle f, 1-i, t\rangle$,
    \item for all $\langle t_1, i_1, f_1\rangle, \langle t_2, i_2, f_2\rangle\in (\mathbf{Z}_p)^3$, $\langle t_1$, $i_1$, $f_1\rangle \rightarrow_{INL} \langle t_2$, $i_2$, $f_2\rangle=\langle N_{max}- \max(t_1,t_2)+t_2$, $\max (0, i_2 - i_1), \max (0, f_2 - f_1)\rangle$,
    \item for all $\langle t_1$, $i_1$, $f_1\rangle$, $\langle t_2$, $i_2$, $f_2\rangle\in (\mathbf{Z}_p)^3$, $\langle t_1$, $i_1$, $f_1\rangle \wedge_{INL} \langle t_2$, $i_2$, $f_2\rangle$ = $\langle \min (t_1, t_2), \max (i_1,i_2), \max (f_1,f_2)\rangle$,
    \item for all $\langle t_1$, $i_1$, $f_1\rangle$, $\langle t_2$, $i_2$, $f_2\rangle\in (\mathbf{Z}_p)^3$, $\langle t_1$, $i_1$, $f_1\rangle \vee_{INL} \langle t_2$, $i_2$, $f_2\rangle$ = $\langle \max (t_1, t_2), \min (i_1,i_2), \min (f_1,f_2)\rangle$,
    \item for a subset $\langle M_1$, $M_2$, $M_3 \rangle\subseteq (\mathbf{Z}_p)^3$, $\widetilde{\exists}(\langle M_1$, $M_2$, $M_3 \rangle) = \langle \max (M_1)$, $\min (M_2), \min (M_3) \rangle$,
    \item for a subset $\langle M_1$, $M_2$, $M_3 \rangle \subseteq (\mathbf{Z}_p)^3$, $\widetilde{\forall}(\langle M_1$, $M_2$, $M_3 \rangle) = \langle \min (M_1)$, $\max (M_2), \max (M_3) \rangle$,
    \item $ \{\langle N_{max}$, $0$,
$0\rangle\} $ is the set of designated truth values.
    \end {enumerate}

As we see, interval neutrosophic matrix logic $INL$ is an
extension of the non-Archimedean valued \textsc{{\L}ukasiewicz}
matrix logic.

\section{Hilbert's type calculus for interval neutrosophic propositional logic}

Interval neutrosophic calculus denoted by
$\mathbf{INL}$\index{\textsc{Hilbert}'s type calculus for interval
neutrosophic propositional logic} is built in the framework of the
language $\mathcal{L}'$ considered in section 10.4, but its
semantics is different.\\

An interpretation is defined in the standard way. Extend the
valuation of $\mathcal{L}'$ to the valuation for interval
neutrosophic calculus as follows.

\begin {definitions} Given an interpretation $I = \langle
\mathcal{M}, s\rangle $ and a valuation $\mathrm{val}_I^\infty$ of
$\mathcal{L}'$, we define the hyper-valued interval neutrosophic
valuation\index{hyper-valued interval neutrosophic valuation}
$\mathrm{val}_I^{\infty, INL}(\cdot)$ to be a mapping from
formulas of the form $\varphi^\infty$ of $\mathcal{L}'$ to
interval neutrosophic matrix logic $INL$ as follows:

$$\mathrm{val}_I^{\infty, INL}(\varphi^\infty) = \langle
\mathrm{val}_I^\infty(\varphi^\infty)=t(\varphi^\infty),
i(\varphi^\infty), f(\varphi^\infty)\rangle \in ({}^*[0,1])^3.$$

\end {definitions}

\begin {definitions} Given an interpretation $I = \langle
\mathcal{M}, s\rangle $ and a valuation $\mathrm{val}_I^\infty$ of
$\mathcal{L}'$, we define the $p$-adic valued interval
neutrosophic valuation\index{$p$-adic valued interval neutrosophic
valuation} $\mathrm{val}_I^{\infty, INL}(\cdot)$ to be a mapping
from formulas of the form $\varphi^\infty$ of $\mathcal{L}'$ to
interval neutrosophic matrix logic $INL$ as follows:

$$\mathrm{val}_I^{\infty, INL}(\varphi^\infty) = \langle
\mathrm{val}_I^\infty(\varphi^\infty)=t(\varphi^\infty),
i(\varphi^\infty), f(\varphi^\infty)\rangle \in
(\mathbf{Z}_p)^3.$$

\end {definitions}

We say that an $INL$-structure $\mathcal{M}$ is a {\it model} of
an $\mathbf{INL}$-theory $T$ iff
$$\mathrm{val}_I^{\infty,INL}(\varphi ^\infty) = \langle {}^*1,
{}^*0, {}^*0\rangle$$ on $\mathcal{M}$ for each $\varphi^\infty\in
T$.

\begin{propositions} In the matrix logic $INL$, modus ponens is preserved, i.e.\ if $\varphi^\infty$ and $\varphi^\infty\rightarrow_{INL}\psi^\infty$ are $INL$-tautologies, then $\psi^\infty$ is also
an $INL$-tautology.
\end{propositions}

\noindent \emph{Proof}. Consider the hyper-valued case. Since
$\varphi^\infty$ and $\varphi^\infty\rightarrow_{INL}\psi^\infty$
are $INL$-tautologies, then
$$\mathrm{val}_I^{\infty, INL}(\varphi) =
\mathrm{val}_I^{\infty,
INL}(\varphi^\infty\rightarrow_{INL}\psi^\infty) = \langle {}^*1,
{}^*0, {}^*0\rangle,$$ that is $\mathrm{val}_I^{\infty,
INL}(\varphi^\infty) = \langle {}^*1, {}^*0, {}^*0\rangle$,
$\mathrm{val}_I^{\infty,
INL}(\varphi^\infty\rightarrow_{INL}\psi^\infty) = \langle
\min({}^*1, {}^*1 - t(\varphi^\infty) + t(\psi)) = {}^*1,
\max({}^*0, i(\psi^\infty) - i(\varphi^\infty)) = {}^*0,
\max({}^*0, f(\psi^\infty) - f(\varphi^\infty)) = {}^*0\rangle$.
Hence, $t(\psi^\infty) = {}^*1$, $i(\psi^\infty) = f(\psi^\infty)
= {}^*0$. So $\psi^\infty$ is an $INL$-tautology.\hfill $\Box$\\

The following axiom schemata for $\mathbf{INL}$ were regarded in
\cite{Wang}.

\begin{eqnarray} \psi^\infty \rightarrow_{INL} (\varphi^\infty \rightarrow_{INL} \psi^\infty),
\end{eqnarray}
\begin{eqnarray} (\psi^\infty\wedge_{INL}\varphi^\infty) \rightarrow_{INL} \varphi^\infty,
\end{eqnarray}
\begin{eqnarray} \psi^\infty\rightarrow_{INL} (\psi^\infty \vee_{INL} \varphi^\infty),
\end{eqnarray}
\begin{eqnarray} \psi^\infty \rightarrow_{INL} (\varphi^\infty \rightarrow_{INL} (\psi^\infty\wedge_{INL}\varphi^\infty)),
\end{eqnarray}
\begin{eqnarray}(\psi^\infty \rightarrow_{INL} \chi^\infty) \rightarrow_{INL} ((\varphi^\infty \rightarrow_{INL} \chi^\infty) \rightarrow_{INL} \\ {}\nonumber((\psi^\infty \vee_{INL} \varphi^\infty)\rightarrow_{INL} \chi^\infty)), \end{eqnarray}
\begin{eqnarray}
((\psi^\infty \vee_{INL} \varphi^\infty) \rightarrow_{INL}
\chi^\infty) \leftrightarrow_{INL} \\ {}\nonumber((\psi^\infty
\rightarrow_{INL} \chi^\infty)\wedge_{INL} (\varphi^\infty
\rightarrow_{INL} \chi^\infty)),
\end{eqnarray}
\begin{eqnarray}(\psi^\infty \rightarrow_{INL} \varphi^\infty) \leftrightarrow_{INL} (\neg_{INL} \varphi^\infty \rightarrow_{INL} \neg_{INL}
\psi^\infty),
\end{eqnarray}
\begin{eqnarray} (\psi^\infty \rightarrow_{INL} \varphi^\infty) \wedge_{INL} (\varphi^\infty \rightarrow_{INL} \chi^\infty) \rightarrow_{INL} (\psi^\infty
\rightarrow_{INL} \chi^\infty), \end{eqnarray}
\begin{eqnarray} (\psi^\infty \rightarrow_{INL} \varphi^\infty) \leftrightarrow_{INL} \\ {}\nonumber(\psi^\infty \leftrightarrow_{INL} (\psi^\infty \wedge_{INL} \varphi^\infty)) \leftrightarrow_{INL} \\ {}\nonumber(\varphi^\infty \rightarrow_{INL} (\psi^\infty \vee_{INL} \varphi^\infty)).\end{eqnarray}

The only inference rule of $\mathbf{INL}$ is modus ponens.\\

We can take also the non-Archimedean case of axiom schemata
\eqref{shuma1} -- \eqref{shuma2} for the axiomatization of
$\mathbf{INL}$, because $\mathbf{INL}$ is a generalization of
non-Archimedean valued \textsc{{\L}ukasiewicz}'s logic (see the
previous section). This means that we can also set $\mathbf{INL}$
as generalization of non-Archimedean valued \textsc{G\"{o}del}'s,
Product or \textsc{Post}'s logics.

\chapter{Conclusion}

The informal sense of \textsc{Archimedes}' axiom is that anything
can be measured by a ruler. The negation of this axiom allows to
postulate infinitesimals and infinitely large integers and, as a
result, to consider non-wellfounded and neutrality phenomena. In
this book we examine the non-Archimedean fuzziness, i.e.\
fuzziness that runs over the non-Archimedean number systems. We
show that this fuzziness is constructed in the framework of the
t-norm based approach. We consider two cases of the
non-Archimedean fuzziness: one with many-validity in the interval
$[0,1]$ of hypernumbers and one with many-validity in the ring
$\mathbf{Z}_p$ of $p$-adic integers. This fuzziness has a lot of
practical applications, e.g. it can be
used in non-Kolmogorovian approaches to probability theory.\\

Non-Archimedean logic is constructed on the base of infinite DSm
models. Its instances are the following multi-valued logics:

\begin{itemize}
    \item hyperrational valued \textsc{{\L}u\-kasiewicz}'s,
\textsc{G\"{o}\-del}'s, and Product logics\index{hyperrational
valued \textsc{{\L}u\-kasiewicz}'s logic}\index{hyperrational
valued \textsc{G\"{o}\-del}'s logic}\index{hyperrational valued
Product logic},
    \item hyperreal valued
\textsc{{\L}u\-kasiewicz}'s\index{hyperreal valued
\textsc{{\L}u\-kasiewicz}'s logic},
\textsc{G\"{o}\-del}'s\index{hyperreal valued
\textsc{G\"{o}\-del}'s logic}, and Product logics\index{hyperreal
valued Product logic},
    \item $p$-adic valued \textsc{{\L}u\-kasie\-wicz}'s\index{$p$-adic
valued  \textsc{{\L}u\-kasiewicz}'s logic},
\textsc{G\"{o}\-del}'s\index{$p$-adic valued
\textsc{G\"{o}\-del}'s logic}, and \textsc{Post}'s
logics\index{$p$-adic valued \textsc{Post}'s logic}.
\end{itemize}

These systems can be used in probabilistic and fuzzy reasoning.\\

Hyper-valued (resp.\ $p$-adic valued) interval neutrosophic logic
$\mathbf{INL}$ by which we can describe neutrality phenomena is an
extension of non-Archimedean valued fuzzy logic that is obtained
by adding a truth triple $\langle t, i, f\rangle\in ({}^*[0,1])^3$
(resp.\ $\langle t, i, f\rangle\in (\mathbf{Z}_p)^3$) instead of
one truth value $t\in {}^*[0,1]$ (resp.\ $t\in\mathbf{Z}_p$) to
the formula valuation, where $t$ is a truth-degree, $i$ is an
indeterminacy-degree, and $f$ is
a falsity-degree.\\

\printindex


\end{document}